\documentclass{article}%
\usepackage{amssymb}
\usepackage{amsmath}
\usepackage{amsfonts}
\usepackage{graphicx}%
\input{diagrams.tex}
\begin{document}

\title{Poisson-Lie T-Duality and non trivial monodromies}
\author{A. Cabrera *, H. Montani$^{\dag}$ \& M. Zuccalli*\\\\
\and * Departamento de Matem\'{a}ticas, Universidad de La Plata,
\and Calle 50 esq. 115 (1900) La Plata, Argentina
\and $\dag$ Centro At\'{o}mico Bariloche and Instituto Balseiro\\(8400) S. C. de Bariloche, R\'{\i}o Negro, Argentina}
\maketitle

\begin{abstract}
We describe a general framework for studying duality among different phase
spaces which share the same symmetry group $\mathrm{H}$. Solutions
corresponding to collective dynamics become dual in the sense that they are
generated by the same curve in $\mathrm{H}$. Explicit examples of phase spaces
which are dual with respect to a common non trivial coadjoint orbit
$\mathcal{O}_{c,0}\left(  \mathbf{\alpha},1\right)  \subset\mathfrak{h}^{\ast
}$ are constructed on the cotangent bundles of the factors of a double Lie
group $\mathrm{H}=\mathrm{N}\Join\mathrm{N}^{\ast}$. In the case
$\mathrm{H}=LD$, the loop group of a Drinfeld double Lie group $D$, a
hamiltonian description of Poisson-Lie T-duality for non trivial monodromies
and its relation with non trivial coadjoint orbits is obtained.

\end{abstract}

\vskip 1cm

\tableofcontents

%\tableofcontents

\section{Introduction}

Poisson-Lie T-duality \cite{KS-1} refers to a non-Abelian duality between two
$1+1$ dimensional $\sigma$-models describing the motion of a string on targets
which are a dual pair of Poisson-Lie groups. The lagrangians of the models are
written in terms of the underlying bialgebra structure of the Lie groups, and
Poisson-Lie T-duality stems from the self dual character the Drinfeld double.
Indeed, classical T-duality transformation relates some \emph{dualizable
}subspaces of the associated phase spaces, mapping solutions reciprocally.
Many different approaches have revealed the canonical character of these
transformations \cite{KS-1} \cite{Lozano} \cite{Alvarez-npb} \cite{Sfetsos}
\cite{Stern-1}. There are also WZNW-type models with target on the associated
Drinfeld double group $D$ whose dynamics encodes in some way the equivalent
sigma models on the factors \cite{KS-1} \cite{Alek-Klim}.

In reference \cite{Ale-Hugo}, most of the known facts of T-duality were
embodied in a purely hamiltonian approach, offering a unified description of
its classical aspects based on the symplectic geometry of the underlying loop
group phase spaces.\ There, PL T-duality is realized via momentum maps from
the cotangent bundles of both the factor Lie groups to a pure central
extension coadjoint orbit of the Drinfeld double. These moment maps are
associated to the dressing symmetry inherited from the double Lie group
structure. Since the hamiltonians corresponding to the $\sigma$-models are in
\emph{collective form}, the pre-images of coadjoint orbits through the moment
maps allows us to identify the \emph{dualizable }spaces that are preserved by
the dynamics. On the other hand, this orbits are symplectomorphic to the
reduced phase space of one of the chiral sectors of the WZNW model with
trivial monodromy, making clear the geometric content of the relation between
the cotangent bundle of $D$ and its factors. These facts are encoded in the
commutative diagram%
\begin{equation}
\begin{diagram}[h=1.9em] &&& &(L{\mathfrak{d}}_{\Gamma }^{\ast};\;\{,\}_{KK})&&&&&\\ &&\ruTo^{\mu} && &&\luTo^{\tilde{\mu}} && &\\ (T^{\ast }LG;\omega _{o}) &&& &\uTo^{\hat{\Phi}}& &&& (T^{\ast }{LG^{\ast }};\;\tilde{\omega} _{o})&\\ &&\luTo &&&& \ruTo&& &\\ & &&& (\Omega D;\omega_{\Omega D} )&&& &&\\ \end{diagram} \label{t-duality-trivial mon}%
\end{equation}
where the left and right vertices are the phases spaces of the $\sigma
$-models, with the canonical Poisson (symplectic) structures, $L\mathfrak{d}%
_{\Gamma}^{\ast}$ is the dual of the centrally extended Lie algebra of $LD$
with the Kirillov-Kostant Poisson structure, and $\Omega D$ is the symplectic
manifold of based loops. In particular, $\mu$ and $\tilde{\mu}$ are derived as
momentum maps associated to hamiltonian actions of the centrally extended loop
group $LD_{c,0}$ on the $\sigma$-model phase spaces. The subsets which can
related by T-duality are the pre images under $\mu$ and $\tilde{\mu}$ of the
coadjoint orbit $\mathcal{O}(0,1)\simeq\Omega D$ where both momentum maps
intersect and the \emph{dualizable subspaces} are identified as the orbits of
$\Omega D$, their symplectic foliation. From this setting, we were able to
build dual hamiltonian models by taking any suitable hamiltonian function on
the loop algebra of the double and lifting it in a\emph{ collective form}
\cite{Guill-Sten}. For particular choices, the lagrangian formalism is
reconstructed obtaining the known dual sigma and WZNW-like model models.

The present work is devoted to the extension of the framework developed in
\cite{Ale-Hugo} in order to include T-duality based on non trivial coadjoint
orbits of the form $\mathcal{O}(\alpha,1)$, leading to meaningful changes as a
consequence of the non trivial monodromies.

To that end, we describe a general setting for studying duality between
different hamiltonian systems with phase spaces $P$ and $\tilde{P}$ linked by
a common hamiltonian $\mathrm{H}$-action. This duality is \emph{supported} on
some coadjoint orbit $\mathcal{O}\subset\mathfrak{h}^{\ast}$ where the
corresponding momentum maps intersect cleanly. Thus, diagram $\left(
\ref{t-duality-trivial mon}\right)  $ becomes a special case of this situation.

Motivated by the previous T-duality investigations, we work out examples built
on the cotangent bundles of a double Lie group $\mathrm{H}=\mathrm{N}%
\Join\mathrm{N}^{\ast}$ and on its factors $\left(  \mathrm{N},\mathrm{N}%
^{\ast}\right)  $. As a corner stone of the T-duality scheme we choose a non
trivial coadjoint orbit $\mathcal{O}(\alpha,1)$ of the centrally extended
group $\mathrm{H}$, with $\alpha\mathbf{\ }$in either $\mathfrak{n}$ or
$\mathfrak{n}^{\ast}$. This orbit, in turn, can be related to a coboundary
shifted trivial one. We find out some actions of $\mathrm{H}$ on the cotangent
bundles of the factors $\left(  \mathrm{N},\mathrm{N}^{\ast}\right)  $ whose
associated equivariant momentum maps serve as the linking arrows with the
coadjoint orbits mentioned before. This gives the basic structure underlying
T-duality. Hence, collective dynamics completes the approach introducing the
appropriate dynamics. However, we shall see that since the symmetric role
played by the factors in the case described by diagram $\left(
\ref{t-duality-trivial mon}\right)  $ no longer holds, richer T-dual models
emerge. Later, all these is applied in the case $\mathrm{H}=LD$ with $D=G\Join
G^{\ast}$, driving in a constructive way to some previously studied models.

This work is organized as follows: in Section \ref{sec: setting}, we describe
a general setting for studying duality based on a common hamiltonian
$G$-action and give some simple examples; in Section
\ref{sec: centrally exended Lie}, we present the general geometric framework
linking coadjoint orbits on $\mathrm{H}$ and the phase spaces on its factors
$\left(  \mathrm{N},\mathrm{N}^{\ast}\right)  $. The role played by central
extensions is also analyzed, and the symmetry actions leading to the relevant
momentum maps are constructed. In Section \ref{subsec: duality}, all the
kinematic aspects studied before are condensed into a new PL T-duality scheme
based on collective dynamics. The construction of the resulting $\mathrm{H}%
$-hamiltonian systems and the associated lagrangians is addressed in Section
\ref{sec: hamiltonian asociados}. In Section \ref{sec: Loops}, we illustrate
the previous developments for $\mathrm{H}=LD$, with $D=G\Join G^{\ast}$,
discussing some properties of the models. Finally, some conclusions and
comments are condensed in the last Section \ref{sec: conclusions}.

\section{Setting for duality and the diagram}

\label{sec: setting}In this section, we generalize the geometrical framework
behind $T$-duality which was described in \cite{Ale-Hugo}.

Let us consider a Poisson manifold $(P,\{,\}_{P})$ on which a Lie group $G$
acts by canonical transformations. Suppose further that the action is
hamiltonian with $Ad^{\ast}$-equivariant moment map $J:P\longrightarrow
\mathfrak{g}^{\ast}$, where $\mathfrak{g}$ denotes the Lie algebra of $G$.
Recall that $J:P\longrightarrow\mathfrak{g}^{\ast}$ is a Poisson map for the
($+$) Kirillov-Kostant Poisson bracket $\{,\}_{KK}$ on $\mathfrak{g}^{\ast}$.
Following \cite{Guill-Sten}, we define

\begin{description}
\item[Definition:] \textit{We say that the }$G$-\textit{hamiltonian system
}$(P,\{,\}_{P},G,J,H)$\textit{, with Hamilton function }$\mathcal{H}%
:P\longrightarrow\mathbb{R}$\textit{, has dynamics of \emph{collective motion}
type if }$\mathcal{H}$\textit{ is given by the composition}%
\[
\mathcal{H}=\mathsf{h}\circ J
\]
\textit{with }$\mathsf{h}:\mathfrak{g}^{\ast}\longrightarrow\mathbb{R}%
$\textit{.}
\end{description}

For this kind of systems, the dynamics is confined in a coadjoint orbit, as it
is stated in the next theorem.

\begin{description}
\item[Theorem:] \textit{Let} $(P,\{,\}_{P},G,J,\mathsf{h}\circ J)$\textit{ be
a collective }$G$-\textit{hamiltonian system. Then for the initial value}
$p_{0}\in P$ \textit{s.t.} $J(p_{0})=J_{0}\in\mathfrak{g}^{\ast}$ \textit{the
solution} $p(t)$ \textit{of the Hamilton equations of motion in} $P$
\textit{is given by}%
\[
p(t)=g(t)\cdot p_{0}%
\]
\textit{with} $g(t)\in G$ \textit{such that}%
\begin{align*}
\dot{g}g^{-1}  &  =d\mathsf{h}_{\xi(t)}\\
g(0)  &  =e
\end{align*}
\textit{where} $\xi(t)\in\mathfrak{g}^{\ast}$ \textit{is the solution of the
Hamilton equations on} $\mathfrak{g}^{\ast}$
\begin{align*}
\dot{\xi}(t)  &  =-ad_{d\mathsf{h}_{\xi(t)}}^{\ast}\xi(t)\\
\xi(0)  &  =J_{0}.
\end{align*}

\end{description}

The above result can be summarized in the following diagram%

\begin{equation}
\begin{diagram}[h=1.9em] (P,\{,\}_{P},\mathsf{h}\circ J)&&&& (T^{\ast}G,\omega_{o},\mathsf{h}\circ s) &\\ &\rdTo(4,3)_{J}&&&\dTo_{s}&\\ &&&&\\ &&&& (\mathfrak{g}^{\ast},\{,\}_{KK},\mathsf{h})&\\ \end{diagram} \label{Eq: diag P-g}%
\end{equation}
where $s(\alpha_{g})=R_{g}^{\ast}\alpha_{g}$ for $\alpha_{g}\in T_{g}^{\ast}G$
is the momentum map associated to the lifting to $T^{\ast}G$ of the left
action $L_{g}$ of $G$ on itself and $R_{g}$ denotes the right translation in
$G$. The map $s$ is a Poisson map in relation with the Kirillov-Kostant
Poisson bracket on $\mathfrak{g}^{\ast}$ defined as $\{F,H\}_{KK}\left(
\eta\right)  =+\left\langle \eta,\left[  dF,dH\right]  \right\rangle $.

\begin{description}
\item[Remark:] \emph{(Groupoid actions)}\ \textit{In fact, }$T^{\ast
}G\rightrightarrows\mathfrak{g}^{\ast}$\textit{ is a symplectic groupoid
integrating the Poisson manifold }$(\mathfrak{g}^{\ast},\{,\}_{KK})$\textit{
(\cite{Grupoids}) and any such diagram as above with complete Poisson }%
$J$\textit{, defines a }$T^{\ast}G$-\textit{groupoid action on }$P$\textit{.
In this case, it coincides with a usual }$G$-\textit{action. As in the above
proposition, solutions in }$P$\textit{ for collective }$\mathcal{H}$\textit{
are given by this groupoid action }$\alpha_{g}(t)\cdot p_{0}$\textit{ with
}$\alpha_{g}(t)$\textit{ a solution to the corresponding collective
hamiltonian eqs. on }$(T^{\ast}G,\omega_{0},\mathsf{h}\circ s)$.
\end{description}

Notice that, if $(\tilde{P},\{,\}_{\tilde{P}},G,\tilde{J},\mathsf{h}%
\circ\tilde{J})$ is another collective hamiltonian $G$-system, we would have
the analogous diagram to $\left(  \ref{Eq: diag P-g}\right)  $ so we can glue
both of them yielding
\begin{equation}
\begin{diagram}[h=1.9em] (P,\{,\}_{P},\mathsf{h}\circ J) &&&(T^{\ast}G,\omega_{0},\mathsf{h}\circ s) &&& (\tilde{P},\{,\}_{\tilde{P}},\mathsf{h}\circ\tilde{J})\\ &\rdTo_{J} &&\dTo_{s}&& \ldTo_{\tilde{J}}& \\ &&&&&&\\ &&& (\mathfrak{g}^{\ast},\{,\}_{KK},\mathsf{h})&&&\\ \end{diagram} \label{Diag-gen}%
\end{equation}
If both systems share some non empty set in the images of the corresponding
momentum maps, then part of its dynamics can be described in a unified way, as
stated in the next proposition.

\begin{description}
\item[Proposition:] \textit{Let} $(P,\{,\}_{P},G,J,\mathsf{h}\circ J)$
\textit{and} $(\tilde{P},\{,\}_{\tilde{P}},G,\tilde{J},\mathsf{h}\circ
\tilde{J})$ \textit{be collective} $G-$hamiltonian\textit{ systems such that}
$ImJ\cap Im\tilde{J}\neq\emptyset$. \textit{Then, for} $J_{0}\in ImJ\cap
Im\tilde{J}\subset\mathfrak{g}^{\ast}$ \textit{and} $p_{0}\in J^{-1}(J_{0}),$
$\tilde{p}_{0}\in\tilde{J}^{-1}(J_{0})$, \textit{the solutions }$p(t)$
\textit{and} $\tilde{p}(t)$ \textit{corresponding to the initial values}
$p_{0}$, $\tilde{p}_{0}$ \textit{for the hamiltonian equations on}
$(P,\{,\}_{P},\mathsf{h}\circ J)$ \textit{and} $(\tilde{P},\{,\}_{\tilde{P}%
},\mathsf{h}\circ\tilde{J})$\textit{, respectively, are given by}%
\begin{align*}
p(t)  &  =g(t)\cdot p_{0},\\
\tilde{p}(t)  &  =g(t)\cdot\tilde{p}_{0}%
\end{align*}
\textit{where} $g(t)\in G$ \textit{is the curve solution to}%
\begin{align}
\dot{g}g^{-1}  &  =d\mathsf{h}_{\xi(t)}\label{Eq: g(t)}\\
g(0)  &  =e\nonumber
\end{align}
\textit{with }$\xi(t)\in\mathfrak{g}^{\ast}$ \textit{the solution of the
hamiltonian equations on} $\mathfrak{g}^{\ast}$
\begin{align*}
\dot{\xi}  &  =-ad_{d\mathsf{h}_{\xi(t)}}^{\ast}\xi(t)\\
\xi(0)  &  =J_{0}.
\end{align*}

\end{description}

Thus both solutions on $P$ and $\tilde{P}$, with compatible initial
conditions, are obtained from the \emph{same} curve $g(t)$ in $G$. So, if we
had one of them (say $p(t)$) we can map it through $J$ to $\mathfrak{g}^{\ast
}$ and then, by solving eq. $\left(  \ref{Eq: g(t)}\right)  $, obtain the
other solution $\tilde{p}(t)=g(t)\cdot\tilde{p}_{0}$. Motivated by field
theory applications we do the following definition:

\begin{description}
\item[Definition:] \textit{Two collective }$G$-\textit{hamiltonian systems}
$(P,\{,\}_{P},G,J,\mathsf{h}\circ J)$ \textit{and} $(\tilde{P},\{,\}_{\tilde
{P}},G,\tilde{J},\mathsf{h}\circ\tilde{J})$ \textit{with the same collective}
\textit{function }$\mathsf{h}$\textit{ and such that there exist a subspace}
$\mathcal{O}\subset ImJ\cap Im\tilde{J}$, \textit{are said to be
}\textbf{dual} \textit{to each other }\textbf{with respect to} $\mathcal{O}$.
\end{description}

When $ImJ\cap Im\tilde{J}=\emptyset$, then duality is trivial. In general,
$ImJ\cap Im\tilde{J}$ is a disjoint union of \emph{coadjoint orbits} in
$\mathfrak{g}^{\ast}$.

\begin{description}
\item[Remark: \emph{(Transitivity)}] \textit{Note that being} \emph{dual with
respect to }\textit{a certain fixed subspace} $\mathcal{O}\subset
\mathfrak{g}^{\ast}$ \textit{is a transitive property.}
\end{description}

In the following sections, we shall apply this setting to systems coming from
classical field theory. Meanwhile, we show some applications in simple examples.

\begin{description}
\item[Example: \emph{(Rigid body dual to a Pendulum)}] \textbf{ }\textit{This
example comes from} \textit{\cite{MR}}. \textit{The hamiltonian system}
$(T^{\ast}SE(2),\omega_{0},\mathsf{h}\circ J)$ \textit{describes the motion of
a rigid body where }$\omega_{0}$\textit{ denotes the standard symplectic
structure on }$T^{\ast}SE(2)$\textit{ and }$J$\textit{ the moment map
corresponding to the lifted left }$SE(2)$\textit{ action. The collective
function }$\mathsf{h}:se(2)^{\ast}\longrightarrow\mathbb{R}$\textit{ is}
\[
\mathsf{h}(\beta)=\frac{1}{2}\left\langle \beta,\mathbf{N}\beta\right\rangle
\]
\textit{for} $\mathbf{N}:se(2)^{\ast}\simeq\mathbb{R}^{3}\longrightarrow
\mathbb{R}^{3}$ \textit{given by the matrix}%
\[
\mathbf{N}=\left(
\begin{array}
[c]{ccc}%
0 & 0 & 0\\
0 & c\left(  \frac{1}{I_{1}}-\frac{1}{I_{2}}\right)  & 0\\
0 & 0 & 1
\end{array}
\right)
\]
\textit{where} $c=\left(  \frac{1}{I_{1}}-\frac{1}{I_{3}}\right)  ^{-1}$
\textit{and }$I_{1}<I_{2}<I_{3}$ \textit{are the principal moments of inertia
of the underlying rigid body. Now, consider the hamiltonian system} $(T^{\ast
}S^{1},\tilde{\omega},\mathsf{h}\circ\tilde{J})$ \textit{where }$\tilde
{\omega}$ \textit{denotes} $(k_{1}k_{2})$ \textit{times the standard
symplectic structure on }$T^{\ast}S^{1}$, \textit{with the momentum map
}$\tilde{J}:T^{\ast}S^{1}\longrightarrow se(2)^{\ast}$\textit{ being} \textit{
}%
\[
\tilde{J}(\theta,p)=\left(  r\ sin\theta,r\ cos\theta,k_{1}k_{2}\ p\right)
\]
\textit{Here}
\[%
\begin{array}
[c]{ccccc}%
\dfrac{1}{k_{1}^{2}}=\dfrac{1}{I_{1}}-\dfrac{1}{I_{3}} & ~~,~~ & \dfrac
{1}{k_{2}^{2}}=\dfrac{1}{I_{2}}-\dfrac{1}{I_{3}} & ~~,~~ & \dfrac{1}{k_{3}%
^{2}}=c\left(  \dfrac{1}{I_{1}}-\dfrac{1}{I_{2}}\right)
\end{array}
\]
\textit{and} $r=\sqrt{2K}$\textit{ denotes the constant value of the
function}
\[
K(\beta)=\frac{1}{2}\left(  \frac{1}{I_{1}}-\frac{1}{I_{3}}\right)  \beta
_{1}^{2}+\frac{1}{2}\left(  \frac{1}{I_{2}}-\frac{1}{I_{3}}\right)  \beta
_{2}^{2}\mathit{.}%
\]
\textit{ From the dual Hamilton equations on }$T^{\ast}S^{1}$\textit{, one
easily arrives to the following equation for }$\theta(t)$
\[
\frac{d^{2}}{dt^{2}}\theta=-K\left(  \frac{1}{I_{1}}-\frac{1}{I_{2}}\right)
sin(2\theta)
\]
\textit{which is the equation for the motion of a pendulum with angle }%
$\theta$\textit{. It thus follows from the above considerations that the rigid
body hamiltonian system is \emph{dual} to a pendulum hamiltonian system}
\emph{with respect to} $\mathcal{O}:=Im\tilde{J}\subset se(2)^{\ast}$.
\end{description}

\begin{description}
\item[Example:\emph{(Coadjoint orbits)}] \emph{ }\textit{Generalizing the
previous example, suppose that} $(P,\{,\}_{P},G,J,\mathsf{h}\circ
J)\ $\textit{is a} $G$\textit{-hamiltonian system such that the coadjoint
orbit} $\mathcal{O}_{\mu}\subset ImJ$. \textit{Then the following diagram}%
\[
\begin{diagram}[h=1.9em]
(P,\{,\}_{P},\mathsf{h}\circ J) &&&& (\mathcal{O}_{\mu},\omega_{\mu}%
^{KK},\mathsf{h}\circ i)\\
%&&&&\\
&\rdTo_{J} && \ldTo_{i}& \\
%&&&&\\
&& (\mathfrak{g}^{\ast},\{,\}_{KK},\mathsf{h})&&\\ \end{diagram}
\]
\textit{says that} $(P,\{,\}_{P},\mathsf{h}\circ J)$ \textit{is} \emph{dual to
}$(\mathcal{O}_{\mu},\omega_{\mu},\mathsf{h}\circ i)$ \emph{with respect to
}$\mathcal{O}_{\mu}\subset\mathfrak{g}^{\ast}$\textit{, where} $i:\mathcal{O}%
_{\mu}\hookrightarrow\mathfrak{g}^{\ast}$ \textit{denotes the inclusion and
}$\omega_{\mu}^{KK}$ \textit{the Kirillov-Kostant symplectic structure on the
coadjoint orbit }$\mathcal{O}_{\mu}$ \textit{corresponding to} $\{,\}_{KK}$ in
$\mathfrak{g}^{\ast}$.

\item[Example: \emph{(Dual groups)}] \label{ex: dual groups}\textbf{
}\textit{Suppose that }$(\mathrm{H},\mathrm{N},\mathrm{N}^{\ast})$ \textit{are
a triple of Lie groups for which} $\mathrm{H}$ \textit{is the corresponding
perfect Drinfeld double of the Poisson-Lie groups }$\mathrm{N}$ \textit{and}
$\mathrm{N}^{\ast}$ \textit{\cite{Drinfeld-1}. Moreover, suppose that there is
a one cocycle}%
\[
C:\mathrm{H}\longrightarrow\mathfrak{\mathrm{h}}^{\ast}%
\]
i.e. a map \textit{satisfying}
\[
C\left(  lk\right)  =Ad_{l^{-1}}^{\mathrm{H}\ast}C\left(  k\right)  +C\left(
l\right)  ,\ \ l,k\in\mathrm{H}%
\]
with the additional property $C(\mathrm{N}\subseteq\mathrm{H})\subseteq
\mathfrak{\mathrm{n}}^{\circ}:=\mathfrak{\mathrm{n}}$\textit{-annihilator in}
$\mathfrak{\mathrm{h}}^{\ast}$ and $C(\mathrm{N}^{\ast}\subseteq
\mathrm{H})\subseteq\mathfrak{\mathrm{n^{\ast}}}^{\circ}:=\mathfrak{\mathrm{n}%
}^{\ast}$\textit{-annihilator in }$\mathfrak{\mathrm{h}}^{\ast}$\textit{.
Then, we can consider the following diagram corresponding to} $\left(
\ref{Diag-gen}\right)  $%
\[
\begin{diagram}[h=1.9em]
(T^{\ast}N,\omega_{o},\mathsf{h}\circ J) &&&(T^{\ast}H,\omega
_{C},\mathsf{h}\circ s) &&& (T^{\ast}N^{\ast},\tilde{\omega}_{o}%
,\mathsf{h}\circ\tilde{J})\\
%&&&&&&&&\\
&\rdTo_{J} &&\dTo_{s}&& \ldTo_{\tilde{J}}& \\
&&&&&&\\
&&& (\mathfrak{h}^{\ast},\{,\}_{C},\mathsf{h})&&&\\ \end{diagram}
\]
The maps involved are $J(g,\alpha)=C(g)+Ad_{g}^{\mathrm{H}\ast}\alpha$ and
$\tilde{J}(\tilde{g},X)=C(\tilde{g})+Ad_{\tilde{g}}^{\mathrm{H}\ast}X$. The
structures are: $\omega_{C}$ \textit{the right invariant} $C$-\textit{modified
symplectic structure \cite{Harnad-1},} $\{,\}_{C}$ \textit{the affine Poisson
structure on }$\mathfrak{\mathrm{h}}^{\ast}$ \textit{defined by }$C$\textit{.}
\textit{With these, }$s(d,\xi)=\xi$ is \textit{the (source) Poisson map.
Notice that the orbit of the }$C$\textit{-affine coadjoint action on
}$\mathfrak{\mathrm{h}}^{\ast}$\textit{ through} $0$ \textit{is }%
$C(\mathrm{H})\subset\mathfrak{\mathrm{h}}^{\ast}$ (see below)\textit{ and
gives the intersection }$ImJ\cap Im\tilde{J}$. \textit{So,} $(T^{\ast
}\mathrm{N},\omega_{o},\mathsf{h}\circ J)$ \textit{and }$(T^{\ast
}\mathrm{N^{\ast}},\tilde{\omega}_{o},\mathsf{h}\circ\tilde{J})$ \textit{are
}\emph{duals} \textit{to each other (and, hence, also to} $(T^{\ast}%
\mathrm{H},\omega_{C},\mathsf{h}\circ s)$) \emph{with respect to}
$\mathcal{O}:=C(\mathrm{H})$. \textit{See also \cite{Ale-Hugo}.}
\end{description}

\section{Phase spaces on Lie groups, central extensions and double Lie
groups.}

\label{sec: centrally exended Lie}In this section, we elaborate on the
structure proposed in the previous section (the last example above) for
describing duality on non-trivial coadjoint orbits. Dual phase spaces are
built from the factors $\mathrm{N\ }$and $\mathrm{N}^{\ast}$ of a (perfect)
double Lie group $\mathrm{H}=\mathrm{N}\Join\mathrm{N}^{\ast}$, and the
corresponding $\mathrm{H}$-action on them is constructed by means a certain
Lie algebra $\mathfrak{h}$-cocycle. In contrast with ref. \cite{Ale-Hugo}, the
symmetric role that the factors $\mathrm{N\ }$and $\mathrm{N}^{\ast}$ played
in the duality formulation is \emph{broken} by considering solutions
associated to an element\footnote{The case studied in \cite{Ale-Hugo}
corresponds to the $\alpha=0$, i.e., \emph{trivial monodromy solutions }case.}
$\alpha\in\mathfrak{n}^{\ast}$.

\subsection{Chiral WZNW type phase spaces}

\label{sec: WZW phase space}Let us begin recalling some results of
\cite{Harnad-1} with explicit considerations for coboundary modified cocycles.
Let $\mathrm{H}$ be a Lie group and $T^{\ast}\mathrm{H}\sim\mathrm{H}%
\times\mathfrak{h}^{\ast}$ its cotangent bundle trivialized by left
translations. We consider on it the canonical $1$-form $\vartheta_{o}$ and the
symplectic form $\omega_{o}=-d\vartheta_{o}$, which on vectors $(\mathbf{v}%
,\mathbf{\xi}),(\mathbf{w},\mathbf{\lambda})\in T_{(l,\mathbf{\eta})}\left(
\mathrm{H}\times\mathfrak{h}^{\ast}\right)  =T_{l}\mathrm{H}\times
\mathfrak{h}^{\ast}$ is
\begin{equation}
\langle\mathbf{\omega}_{o},(\mathbf{v},\mathbf{\xi})\otimes(\mathbf{w}%
,\mathbf{\lambda})\rangle_{(l,\mathbf{\eta})}=-\langle\mathbf{\xi}%
,l^{-1}\mathbf{w}\rangle+\langle\mathbf{\lambda},l^{-1}\mathbf{v}%
\rangle+\langle\mathbf{\eta},[l^{-1}\mathbf{v},l^{-1}\mathbf{w}]\rangle
\label{canon}%
\end{equation}

A new symplectic structure can be obtained by adding a two cocycle
$c:\mathfrak{h}\otimes\mathfrak{h}\longrightarrow\mathbb{R}$, derived from an
$Ad^{\ast}$-cocycle $C:\mathrm{H}\longrightarrow\mathfrak{h}^{\ast}$,
characterized by $c\left(  Ad_{g}\mathbf{X},Ad_{g}\mathbf{Y}\right)  =c\left(
\mathbf{X},\mathbf{Y}\right)  +\left\langle C\left(  g^{-1}\right)  ,\left[
\mathbf{X},\mathbf{Y}\right]  \right\rangle $ for all $\mathbf{X}%
,\mathbf{Y}\in\mathfrak{h}$. Also recall that $C\left(  lk\right)
=Ad_{l^{-1}}^{\mathrm{H}\ast}C\left(  k\right)  +C\left(  l\right)  $ and
$\hat{c}\equiv-\left.  dC\right\vert _{e}:\mathfrak{h}\longrightarrow
\mathfrak{h}^{\ast}$ produces $c\left(  \mathbf{X},\mathbf{Y}\right)
\equiv\left\langle \hat{c}\left(  \mathbf{X}\right)  ,\mathbf{Y}\right\rangle
$. In the remaining, we fix $C$ and consider its \emph{shifting} by a
\emph{coboundary} $B_{\mathbf{\theta}}$ defined by $\mathbf{\theta}%
\in\mathfrak{h}^{\ast}$ as
\[
B_{\mathbf{\theta}}\left(  l\right)  =Ad_{l^{-1}}^{\ast}\mathbf{\theta
}-\mathbf{\theta}%
\]
defining the following shifted cocycle $C_{\mathbf{\theta}}$ and two cocycle
$c_{\mathbf{\theta}}$,%
\begin{align*}
C_{\mathbf{\theta}}\left(  l\right)   &  =C\left(  l\right)  -Ad_{l^{-1}%
}^{\mathrm{H}\ast}\mathbf{\theta}+\mathbf{\theta}\\
c_{\mathbf{\theta}}\left(  \mathbf{X},\mathbf{Y}\right)   &  =c\left(
\mathbf{X},\mathbf{Y}\right)  -\left\langle \mathbf{\theta},\left[
\mathbf{X},\mathbf{Y}\right]  \right\rangle .
\end{align*}
The \emph{extended }symplectic form $\omega_{c,\mathbf{\theta}}$ on $T^{\ast
}\mathrm{H}$, for $(\mathbf{v},\mathbf{\xi}),(\mathbf{w},\mathbf{\lambda})\in
T_{(l,\mathbf{\eta})}\left(  \mathrm{H}\times\mathfrak{h}^{\ast}\right)
=T_{l}\mathrm{H}\times\mathfrak{h}^{\ast}$ is given by%
\begin{equation}
\langle\mathbf{\omega}_{c,\mathbf{\theta}},(\mathbf{v},\mathbf{\xi}%
)\otimes(\mathbf{w},\mathbf{\lambda})\rangle_{(l,\mathbf{\eta})}%
=\langle\mathbf{\omega}_{o},(\mathbf{v},\mathbf{\xi})\otimes(\mathbf{w}%
,\mathbf{\lambda})\rangle_{(l,\mathbf{\eta})}-c_{\mathbf{\theta}}\left(
\mathbf{v}l^{-1}\,,\mathbf{w}l^{-1}\right)  \label{w-c-2}%
\end{equation}
which is invariant under right translations of $\mathrm{H}$. This symplectic
manifold $\left(  \mathrm{H}\times\mathfrak{h}^{\ast},\mathbf{\omega
}_{c,\mathbf{\theta}}\right)  $ is related to the phase space of chiral modes
of the WZNW models when loops groups are considered.

Now, consider the \emph{extended coadjoint action} $\widehat{Ad}%
_{\mathbf{\theta};}^{\mathrm{H}\ast}$of the corresponding centrally extended
group $\mathrm{H}_{c,\mathbf{\theta}}$ on $\mathfrak{h}_{c,\mathbf{\theta}%
}^{\ast}$, the dual of the central extended Lie algebra $\mathfrak{h}%
_{c,\mathbf{\theta}}$ of $\mathfrak{h}$ by the cocycle $c_{\mathbf{\theta}}$.
It is given by%
\[
\widehat{Ad}_{\mathbf{\theta};l^{-1}}^{\mathrm{H}\ast}\left(  \mathbf{\xi
},b\right)  =\left(  Ad_{l^{-1}}^{\ast}\mathbf{\xi}+bC_{\mathbf{\theta}%
}\left(  l\right)  \,,b\right)
\]
and the linear Poisson bracket $\left\{  \mathbf{,}\right\}
_{c,\mathbf{\theta}}$ on $\mathfrak{h}_{c,\mathbf{\theta}}^{\ast}$ by%
\[
\left\{  \left\langle \mathbf{X},-\right\rangle ,\left\langle \mathbf{Y}%
,-\right\rangle \right\}  _{c,\mathbf{\theta}}\left(  \mathbf{\xi},1\right)
=\left\langle \mathbf{\xi,}\left[  \mathbf{X},\mathbf{Y}\right]
_{\mathfrak{h}}\right\rangle -c_{\mathbf{\theta}}\left(  \mathbf{X}%
,\mathbf{Y}\right)
\]
for $\mathbf{X},\mathbf{Y}\in\mathfrak{h}$. Its symplectic leaves are the
$\widehat{Ad}_{\mathbf{\theta};}^{\mathrm{H}\ast}-$coadjoint orbits equipped
with the Kirillov-Kostant symplectic structure.

The $\widehat{Ad}_{\mathbf{\theta};}^{\mathrm{H}\ast}$-equivariant momentum
map $\hat{J}_{c,\mathbf{\theta}}^{R}:\mathrm{H}\times\mathfrak{h}^{\ast
}\rightarrow\mathfrak{h}_{c,\mathbf{\theta}}^{\ast}$ associated to the induced
symplectic $\mathrm{H}_{c,\mathbf{\theta}}$-action on $\left(  \mathrm{H}%
\times\mathfrak{h}^{\ast},\mathbf{\omega}_{c,\mathbf{\theta}}\right)  $ is%
\begin{equation}
\hat{J}_{c,\mathbf{\theta}}^{R}\left(  l,\mathbf{\eta}\right)  =\left(
\mathbf{\eta}-Ad_{l}^{\ast}C_{\mathbf{\theta}}\left(  l\right)  ,1\right)
\label{R-momentum}%
\end{equation}

\begin{description}
\item[Remark] \emph{(Affine coadjoint action)} \textit{We observe that
everything that follows can be carried out by means of the affine coadjoint
action of the group }$\mathrm{H},$\textit{without extension,} \textit{on}
$\mathfrak{h}^{\ast}$
\[
\widehat{Ad}_{\mathbf{Aff};l^{-1}}^{\mathrm{H}\ast}\mathbf{\xi}=Ad_{l^{-1}%
}^{\ast}\mathbf{\xi}+C_{\mathbf{\theta}}\left(  l\right)
\]
\textit{instead of the extended one} $\widehat{Ad}_{\mathbf{\theta}%
;}^{\mathrm{H}\ast}$of $\mathrm{H}_{c,\mathbf{\theta}}$ on $\mathfrak{h}%
_{c,\mathbf{\theta}}^{\ast}$. \textit{This affine action gives rise to the
coadjoint affine orbits }$\mathcal{O}_{\alpha}^{Aff}$ \textit{and the
corresponding affine Poisson bracket on }$\mathfrak{h}^{\ast}$\textit{,
without further reference to central extensions. However, we keep the central
extension framework for simplicity.}
\end{description}

The phase spaces $\left(  \mathrm{H}\times\mathfrak{h}^{\ast},\mathbf{\omega
}_{c,\mathbf{\theta}}\right)  $ play a central role in our $T$-duality scheme:
most of its features rely on their symmetry properties and the corresponding
reduced spaces are the bridge connecting $T$-dual systems. So let us work out
a couple of related phase spaces which we shall be concerned with.

\begin{description}
\item[S1-] For $\mathbf{\theta}=0$, the Marsden-Weinstein \cite{Mars-Wein}
reduction procedure can be applied to a regular value of the form $\left(
\mathbf{\alpha},1\right)  \in\mathfrak{h}_{c,0}^{\ast}$ within the phase space
$\left(  \mathrm{H}\times\mathfrak{h}^{\ast},\mathbf{\omega}_{c,0}\right)  $.
We get that $\left[  \hat{J}_{c,0}^{R}\right]  ^{-1}\left(  \mathbf{\alpha
},1\right)  \simeq\mathrm{H}$ is a presymplectic manifold with the restricted
$2$-form%
\begin{equation}
\mathbf{\tilde{\omega}}_{-\mathbf{\alpha}}\left(  \mathbf{v},\mathbf{w}%
\right)  :=\left.  \omega_{c,0}\right\vert _{\left[  \hat{J}_{c,0}^{R}\right]
^{-1}\left(  \mathbf{\alpha},1\right)  }\left(  \mathbf{v},\mathbf{w}\right)
=c_{-\mathbf{\alpha}}\left(  l^{-1}\mathbf{v},l^{-1}\mathbf{w}\right)
\label{red-presympl-alpha}%
\end{equation}
for $\mathbf{v},\mathbf{w\in}T_{l}\mathrm{H}$. Its null distribution is
spanned by the infinitesimal generators of the action of the subgroup
$\mathrm{H}_{\mathbf{\alpha}}:=\ker C_{-\mathbf{\alpha}}$, so that the reduced
symplectic space is
\[
M_{c,0}^{\left(  \mathbf{\alpha},1\right)  }:=\frac{\left[  \hat{J}_{c,0}%
^{R}\right]  ^{-1}\left(  \mathbf{\alpha},1\right)  }{\mathrm{H}%
_{\mathbf{\alpha}}}\simeq\frac{\mathrm{H}}{\mathrm{H}_{\mathbf{\alpha}}}%
\]
where the right action of $\mathrm{H}_{\mathbf{\alpha}}$ on $\mathrm{H}$ is
considered.\ Denoting the fiber bundle $\mathrm{H}\overset{\Pi_{\mathrm{H}%
/\mathrm{H}_{\mathbf{\alpha}}}}{\longrightarrow}\mathrm{H}/\mathrm{H}%
_{\mathbf{\alpha}}$, then the base $\mathrm{H}/\mathrm{H}_{\mathbf{\alpha}}$
is endowed with a symplectic form $\mathbf{\omega}_{R}$ defined by
$\Pi_{\mathrm{H}/\mathrm{H}_{\mathbf{\alpha}}}^{\ast}\mathbf{\omega}%
_{R}=\mathbf{\tilde{\omega}}_{-\mathbf{\alpha}}$. This form is invariant under
the\ residual left action of $\mathrm{H}$\ on $\mathrm{H}/\mathrm{H}%
_{\mathbf{\alpha}}$ and has associated momentum map $\hat{\Phi}_{c,0}%
:\mathrm{H}/\mathrm{H}_{\mathbf{\alpha}}\longrightarrow\mathfrak{h}%
_{c,0}^{\ast}$%
\[
\hat{\Phi}_{c,0}:\mathrm{H}/\mathrm{H}_{\mathbf{\alpha}}\longrightarrow
\mathfrak{h}_{c,0}^{\ast}~/~\hat{\Phi}_{c,0}\left(  l\cdot\mathrm{H}%
_{\mathbf{\alpha}}\right)  =\left(  C_{-\mathbf{\alpha}}\left(  l\right)
+\mathbf{\alpha},1\right)
\]
which is $\widehat{Ad}_{\mathbf{\theta=0}}^{\mathrm{H}\ast}$-equivariant and
gives a local symplectic diffeomorphism from $\left(  \mathrm{H}%
/\mathrm{H}_{\mathbf{\alpha}},\mathbf{\omega}_{R}\right)  $\ to the coadjoint
orbit $\mathcal{O}_{c,0}\left(  \mathbf{\alpha},1\right)  $\ equipped with the
Kirillov-Kostant symplectic structure $\mathbf{\omega}_{c,0}^{{\small KK}}.$
Notice that the subgroup $\mathrm{H}_{\mathbf{\alpha}}$ coincides a with the
stabilizer subgroup $\left[  \mathrm{H}_{c,0}\right]  _{\left(  \mathbf{\alpha
},1\right)  }$ of the point $\left(  0,1\right)  \in\mathfrak{h}%
_{c,\mathbf{-\alpha}}^{\ast}$.

\item[S2-] For arbitrary $\mathbf{\theta=-\alpha\in}\mathfrak{h}^{\ast}$, the
reduction procedure can be applied to the regular value $\left(  0,1\right)
\in\mathfrak{h}_{c,\mathbf{-\alpha}}^{\ast}$ of $Im\hat{J}_{c,\mathbf{-\alpha
}}^{R}$. The level set $\left[  \hat{J}_{c,\mathbf{-\alpha}}^{R}\right]
^{-1}\left(  0,1\right)  \cong\mathrm{H}$ is again a presymplectic manifold
with restricted $2$-form%
\begin{equation}
\mathbf{\tilde{\omega}}_{\mathbf{-\alpha}}\left(  \mathbf{v},\mathbf{w}%
\right)  :=\left.  \mathbf{\omega}_{c,\mathbf{-\alpha}}\right\vert _{\left[
\hat{J}_{c,\mathbf{-\alpha}}^{R}\right]  ^{-1}\left(  0,1\right)  }\left(
\mathbf{v},\mathbf{w}\right)  =c_{\mathbf{-\alpha}}\left(  l^{-1}%
\mathbf{v},l^{-1}\mathbf{w}\right)  \label{red-presympl-0}%
\end{equation}
for $\mathbf{v},\mathbf{w\in}T_{l}\mathrm{H}$. The null distribution of
$\mathbf{\tilde{\omega}}_{\mathbf{-\alpha}}$ is spanned by the infinitesimal
generators of the (right) action of the subgroup $\mathrm{H}_{\mathbf{\alpha}%
}$. Hence, the reduced symplectic space is
\[
M_{c,\mathbf{-\alpha}}^{\left(  0,1\right)  }:=\frac{\left[  \hat
{J}_{c,\mathbf{-\alpha}}^{R}\right]  ^{-1}\left(  0,1\right)  }{\mathrm{H}%
_{\mathbf{\alpha}}}\cong\frac{\mathrm{H}}{\mathrm{H}_{\mathbf{\alpha}}}%
\]
again. The symplectic form $\mathbf{\omega}_{R}$ is defined by $\Pi
_{\mathrm{H}/\mathrm{H}_{\mathbf{-\alpha}}}^{\ast}\mathbf{\omega}%
_{R}=\mathbf{\tilde{\omega}}_{\mathbf{-\alpha}}$ as before. Recall that
$\mathbf{\omega}_{R}$\textit{\ }is invariant under the\ residual left action
and that the associated $\widehat{Ad}_{\mathbf{-\alpha}}^{\mathrm{H}\ast}%
$-equivariant momentum map $\hat{\Phi}_{c,\mathbf{-\alpha}}:\mathrm{H}%
/\mathrm{H}_{\mathbf{\alpha}}\longrightarrow\mathfrak{h}_{c,\mathbf{-\alpha}%
}^{\ast}\,\,$now reads%
\begin{equation}
\hat{\Phi}_{c,\mathbf{-\alpha}}\left(  l\cdot\mathrm{H}_{\mathbf{\alpha}%
}\right)  =\left(  C_{\mathbf{-\alpha}}\left(  l\right)  ,1\right)
\label{phi-1}%
\end{equation}
Moreover, $\hat{\Phi}_{c,\mathbf{-\alpha}}$ is a local symplectic
diffeomorphism onto the coadjoint orbit $\mathcal{O}_{c,\mathbf{-\alpha}%
}\left(  0,1\right)  \subset\mathfrak{h}_{c,\mathbf{-\alpha}}^{\ast}$ equipped
with the Kirillov-Kostant symplectic structure $\mathbf{\omega}%
_{c,\mathbf{-\alpha}}^{{\small KK}}$. Notice that the subgroup $\mathrm{H}%
_{\mathbf{\alpha}}$ coincides a with the stabilizer subgroup $\left[
\mathrm{H}_{c,\mathbf{-\alpha}}\right]  _{\left(  0,1\right)  }$ of the point
$\left(  0,1\right)  \in\mathfrak{h}_{c,\mathbf{-\alpha}}^{\ast}$.
\end{description}

\subsubsection{Symplectic equivalence between $\mathcal{O}_{c,0}\left(
\mathbf{\alpha},1\right)  $ and $\mathcal{O}_{c,\mathbf{-\alpha}}\left(
0,1\right)  $}

\label{subsec: sympl equiv thetas}

The above described reduced spaces can be linked through the \emph{shifting
trick }as follows. The orbits $\mathcal{O}_{c,0}\left(  \mathbf{\alpha
},1\right)  \subset\mathfrak{h}_{c,0}^{\ast}$ and $\mathcal{O}%
_{c,-\mathbf{\alpha}}\left(  0,1\right)  \subset\mathfrak{h}%
_{c,-\mathbf{\alpha}}^{\ast}$ are both isomorphic to $\mathrm{H}%
/\mathrm{H}_{\mathbf{\alpha}}$ as symplectic manifolds and, moreover,

\begin{description}
\item[Proposition:] \textit{The map }$\varphi:\mathfrak{h}_{c,-\mathbf{\alpha
}}^{\ast}$ $\longrightarrow$ $\mathfrak{h}_{c,0}^{\ast}$\textit{ }%
\begin{equation}
\varphi\left(  \mathbf{\eta},1\right)  =\left(  \mathbf{\eta}+\mathbf{\alpha
},1\right)  \label{eq: iso shift}%
\end{equation}
\textit{is a Poisson diffeomorphism. When restricted to }$\varphi
:\mathcal{O}_{c,-\mathbf{\alpha}}\left(  0,1\right)  \subset\mathfrak{h}%
_{c,-\mathbf{\alpha}}^{\ast}$ $\longrightarrow$ $\mathcal{O}_{c,0}\left(
\mathbf{\alpha},1\right)  \subset\mathfrak{h}_{c,0}^{\ast}$\textit{ it becomes
a symplectic diffeomorphism.}
\end{description}

\textbf{Proof: }The underlying vector space of $\mathfrak{h}%
_{c,-\mathbf{\alpha}}^{\ast}$, $\mathfrak{h}_{c,0}^{\ast}$ is the same
$\mathfrak{h}^{\ast}\oplus\mathbb{R}$. Introducing the Legendre transform
$\mathcal{L}_{\mathsf{f}}:\mathfrak{h}^{\ast}\oplus\mathbb{R}\longrightarrow
\mathfrak{h}$ of some function $\mathsf{f}\in C^{\infty}\left(  \mathfrak{h}%
_{c,\beta}^{\ast}\right)  $ as%
\[
\left\langle \mathcal{L}_{\mathsf{f}}\left(  \mathbf{\eta},1\right)
,\mathbf{\xi}\right\rangle =\left.  \frac{d}{dt}\mathsf{f}\left(
\mathbf{\eta+}t\mathbf{\xi},1\right)  \right\vert _{t=0}%
\]
The Poisson structures on the generic $\mathfrak{h}_{c,\beta}^{\ast}$ is%
\[
\left\{  \mathsf{f},\mathsf{h}\right\}  _{c,\beta}\left(  \mathbf{\eta
},1\right)  =\left\langle \mathbf{\eta}-\mathbf{\beta},\left[  \mathcal{L}%
_{\mathsf{f}}\left(  \mathbf{\eta},1\right)  ,\mathcal{L}_{\mathsf{h}}\left(
\mathbf{\eta},1\right)  \right]  \right\rangle +c\left(  \mathcal{L}%
_{\mathsf{f}}\left(  \mathbf{\eta},1\right)  ,\mathcal{L}_{\mathsf{h}}\left(
\mathbf{\eta},1\right)  \right)
\]
Then, from this expression and having in mind that $\mathcal{L}_{\mathsf{f}%
}\left(  \varphi\left(  \mathbf{\eta},1\right)  \right)  =\mathcal{L}%
_{\mathsf{f}\circ\varphi}\left(  \mathbf{\eta},1\right)  $, it is immediate to
see that%
\[
\left\{  \mathsf{f},\mathsf{h}\right\}  _{c,0}\left(  \varphi\left(
\mathbf{\eta},1\right)  \right)  =\left\{  \mathsf{f}\circ\varphi
,\mathsf{h}\circ\varphi\right\}  _{c,-\alpha}\left(  \mathbf{\eta},1\right)
\text{.}%
\]
Thus the hamiltonian vector fields associated to the the functions
$\mathsf{f}\in C^{\infty}\left(  \mathfrak{h}_{c,0}^{\ast}\right)  $ and
$\mathsf{f}\circ\varphi\in C^{\infty}\left(  \mathfrak{h}_{c,-\mathbf{\alpha}%
}^{\ast}\right)  $ are $\varphi$-related%
\[
\varphi_{\ast}\left[  \widehat{ad}_{\left(  \mathcal{L}_{\mathsf{f}%
\circ\varphi}\left(  \mathbf{\eta},1\right)  ,1\right)  }^{c,-\mathbf{\alpha
}\ast}\left(  \mathbf{\eta},1\right)  \right]  =\widehat{ad}_{\left(
\mathcal{L}_{\mathsf{f}}\left(  \mathbf{\eta}+\mathbf{\alpha},1\right)
,1\right)  }^{c,0\ast}\left(  \mathbf{\eta}+\mathbf{\alpha},1\right)
\]

The orbits $\mathcal{O}_{c,-\mathbf{\alpha}}\left(  0,1\right)  \subset
\mathfrak{h}_{c,-\mathbf{\alpha}}^{\ast}$ and $\mathcal{O}_{c,0}\left(
\mathbf{\alpha},1\right)  \subset\mathfrak{h}_{c,0}^{\ast}$ are
\begin{align*}
\mathcal{O}_{c,-\mathbf{\alpha}}\left(  0,1\right)   &  =\left\{  \left(
C_{-\mathbf{\alpha}}\left(  l^{-1}\right)  ,1\right)  /l\in\mathrm{H}\right\}
=\frac{\mathrm{H}}{\mathrm{H}_{\left(  0,1\right)  }^{c,-\mathbf{\alpha}}}\\
\mathcal{O}_{c,0}\left(  \mathbf{\alpha},1\right)   &  =\left\{  \left(
C\left(  l^{-1}\right)  +Ad_{l}^{\ast}\mathbf{\alpha},1\right)  /l\in
\mathrm{H}\right\}  =\frac{\mathrm{H}}{\mathrm{H}_{\left(  \mathbf{\alpha
},1\right)  }^{c,0}}%
\end{align*}
where $\mathrm{H}_{\left(  0,1\right)  }^{c,-\mathbf{\alpha}}=\mathrm{H}%
_{\left(  \mathbf{\alpha},1\right)  }^{c,0}=\mathrm{H}_{\mathbf{\alpha}}%
\equiv\ker C_{-\mathbf{\alpha}}$ are the stabilizer subgroups of $\left(
0,1\right)  \in\mathfrak{h}_{c,-\mathbf{\alpha}}^{\ast}$ and $\left(
\mathbf{\alpha},1\right)  \in\mathfrak{h}_{c,0}^{\ast}$. The restriction of
the above Poisson structures to these orbits endow them with the corresponding
Kirillov-Kostant symplectic forms $\omega_{c,-\mathbf{\alpha}}^{{\small KK}%
},\omega_{c,0}^{{\small KK}}$, and the diffeomorphism $\varphi:\mathcal{O}%
_{c,-\mathbf{\alpha}}\left(  0,1\right)  \longrightarrow\mathcal{O}%
_{c,0}\left(  \mathbf{\alpha},1\right)  $, $\varphi\left(  C_{-\mathbf{\alpha
}}\left(  l^{-1}\right)  \right)  =C_{-\mathbf{\alpha}}\left(  l^{-1}\right)
+\mathbf{\alpha}$, becomes a symplectic one. In fact, for $\eta
=C_{-\mathbf{\alpha}}\left(  l^{-1}\right)  $ and after a direct computation,
one recovers
\begin{align*}
&  \left\langle \omega_{c,0}^{{\small KK}},\widehat{ad}_{\left(
\mathcal{L}_{\mathsf{f}}\left(  \eta+\mathbf{\alpha},1\right)  ,1\right)
}^{c,0\ast}\left(  \eta+\mathbf{\alpha},1\right)  \otimes\widehat{ad}_{\left(
\mathcal{L}_{\mathsf{f}}\left(  \eta+\mathbf{\alpha},1\right)  ,1\right)
}^{c,0\ast}\left(  \eta+\mathbf{\alpha},1\right)  \right\rangle _{\left(
\eta+\mathbf{\alpha},1\right)  }\\
&  =\left\langle \omega_{c,-\mathbf{\alpha}}^{{\small KK}},\widehat
{ad}_{\left(  \mathcal{L}_{\mathsf{f}\circ\varphi}\left(  \eta,1\right)
,1\right)  }^{c,-\mathbf{\alpha}\ast}\left(  \eta,1\right)  \otimes
\widehat{ad}_{\left(  \mathcal{L}_{\mathsf{h}\circ\varphi}\left(
\eta,1\right)  ,1\right)  }^{c,-\mathbf{\alpha}\ast}\left(  \eta,1\right)
\right\rangle _{\left(  \eta,1\right)  }%
\end{align*}
showing that $\varphi^{\ast}\omega_{c,0}^{{\small KK}}=\omega
_{c,-\mathbf{\alpha}}^{{\small KK}}$.$\blacksquare$

Hence, all the above maps can be resumed in the following diagram
\begin{equation}
\begin{diagram}[h=1.9em] \left( M_{c,0}^{\left( \alpha,1\right) },\tilde{\omega}_{-\alpha}\right) &\rTo& \left(\frac {H}{H_{\alpha}},\omega^R\right)&\lTo&\left( M_{c,-\alpha}^{\left( 0,1\right) },\tilde{\omega}_{-\alpha }\right)\\ \dTo&&&&\dTo\\ {}&\ldTo&&\rdTo&{}\\ &&&&\\ \left( \mathcal{O}_{c,0}\left( \alpha,1\right) ,\omega_{c,0}^{{\small KK}}\right)&&\lTo_{\varphi}&&\left( \mathcal{O}_{c,-\alpha}\left( 0,1\right) ,\omega_{c,-\alpha}^{{\small KK}}\right)\\ \end{diagram} \label{wzw-t-d}%
\end{equation}
where all the arrows are symplectic isomorphisms. This result shall be used
for describing a WZNW-type model on a double Lie group $\mathrm{H}$ as
described in the next section.

\subsection{Double Lie groups and sigma model phase spaces}

\label{sec: double and sigma}We assume now that $\mathrm{H}$ is a Drinfeld
double Lie group \cite{Drinfeld-1,Lu-We}, $\mathrm{H}=\mathrm{N}%
\Join\mathrm{N}^{\ast}$ with tangent Lie bialgebra $\mathfrak{h}%
=\mathfrak{n}\oplus\mathfrak{n}^{\ast}$. This bialgebra $\mathfrak{h}$ is
naturally equipped with the non degenerate symmetric $Ad$-invariant bilinear
form $\left(  ,\right)  _{\mathfrak{h}}$ provided by the pairing between
$\mathfrak{n}$ and $\mathfrak{n}^{\ast}$ and which turns them into isotropic
subspaces. Let $\psi$ denote the identification between $\mathfrak{h}$ and
$\mathfrak{h}^{\ast}$ induced by this bilinear form $\left(  ,\right)
_{\mathfrak{h}}$. For the sake of brevity, we often omit it from formulas when
there is no danger of confusion.

The aim of the following subsections is to construct dual phase spaces from
the factors $\mathrm{N}$ and $\mathrm{N}^{\ast}$ as described in section
\ref{sec: setting}. T-duality over the \emph{trivial} orbit $\mathcal{O}%
_{c,0}(0,1)$ was considered in ref. \cite{Ale-Hugo} in relation to Poisson-Lie
T-duality for loop groups and trivial monodromies. Now, we focus our attention
into exploring hamiltonian $\mathrm{H}$-actions on phase spaces $T^{\ast
}\mathrm{N}$ and $T^{\ast}\mathrm{N}^{\ast}$ such that they become T-dual over
a \emph{non-trivial }coadjoint orbit $\mathcal{O}_{c,0}(\mathbf{\alpha},1)$,
with $\mathbf{\alpha\ }$in $\mathfrak{n}$ or $\mathfrak{n}^{\ast}$. Notice
that, once $\mathbf{\alpha}$ is chosen in one of the factors, the symmetric
role played by $\mathrm{N}$ and $\mathrm{N}^{\ast}$ in the construction is
broken. By the Poisson isomorphism $\left(  \ref{eq: iso shift}\right)  $,
equivalent systems will be constructed on the orbit $\mathcal{O}%
_{c,-\mathbf{\alpha}}(0,1)$, for $\mathbf{\alpha}$ in $\mathfrak{n}$ or
$\mathfrak{n}^{\ast}$. Hence, we can choose the momentum maps for associated
$\mathrm{H}$-actions to be valued on $\left(  \mathfrak{h}_{c,0}^{\ast
},\left\{  ,\right\}  _{c,0}\right)  \ $or $\left(  \mathfrak{h}%
_{c,-\mathbf{\alpha}}^{\ast},\left\{  ,\right\}  _{c,-\mathbf{\alpha}}\right)
$. Poisson-Lie T-duality for non-trivial monodromies in the loop group case is
described in section \ref{sec: Loops}.

On double Lie groups there exist reciprocal actions between the factors
$\mathrm{N}$ and $\mathrm{N}^{\ast}$ named \emph{dressing actions}
\cite{STS},\cite{Lu-We}. Since every element $l\in\mathrm{H}$ can be written
as $l=g\tilde{h}$, with $g\in\mathrm{N}$ and $\tilde{h}\in\mathrm{N}^{\ast}$,
the product $\tilde{h}g$ in $\mathrm{H}$ can be expressed as $\tilde
{h}g=g^{\tilde{h}}\tilde{h}^{g}$, with $g^{\tilde{h}}\in\mathrm{N}$ and
$\tilde{h}^{g}\in\mathrm{N}^{\ast}$. The dressing action of $\mathrm{N}^{\ast
}$ on $\mathrm{N}$ is then defined as
\[
\mathsf{Dr}:\mathrm{N}^{\ast}\times\mathrm{N}\longrightarrow\mathrm{N}%
\qquad\ |\ \qquad\mathsf{Dr}\left(  \tilde{h},g\right)  =\Pi_{\mathrm{N}%
}\left(  \tilde{h}g\right)  =g^{\tilde{h}}%
\]
where $\Pi_{\mathrm{N}}:\mathrm{H}\longrightarrow\mathrm{N}$ is the projector.
For $\xi\in\mathfrak{n}^{\ast}$, the infinitesimal generator of this action at
$g\in\mathrm{N}$ is
\[
\xi\longrightarrow\mathsf{dr}\left(  \xi\right)  _{g}=-\left.  \dfrac{d}%
{dt}\mathsf{Dr}\left(  e^{t\xi},g\right)  \right\vert _{t=0}%
\]
such that, for $\eta\in\mathfrak{n}^{\ast}$, we have $\left[  \mathsf{dr}%
\left(  \xi\right)  _{g},\mathsf{dr}\left(  \eta\right)  _{g}\right]
=\mathsf{dr}\left(  \left[  \xi,\eta\right]  _{\mathfrak{n}^{\ast}}\right)
_{g}$. It satisfies the relation $Ad_{g^{-1}}^{\mathrm{H}}\xi=-g^{-1}%
\mathsf{dr}\left(  \xi\right)  _{g}+Ad_{g}^{\ast}\xi$, where $Ad_{g^{-1}%
}^{\mathrm{H}}\in Aut\left(  \mathfrak{h}\right)  $ is the adjoint action of
$\mathrm{H}$ on its Lie algebra. Then, using the projector $\Pi_{\mathfrak{n}%
}:\mathfrak{h}\longrightarrow\mathfrak{n}$, we can write $\mathsf{dr}\left(
\xi\right)  _{g}=-g~\Pi_{\mathfrak{n}}Ad_{g^{-1}}^{\mathrm{H}}\xi$.

Let us now consider the action of $\mathrm{H}$ on itself by left translations
$L_{a\tilde{b}}g\tilde{h}=a\tilde{b}g\tilde{h}$. Its projection on the one of
the factors, $\mathrm{N}$ for instance, yields also an action of $\mathrm{H}$
on that factor%
\[
\Pi_{\mathrm{N}}\left(  L_{a\tilde{b}}g\tilde{h}\right)  =\Pi_{\mathrm{N}%
}\left(  a\tilde{b}g\tilde{h}\right)  =ag^{\tilde{b}}%
\]
The projection on the factor $\mathrm{N}^{\ast}$ is obtained by the reversed
factorization of $\mathrm{H}$, namely $\mathrm{N}^{\ast}\times\mathrm{N}$,
such that
\[
\Pi_{\mathrm{N}^{\ast}}\left(  L_{\tilde{b}a}\tilde{h}g\right)  =\Pi
_{\mathrm{N}^{\ast}}\left(  \tilde{b}a\tilde{h}g\right)  =\tilde{b}\tilde
{h}_{a}%
\]

In the next subsections, we lift these actions to $T^{\ast}\mathrm{N}$ and
$T^{\ast}\mathrm{N}^{\ast}$ and twist them using a cocycle. The resulting ones
play a central role in constructing $T$-dual phase spaces out of these
cotangent bundles, turning them in hamiltonian spaces for different central
extensions of the group $\mathrm{H}$.

\subsubsection{Phase spaces on $T^{\ast}\mathrm{N}$}

\paragraph{Hamiltonian $\mathrm{H}_{c,0}$-spaces}

We now consider a phase space $T^{\ast}\mathrm{N}\cong\mathrm{N}%
\times\mathfrak{n}^{\ast}$, trivialized by left translations and equipped with
the canonical symplectic form $\omega_{o}$. We shall realize the symmetry
described above, as it was introduced in \cite{Ale-Hugo}.

We promote this symmetry on $T^{\ast}\mathrm{N}$ to a \emph{centrally
extended} one by means of an $\mathfrak{n}^{\ast}$-valued cocycle
$C^{\mathrm{N}^{\ast}}:\mathrm{N}^{\ast}\longrightarrow\mathfrak{n}^{\ast}$,%
\begin{equation}%
\begin{array}
[c]{l}%
\mathsf{\hat{d}}_{0}^{\mathrm{N}\times\mathfrak{n}^{\ast}}:\mathrm{H}%
_{c,0}\times\left(  \mathrm{N}\times\mathfrak{n}^{\ast}\right)
\longrightarrow\left(  \mathrm{N}\times\mathfrak{n}^{\ast}\right) \\
\mathsf{\hat{d}}_{0}^{\mathrm{N}\times\mathfrak{n}^{\ast}}\left(  a\tilde
{b},\left(  g,\lambda\right)  \right)  =\left(  ag^{\tilde{b}},Ad_{\left(
\tilde{b}^{g}\right)  ^{-1}}^{\mathrm{H}\ast}\lambda+C^{\mathrm{N}^{\ast}%
}\left(  \tilde{b}^{g}\right)  \right)
\end{array}
\label{d-action-cent-ext}%
\end{equation}
Aiming to repeat the same construction on a \emph{dual} phase space, the
adjoint $\mathrm{N}$-cocycle $C^{\mathrm{N}^{\ast}}:\mathrm{N}^{\ast
}\longrightarrow\mathfrak{n}^{\ast}$ is assumed to be the restriction to the
factor $\mathrm{N}$ of an $\mathrm{H}$-coadjoint cocycle on $C:\mathrm{H}%
\longrightarrow\mathfrak{h}^{\ast}$, namely $C^{\mathrm{N}^{\ast}%
}:=C^{\mathrm{H}}|_{\mathrm{N}^{\ast}}$, in such a way that
\begin{equation}%
\begin{array}
[c]{c}%
C|_{\mathrm{N}}:\mathrm{N}\longrightarrow\mathfrak{n}\\
C|_{\mathrm{N}^{\ast}}:\mathrm{N}^{\ast}\longrightarrow\mathfrak{n}^{\ast}%
\end{array}
\label{eq: compat cocycle}%
\end{equation}
We shall say that, in this case, $C$ is \emph{compatible} \emph{with the
factor decomposition}. One of the key ingredients for describing the resulting
duality is that the above $\mathrm{H}$-action is hamiltonian.

\begin{description}
\item[Proposition:] \label{prop: action on N without shift}\textit{Let}
$T^{\ast}\mathrm{N}$ \textit{be identified with} $\mathrm{N}\times
\mathfrak{n}^{\ast}$ \textit{via left translations and endowed with its
canonical symplectic structure. The action }$\mathsf{\hat{d}}_{0}%
^{\mathrm{N}\times\mathfrak{n}^{\ast}}$\textit{, defined in eq.}$\left(
\ref{d-action-cent-ext}\right)  $ \textit{from a cocycle }$C$
\textit{compatible with the factor decomposition,} \textit{is hamiltonian and
the momentum map }$\mu_{0,0}:\left(  \mathrm{N}\times\mathfrak{n}^{\ast
},\omega_{o}\right)  \rightarrow\left(  \mathfrak{h}_{c,0}^{\ast}%
,\{,\}_{c,0}\right)  $\textit{ }%
\[
\mu_{0,0}\left(  g,\lambda\right)  =\widehat{Ad}_{0;g^{-1}}^{\mathrm{H}\ast
}\left(  \psi\left(  \lambda\right)  ,1\right)  =\left(  \psi\left(
Ad_{g}^{\mathrm{H}}\lambda+C^{\mathrm{N}^{\ast}}\left(  g\right)  \right)
,1\right)  ~.
\]
\textit{is }$\widehat{Ad}_{0;}^{\mathrm{H}\ast}$-\textit{equivariant.}
\end{description}

\textbf{Proof}: The infinitesimal generator of the action $\left(
\ref{d-action-cent-ext}\right)  $ associated to $\left(  X,\xi\right)
\in\mathfrak{h}$ is the vector field%
\[
\left.  \left(  X,\xi\right)  _{\mathrm{N}\times\mathfrak{n}^{\ast}%
}\right\vert _{\left(  g,\lambda\right)  }=\left(  Xg-\mathsf{dr}\left(
\xi\right)  _{g},\left[  Ad_{g}^{\ast}\xi\mathbf{,}\lambda\right]  -\hat
{c}\left(  Ad_{g}^{\ast}\xi\right)  \right)
\]
By an straightforward calculation one may see that%
\[
\left.  \imath_{\left(  X,\xi\right)  _{\mathrm{N}\times\mathfrak{n}^{\ast}}%
}\omega_{o}\right\vert _{\left(  g,\lambda\right)  }=d\mathbf{\langle
}Ad_{g^{-1}}^{\mathrm{H}\ast}\lambda+C\left(  g\right)  ,\left(
X\mathbf{,}\xi\right)  \mathbf{\rangle}%
\]
so that $f_{\left(  X\mathbf{,}\xi\right)  }\left(  g,\lambda\right)
\equiv\mathbf{\langle}Ad_{g^{-1}}^{\mathrm{H}\ast}\lambda+C\left(  g\right)
,\left(  X\mathbf{,}\xi\right)  \mathbf{\rangle}$ is the hamiltonian function
associated to the vector field $\left(  X,\xi\right)  _{\mathrm{N}%
\times\mathfrak{n}^{\ast}}$. Then, $\mu_{0,0}:$ $\mathrm{N}\times
\mathfrak{n}^{\ast}\longrightarrow\mathfrak{h}_{c,0}^{\ast}$ defined as
\[
\mu_{0,0}(g,\lambda)=\left(  \psi\left(  Ad_{g^{-1}}^{\mathrm{H}\ast}%
\lambda+C\left(  g\right)  \right)  ,1\right)
\]
is the momentum map associated to the action $\left(  \ref{d-action-cent-ext}%
\right)  $.

Hence, since $\left(  X,\xi\right)  _{\mathrm{N}\times\mathfrak{n}^{\ast}}$ is
hamiltonian for all $\left(  \left(  X,\xi\right)  ,s\right)  \in
\mathfrak{h}_{c,0}$, and $\mathsf{\hat{d}}_{0}^{\mathrm{N}\times
\mathfrak{n}^{\ast}}$ \emph{leaves the canonical symplectic form invariant}.
Furthermore, $\mu_{0,0}$ is $\widehat{Ad}$ -equivariant%
\begin{align*}
\mu_{0,0}\left(  \mathsf{\hat{d}}_{0}^{\mathrm{N}\times\mathfrak{n}^{\ast}%
}\left(  a\tilde{b},\left(  g,\lambda\right)  \right)  \right)   &  =\mu
_{0,0}\left(  ag^{\tilde{b}},Ad_{\left(  \tilde{b}^{g}\right)  ^{-1}%
}^{\mathrm{H}\ast}\lambda+C^{\mathrm{N}^{\ast}}\left(  \tilde{b}^{g}\right)
\right) \\
&  =\left(  Ad_{\left(  a\tilde{b}g\right)  ^{-1}}^{\mathrm{H}\ast}%
\lambda+C^{\mathrm{N}^{\ast}}\left(  a\tilde{b}g\right)  ,1\right)
\end{align*}
then%
\[
\mu_{0,0}\left(  \mathsf{\hat{d}}_{0}^{\mathrm{N}\times\mathfrak{n}^{\ast}%
}\left(  a\tilde{b},\left(  g,\lambda\right)  \right)  \right)  =\widehat
{Ad}_{0;\left(  a\tilde{b}\right)  ^{-1}}^{\mathrm{H}\ast}\mu_{0,0}\left(
g,\lambda\right)
\]
as required.$\blacksquare$

As stated in the introduction, for constructing \emph{dualizable subspaces},
we need to fix an $\alpha\in\mathfrak{n}^{\ast}$ and study the $\mu_{0,0}$-pre
image of the orbit $\mathcal{O}_{c,0}\left(  \alpha,1\right)  \subset
\mathfrak{h}_{c,0}^{\ast}$.

\paragraph{Hamiltonian $\mathrm{H}_{c,-\alpha}$-spaces}

In view of the equivalence stated in section \ref{subsec: sympl equiv thetas},
we can think of $T^{\ast}\mathrm{N}\simeq\mathrm{N}\times\mathfrak{n}^{\ast}$
as phase space linked by a momentum map valued on$\ \mathfrak{h}%
_{c,\mathbf{-\alpha}}^{\ast}$. This is attained by considering the coboundary
shifted cocycle
\[
C_{-\alpha}^{\mathrm{N}^{\ast}}\left(  \tilde{h}\right)  =C^{\mathrm{N}^{\ast
}}\left(  \tilde{h}\right)  +Ad_{\tilde{h}}\alpha-\alpha
\]
for some $\alpha\in\mathfrak{n}^{\ast}$, thus enabling to introduce an
$\mathrm{H}_{c,-\alpha}$-action on $T^{\ast}\mathrm{N}\simeq\mathrm{N}%
\times\mathfrak{n}^{\ast}$ defined as%
\begin{align}
\mathsf{\hat{d}}_{\alpha}^{\mathrm{N}\times\mathfrak{n}^{\ast}}:\mathrm{H}%
_{c,-\alpha}\times &  \left(  \mathrm{N}\times\mathfrak{n}^{\ast}\right)
\longrightarrow\left(  \mathrm{N}\times\mathfrak{n}^{\ast}\right) \nonumber\\
& \label{dr-LG-E-1}\\
\mathsf{\hat{d}}_{\alpha}^{\mathrm{N}\times\mathfrak{n}^{\ast}}\left(
a\tilde{b},\left(  g,\eta\right)  \right)   &  =\left(  ag^{\tilde{b}%
},Ad_{\tilde{b}^{g}}\eta+C_{-\alpha}^{\mathrm{N}^{\ast}}\left(  \tilde{b}%
^{g}\right)  \,\right) \nonumber
\end{align}
for $a\tilde{b}\in\mathrm{H}=\mathrm{N}\Join\mathrm{N}^{\ast}$ and $\left(
g,\eta\right)  \in T^{\ast}\mathrm{N}$. It is worth to remark this action is
not a cotangent lift of a transformation on $\mathrm{N}$, and that it is
meaningful just for $\alpha\in\mathfrak{n}^{\ast}$, it does not make sense to
for arbitrary $\mathbf{\alpha}\in\mathfrak{h}^{\ast}$.

The shifted $\mathrm{H}-$cocycle%
\begin{equation}
C_{-\alpha}\left(  l\right)  =C\left(  l\right)  +Ad_{l^{-1}}^{\mathrm{H}\ast
}\alpha-\alpha\label{D-cocycle}%
\end{equation}
does not satisfy property $\left(  \ref{eq: compat cocycle}\right)  $, so the
above action may be not hamiltonian in $\left(  \mathrm{N}\times
\mathfrak{n}^{\ast},\omega_{o}\right)  $ (compare to example
\ref{ex: dual groups}) unless some constraint is imposed on $\alpha$.

\begin{description}
\item[Proposition:] \label{prop: H action on N shifted formulation}\textit{Let
}$\alpha\in\mathfrak{n}^{\ast}$ \textit{then, provided the condition} \textit{
}%
\begin{equation}
\Pi_{\mathfrak{n}^{\ast}}\left[  X,\alpha\right]  =0 \label{loop-22a}%
\end{equation}
\textit{is fulfilled for all }$X\in\mathfrak{n}$, \textit{then the above
defined shifted }$H$\textit{-cocycle }$C_{-\alpha}$\textit{ becomes compatible
with the factor decomposition }$\mathrm{H}=\mathrm{N}\Join\mathrm{N}^{\ast}$.
\textit{Consequently, in this case, the action }$\mathsf{\hat{d}}_{\alpha
}^{\mathrm{N}\times\mathfrak{n}^{\ast}}:\mathrm{H}_{c,-\alpha}\times\left(
\mathrm{N}\times\mathfrak{n}^{\ast},\omega_{o}\right)  \longrightarrow\left(
\mathrm{N}\times\mathfrak{n}^{\ast},\omega_{o}\right)  $ \textit{defined by
}$\left(  \ref{dr-LG-E-1}\right)  $ \textit{is hamiltonian. The associated
}$\widehat{Ad}_{\alpha;}^{\mathrm{H}\ast}$\textit{-equivariant}\emph{
}\textit{momentum map} $\mu_{0,\alpha}:\left(  \mathrm{N}\times\mathfrak{n}%
^{\ast},\omega_{0}\right)  \longrightarrow\left(  \mathfrak{h}_{c,-\alpha
}^{\ast},\{,\}_{c,-\alpha}\right)  $ \textit{is given by}%
\begin{equation}
\mu_{0,\alpha}\left(  g,\eta\right)  =\widehat{Ad}_{\alpha;g^{-1}}%
^{\mathrm{H}\ast}\left(  \psi\left(  \eta\right)  ,1\right)  =\left(
\psi\left(  Ad_{g}^{\mathrm{H}}\eta+C_{-\alpha}\left(  g\right)  \right)
,1\right)  \label{loop-22b}%
\end{equation}

\end{description}

Alternatively, a hamiltonian $\mathrm{H}_{c,-\alpha}$-space on $\mathrm{N}%
\times\mathfrak{n}^{\ast}$ can be retrieved by considering a coboundary
shifted symplectic form $\omega_{\alpha}$ on $T^{\ast}\mathrm{N}%
\cong\mathrm{N}\times\mathfrak{n}^{\ast}$, obtained by adding the coboundary
$b_{\alpha}\left(  X,Y\right)  =\left\langle \alpha,\left[  X,Y\right]
_{\mathfrak{n}}\right\rangle $ to the canonical one so that, in body
coordinates, it is
\[
\langle\omega_{\alpha},(v,\rho)\otimes(w,\xi)\rangle_{(g,\eta)}=-\langle
\rho,g^{-1}w\rangle+\langle\xi,g^{-1}v\rangle+\langle\eta+\alpha
,[g^{-1}v,g^{-1}w]\rangle
\]
for $(v,\rho),(w,\lambda)\in T_{(g,\mu)}\left(  \mathrm{N}\times
\mathfrak{n}^{\ast}\right)  =T_{g}^{\ast}\mathrm{N}\times\mathfrak{n}$.

\begin{description}
\item[Proposition:] \textit{The action }$\mathsf{\hat{d}}_{\alpha}%
^{\mathrm{N}\times\mathfrak{n}^{\ast}}:\mathrm{H}_{c,-\alpha}\times\left(
\mathrm{N}\times\mathfrak{n}^{\ast},\omega_{\alpha}\right)  \longrightarrow
\left(  \mathrm{N}\times\mathfrak{n}^{\ast},\omega_{\alpha}\right)  $
\textit{defined by }$\left(  \ref{dr-LG-E-1}\right)  $\textit{ for }$\alpha
\in\mathfrak{n}^{\ast}$\textit{,} \textit{is hamiltonian.} \textit{It has the
associated }$\widehat{Ad}_{\alpha;}^{\mathrm{H}\ast}$-\textit{equivariant
momentum map} $\mu_{\alpha,\alpha}:\left(  \mathrm{N}\times\mathfrak{n}^{\ast
},\omega_{\alpha}\right)  \longrightarrow\left(  \mathfrak{h}_{c,-\alpha
}^{\ast},\{,\}_{c,-\alpha}\right)  $%
\begin{equation}
\mu_{\alpha,\alpha}\left(  g,\eta\right)  =\widehat{Ad}_{\alpha;g^{-1}%
}^{\mathrm{H}\ast}\left(  \psi\left(  \eta\right)  ,1\right)  =\left(
\psi\left(  Ad_{g}^{\mathrm{H}}\eta+C_{-\alpha}\left(  g\right)  \right)
,1\right)  \label{standard momentum map}%
\end{equation}

\end{description}

Within this formulation, for constructing \emph{dualizable subspaces}, we must
look at the $\mu_{0,\alpha}$-pre-image of the orbit $\mathcal{O}%
_{c,\mathbf{-\alpha}}\left(  0,1\right)  \subset\mathfrak{h}%
_{c,\mathbf{-\alpha}}^{\ast}$.

\subsubsection{Phase spaces on $T^{\ast}\mathrm{N}^{\ast}$}

\label{sec: dual factor phase space}

\paragraph{Hamiltonian $\mathrm{H}_{c,0}$-spaces}

In searching for some $T$-dual partners for the phase spaces on $\mathrm{N}%
\times\mathfrak{n}^{\ast}$ built above, we shall consider the symplectic
manifold $\left(  \mathrm{N}^{\ast}\times\mathfrak{n},\tilde{\omega}%
_{o}\right)  $ where $\tilde{\omega}_{o}$ is the canonical $2$-form in body
coordinates. However, the way is not so direct as in the pure central
extension orbit case \cite{Ale-Hugo} and a different strategy is needed in
order to complete the diagram.

Let us consider $\mathrm{H}$ with the opposite factorization, denoted as
$\mathrm{H}\rightarrow\mathrm{H}^{\top}=\mathrm{N}^{\ast}\bowtie\mathrm{N}$,
so that every element is now written as $\tilde{h}g$ with $\tilde{h}%
\in\mathrm{N}^{\ast}$ and $g\in\mathrm{N}$. From \ref{sec: double and sigma},
we get the action $\mathsf{b}^{G^{\ast}}:\mathrm{H}\times\mathrm{N}^{\ast
}\longrightarrow\mathrm{N}^{\ast}\ $defined as $\mathsf{b}^{G}\left(
\tilde{b}a,\tilde{h}\right)  =\tilde{b}\tilde{h}_{a}$\textit{\ }with
$a\in\mathrm{N}$ and $\tilde{h},\tilde{b}\in\mathrm{N}^{\ast}$.

As we shall see below, the search for a hamiltonian $\mathrm{H}$-action on
$\mathrm{N}^{\ast}\times\mathfrak{n}$ will leads us to meet again the
restriction $\left(  \ref{loop-22a}\right)  $ on $\alpha$. First, let us
consider the arrow
\[
z:(\mathrm{N}^{\ast}\times\mathfrak{n},\tilde{\omega}_{o})\longrightarrow
(\mathfrak{h},\{,\}_{Aff})\subset(\mathfrak{h}_{c,0},\{,\}_{c,0})~,
\]
for $\alpha\in\mathfrak{n}^{\ast}$, defined by the diagram%

\[
\begin{diagram}[h=1.9em]
(\mathrm{H}\times \mathfrak{h},\omega_{c}^{R}) &&\lTo^{i_\alpha}&& (\mathrm{N}^{\ast}\times \mathfrak{n},\tilde{\omega}_{o})&\\
&&&&&\\
\dTo_{s}&&\ldTo_{z}&&&\\
&&&&&\\
(\mathfrak{h},\{,\}_{Aff})&&&&&\\
\end{diagram}
\]
where
\[
i_{\alpha}(\tilde{h},X)=(\tilde{h},C(\tilde{h})+Ad_{\tilde{h}}^{\mathrm{H}%
}(X+\alpha))
\]
and the map $s$ being%
\[
s(l,\mathbf{X})=\mathbf{X}%
\]

In the above diagram, $\mathrm{H}\times\mathfrak{h}^{\ast}$ is regarded as the
trivialization of $T^{\ast}\mathrm{H}$ by \emph{right} \emph{translations}
equipped with the symplectic structure $\mathbf{\omega}_{c}^{R}=\mathbf{\omega
}_{o}^{R}-c\circ R$
\[
\langle\mathbf{\omega}_{c}^{R},(v,Y)\otimes(w,Z)\rangle_{(\tilde{h}%
,X)}=\mathbf{\omega}_{o}^{R}-c\left(  v\tilde{h}^{-1}\,,w\tilde{h}%
^{-1}\right)
\]
and $\mathbf{\omega}_{o}^{R}$ denotes the standard symplectic structure on
$T^{\ast}\mathrm{H}$\textrm{ }in\textrm{ }\emph{space coordinates}. Recall
that $\mathfrak{h}\simeq\mathfrak{h}^{\ast}$ is equipped with a non degenerate
symmetric bilinear form $\left(  ,\right)  _{\mathfrak{h}}:\mathfrak{h}%
\otimes\mathfrak{h}\longrightarrow\mathfrak{h}$. Finally, we recall the affine
Poisson bracket $\left\{  ,\right\}  _{c,0}^{Aff}:C^{\infty}\left(
\mathfrak{h}\right)  \otimes C^{\infty}\left(  \mathfrak{h}\right)
\longrightarrow C^{\infty}\left(  \mathfrak{h}\right)  $
\begin{equation}
\{\left(  \mathbf{X},-\right)  _{\mathfrak{h}},\left(  \mathbf{Y},-\right)
_{\mathfrak{h}}\}_{c,0}^{Aff}(\mathbf{Z})=-\left(  [\mathbf{X},\mathbf{Y}%
],\mathbf{Z}\right)  _{\mathfrak{h}}-c(\mathbf{X},\mathbf{Y})
\label{aff-bracket}%
\end{equation}
with $\mathbf{X},\mathbf{Y,Z}\in\mathfrak{h}$, so that $(\mathfrak{h}%
,\{,\}_{Aff})\subset(\mathfrak{h}_{c,0},\{,\}_{c,0})$ via $\mathbf{X}%
\longmapsto(\mathbf{X},1)$.

\begin{description}
\item[Remark] \emph{(Affine coadjoint action)} \textit{Via the isomorphism
}$\mathfrak{h}\simeq\mathfrak{h}^{\ast}$ \textit{induced by the bilinear form
on }$\mathfrak{h}$\textit{, we can work on }$\left(  \mathfrak{h}%
,\{,\}_{Aff}\right)  $\textit{ by considering the affine coadjoint action }%
\[
Ad_{C_{\theta};l}^{\mathrm{H}}\mathbf{Y}=Ad_{l}^{\mathrm{H}}\mathbf{Y}%
+C_{\theta}(l)
\]
\textit{ on }$\mathfrak{h}$\textit{ instead of the full extended coadjoint
action }$\widehat{Ad}_{\theta;}^{\mathrm{H}\ast}$\textit{ on }$\mathfrak{h}%
_{c,\theta}^{\ast}$\textit{.}
\end{description}

It is not hard to see that the map $i_{\alpha}:$ $(\mathrm{N}^{\ast}%
\times\mathfrak{n},\tilde{\omega}_{o})\longrightarrow(\mathrm{H}^{\ast}%
\times\mathfrak{h},\mathbf{\omega}_{c}^{R})$ is\emph{ symplectic}, i.e.,
$i_{\alpha}^{\ast}\mathbf{\omega}_{c}^{R}=\tilde{\omega}_{o}$ for all
$\alpha\in\mathfrak{n}^{\ast}$, and that the map $s:(\mathrm{H}^{\ast}%
\times\mathfrak{h},\mathbf{\omega}_{c}^{R})\longrightarrow(\mathfrak{h}%
,\{,\}_{Aff})$ is a Poisson map.

Thus, the resulting map $z:(\mathrm{N}^{\ast}\times\mathfrak{n},\tilde{\omega
}_{o})\longrightarrow(\mathfrak{h},\{,\}_{Aff})$ is a suitable candidate to be
the generator of a hamiltonian action of $\mathrm{H}$ on the phase space
$\mathrm{N}^{\ast}\times\mathfrak{n}$ and, moreover, its image $z(\mathrm{N}%
^{\ast}\times\mathfrak{n})$ contains the orbit $\mathcal{O}_{c,0}\left(
\alpha,1\right)  \subset\mathfrak{h}^{\ast}$ as desired. Notice that, if
$\tilde{\mu}_{0,\alpha}:(\mathrm{N}^{\ast}\times\mathfrak{n},\tilde{\omega
}_{o})\longrightarrow(\mathfrak{h}_{c,-\alpha},\{,\}_{c,-\alpha})$ denotes the
dual version of the map\textit{ }$\left(  \ref{loop-22b}\right)  $, then the
map $z$ coincides with the composition $\varphi\circ\tilde{\mu}_{0,\alpha}$,
where the isomorphism $\varphi$ was given in\textit{ }$\left(
\ref{eq: iso shift}\right)  $\textit{. }However, this candidate to momentum
map fails to be a Poisson map for general $\alpha$. This issue is addressed in
the following proposition.

\begin{description}
\item[Proposition:] \textit{Let us define the map }$\tilde{\mu}_{0,\alpha
}^{\varphi}:(\mathrm{N}^{\ast}\times\mathfrak{n},\tilde{\omega}_{o}%
)\longrightarrow(\mathfrak{h},\{,\}_{Aff})$\textit{\ as}\
\begin{equation}
\tilde{\mu}_{0,\alpha}^{\varphi}(\tilde{h},X):=z=Ad_{\tilde{h}}^{\mathrm{H}%
}X+C_{-\alpha}\left(  \tilde{h}\right)  +\alpha=Ad_{\tilde{h}}^{\mathrm{H}%
}\left(  X+\alpha\right)  +C\left(  \tilde{h}\right)  \label{mu-dual}%
\end{equation}
\textit{ } \textit{Then, it} \textit{is Poisson iff condition }$\left(
\ref{loop-22a}\right)  $ \textit{is satisfied: }%
\[
\Pi_{\mathfrak{n}^{\ast}}\left[  \alpha,X\right]  =0
\]
$\forall X\in\mathfrak{n}$. \textit{Here }$\tilde{\omega}_{o}$ \textit{is the
canonical }$2$\textit{-form on }$\mathrm{N}^{\ast}\times\mathfrak{n}$
\textit{trivialized by left translations.}
\end{description}

\textbf{Proof: }Let us sketch the guiding lines of this proof. It must be
proved, for $\mathcal{F},\mathcal{G}\in C^{\infty}\left(  \mathfrak{h}^{\ast
}\right)  $, that%
\[
\left\{  \mathcal{F}\circ\tilde{\mu}_{0,\alpha}^{\varphi},\mathcal{G}%
\circ\tilde{\mu}_{0,\alpha}^{\varphi}\right\}  _{\mathrm{N}^{\ast}%
\times\mathfrak{n}}\left(  \tilde{h},X\right)  =\left\{  \mathcal{F}%
,\mathcal{G}\right\}  _{Aff}\left(  \tilde{\mu}_{0,\alpha}^{\varphi}\left(
\tilde{h},X\right)  \right)
\]
The Poisson bracket $\left\{  ,\right\}  _{\mathrm{N}^{\ast}\times
\mathfrak{n}}$ is the symplectic one corresponding to $\tilde{\omega}_{o}$:
\[
\left\{  \mathcal{F}\circ\tilde{\mu}_{0,\alpha}^{\varphi},\mathcal{G}%
\circ\tilde{\mu}_{0,\alpha}^{\varphi}\right\}  _{\mathrm{N}^{\ast}%
\times\mathfrak{n}}\left(  \tilde{h},X\right)  =\left\langle d\left(
\mathcal{G}\circ\tilde{\mu}_{0,\alpha}^{\varphi}\right)  ,V_{\mathcal{F}%
\circ\tilde{\mu}_{0,\alpha}^{\varphi}}\right\rangle
\]
where the hamiltonian vector field is $V_{f}=\left(  \tilde{h}\delta
f,ad_{\delta f}^{\ast}X-\tilde{h}\mathbf{d}f\right)  $ and $df=\mathbf{d}%
f+\delta f\in\mathfrak{n}^{\ast}\oplus\mathfrak{n}$, for $f\in C^{\infty
}\left(  \mathrm{N}^{\ast}\times\mathfrak{n}\right)  $. The explicit
expression for the differentials is
\begin{align*}
\mathbf{d}\left(  \mathcal{F}\circ\tilde{\mu}_{0,\alpha}^{\varphi}\right)   &
=\tilde{h}^{-1}\Pi_{\mathfrak{n}}ad_{X}^{\mathfrak{h}}Ad_{\tilde{h}%
}^{\mathrm{H}\ast}d\mathcal{F}+\tilde{h}^{-1}\Pi_{\mathfrak{n}}c_{-\alpha
}\left(  Ad_{\tilde{h}}^{\mathrm{H}\ast}d\mathcal{F}\right) \\
\delta\left(  \mathcal{F}\circ\tilde{\mu}_{0,\alpha}^{\varphi}\right)   &
=\Pi_{\mathfrak{n}^{\ast}}Ad_{\tilde{h}}^{\mathrm{H}\ast}d\mathcal{F}%
\end{align*}
that leads to%
\[%
\begin{array}
[c]{l}%
\left\{  \mathcal{F}\circ\tilde{\mu}_{0,\alpha}^{\varphi},\mathcal{G}%
\circ\tilde{\mu}_{0,\alpha}^{\varphi}\right\}  _{\mathrm{N}^{\ast}%
\times\mathfrak{n}}\left(  \tilde{h},X\right) \\
\qquad=\left\langle \Pi_{\mathfrak{n}}ad_{X}^{\mathfrak{h}}Ad_{\tilde{h}%
}^{\mathrm{H}\ast}d\mathcal{G}+\Pi_{\mathfrak{n}}\hat{c}\left(  Ad_{\tilde{h}%
}^{\mathrm{H}\ast}d\mathcal{G}\right)  ,\Pi_{\mathfrak{n}^{\ast}}Ad_{\tilde
{h}}^{\mathrm{H}\ast}d\mathcal{F}\right\rangle \\
\qquad\mathrm{~~~~}+\left\langle \Pi_{\mathfrak{n}}\left[  \Pi_{\mathfrak{n}%
^{\ast}}Ad_{\tilde{h}}^{\mathrm{H}\ast}d\mathcal{G},\alpha\right]
+\Pi_{\mathfrak{n}}\left[  \Pi_{\mathfrak{n}}Ad_{\tilde{h}}^{\mathrm{H}\ast
}d\mathcal{G},\alpha\right]  ,\Pi_{\mathfrak{n}^{\ast}}Ad_{\tilde{h}%
}^{\mathrm{H}\ast}d\mathcal{F}\right\rangle \\
\qquad\mathrm{~~~~}+\left\langle \Pi_{\mathfrak{n}^{\ast}}Ad_{\tilde{h}%
}^{\mathrm{H}\ast}d\mathcal{G},-ad_{\Pi_{\mathfrak{n}^{\ast}}Ad_{\tilde{h}%
}^{\mathrm{H}\ast}d\mathcal{F}}^{\ast}X-\Pi_{\mathfrak{n}}ad_{X}%
^{\mathfrak{h}}Ad_{\tilde{h}}^{\mathrm{H}\ast}d\mathcal{F}-\Pi_{\mathfrak{n}%
}\hat{c}\left(  Ad_{\tilde{h}}^{\mathrm{H}\ast}d\mathcal{F}\right)
\right\rangle \\
\qquad\mathrm{~~~~}-\left\langle \Pi_{\mathfrak{n}^{\ast}}Ad_{\tilde{h}%
}^{\mathrm{H}\ast}d\mathcal{G},\Pi_{\mathfrak{n}}\left[  \Pi_{\mathfrak{n}%
^{\ast}}Ad_{\tilde{h}}^{\mathrm{H}\ast}d\mathcal{F},\alpha\right]
-\Pi_{\mathfrak{n}}\left[  \Pi_{\mathfrak{n}}Ad_{\tilde{h}}^{\mathrm{H}\ast
}d\mathcal{F},\alpha\right]  \right\rangle
\end{array}
\]
If $\hat{c}$ is defined as in eq. $\left(  \ref{canon}\right)  $ the
restrictions $\left.  \hat{c}\right\vert _{\mathfrak{n}}:\mathfrak{n}%
\rightarrow\mathfrak{n}$ and $\left.  \hat{c}\right\vert _{\mathfrak{n}^{\ast
}}:\mathfrak{n}^{\ast}\rightarrow\mathfrak{n}^{\ast}$ are assumed to be
endomorphism of vector spaces. Then we write $\Pi_{\mathfrak{n}^{\ast}}\hat
{c}\left(  Ad_{\tilde{h}}^{\mathrm{H}\ast}d\mathcal{G}\right)  =\hat{c}\left(
\Pi_{\mathfrak{n}^{\ast}}Ad_{\tilde{h}}^{\mathrm{H}\ast}d\mathcal{G}\right)  $
and $\Pi_{\mathfrak{n}^{\ast}}c\left(  Ad_{\tilde{h}}^{\mathrm{H}\ast
}d\mathcal{F}\right)  =c\left(  \Pi_{\mathfrak{n}^{\ast}}Ad_{\tilde{h}%
}^{\mathrm{H}\ast}d\mathcal{F}\right)  $. Besides these, we use also the
relations
\begin{align*}
\left\langle \Pi_{\mathfrak{n}^{\ast}}ad_{Ad_{\tilde{h}}^{\mathrm{H}\ast
}d\mathcal{G}}^{\mathfrak{h}}X,\Pi_{\mathfrak{n}}Ad_{\tilde{h}}^{\mathrm{H}%
\ast}d\mathcal{F}\right\rangle  &  =\left\langle ad_{X}^{\ast}\,\Pi
_{\mathfrak{n}^{\ast}}Ad_{\tilde{h}}^{\mathrm{H}\ast}d\mathcal{G}%
,\Pi_{\mathfrak{n}}Ad_{\tilde{h}}^{\mathrm{H}\ast}d\mathcal{F}\right\rangle \\
& \\
\left\langle \Pi_{\mathfrak{n}}Ad_{\tilde{h}}^{\mathrm{H}\ast}d\mathcal{G}%
,\Pi_{\mathfrak{n}^{\ast}}ad_{Ad_{\tilde{h}}^{\mathrm{H}\ast}d\mathcal{F}%
}^{\mathfrak{h}}X\right\rangle  &  =\left\langle \Pi_{\mathfrak{n}}%
Ad_{\tilde{h}}^{\mathrm{H}\ast}d\mathcal{G},ad_{X}^{\ast}\,\Pi_{\mathfrak{n}%
^{\ast}}Ad_{\tilde{h}}^{\mathrm{H}\ast}d\mathcal{F}\right\rangle
\end{align*}
and the Lie bracket in the double Lie algebra
\[
\left[  \left(  X,\eta\right)  ,\left(  Z,\xi\right)  \right]  _{\mathfrak{h}%
}=\left(  \left[  X,Z\right]  _{\mathfrak{n}}-ad_{\eta}^{\ast}Z+ad_{\xi}%
^{\ast}X,\left[  \eta,\xi\right]  _{\mathfrak{n}^{\ast}}-ad_{X}^{\ast}%
\xi+ad_{Z}^{\ast}\eta\right)
\]
for $\left(  X,\eta\right)  ,\left(  Z,\xi\right)  \in\mathfrak{h}$. After a
tedious but straightforward calculations, one arrives to%

\[%
\begin{array}
[c]{l}%
\left\{  \mathcal{F}\circ\tilde{\mu}_{0,\alpha}^{\varphi},\mathcal{G}%
\circ\tilde{\mu}_{0,\alpha}^{\varphi}\right\}  _{\mathrm{N}^{\ast}%
\times\mathfrak{n}}\left(  \tilde{h},X\right)  -\left\{  \mathcal{F}%
,\mathcal{G}\right\}  _{Aff}\left(  \tilde{\mu}_{0,\alpha}^{\varphi}\left(
\tilde{h},X\right)  \right) \\
~~=\left\langle \Pi_{\mathfrak{n}}Ad_{\tilde{h}}^{\mathrm{H}\ast}%
d\mathcal{G},\Pi_{\mathfrak{n}^{\ast}}\left[  Ad_{\tilde{h}}^{\mathrm{H}\ast
}d\mathcal{F},\alpha\right]  \right\rangle +\left\langle \Pi_{\mathfrak{n}%
}\left[  Ad_{\tilde{h}}^{\mathrm{H}\ast}d\mathcal{G},\alpha\right]
,\Pi_{\mathfrak{n}^{\ast}}Ad_{\tilde{h}}^{\mathrm{H}\ast}d\mathcal{F}%
\right\rangle
\end{array}
\]
implying that $\tilde{\mu}_{0,\alpha}^{\varphi}$ is a Poisson map provided the
right hand side vanish for arbitrary $d\mathcal{G},d\mathcal{F}\in
\mathfrak{h}$. After some manipulations, it reduces to%
\[
\left\langle Ad_{\tilde{h}}^{\ast}\Pi_{\mathfrak{n}}d\mathcal{G}%
,\Pi_{\mathfrak{n}^{\ast}}\left[  Ad_{\tilde{h}}^{\ast}\Pi_{\mathfrak{n}%
}d\mathcal{F},\alpha\right]  \right\rangle =0
\]
that is equivalent to require that%
\[
\Pi_{\mathfrak{n}^{\ast}}\left[  \alpha,X\right]  =0
\]
for all $X\in\mathfrak{n}$. $\blacksquare$

\begin{description}
\item[Example:] \emph{(Lu-Weinstein doubles }\cite{Lu-We}\emph{)}%
\textbf{\ }\textit{When} $\mathrm{N}^{\ast}=K$ \textit{a compact simple real
Lie group, e.g.} $SU(N)$\textit{, then} $\mathrm{H}=AN\times K$ \textit{where}
$\mathrm{N}=AN\;$\textit{and} $\mathrm{N}^{\ast}=K$ \textit{are} \textit{the
subgroups given by the Iwasawa decomposition of} $\mathrm{H}=K^{\mathbb{C}}$.
\textit{For any element} $\alpha\in\mathfrak{t}$\textit{, with} $\mathfrak{t}%
\subset\mathfrak{n}^{\ast}$ \textit{being the Cartan subalgebra of}
${\mathfrak{k}}=Lie(K)$\textit{, condition }$\left(  \ref{loop-22a}\right)
$\textit{ is satisfied (see Appendix 1).}
\end{description}

For $\alpha$ fulfilling condition $\left(  \ref{loop-22a}\right)  $, an
$\mathrm{H}$-hamiltonian action on $\left(  \mathrm{N}^{\ast}\times
\mathfrak{n},\tilde{\omega}_{o}\right)  $ is obtained as stated below.

\begin{description}
\item[Proposition:] \textit{Consider the symplectic manifold }$\left(
\mathrm{N}^{\ast}\times\mathfrak{n},\tilde{\omega}_{o}\right)  $ \textit{where
}$\tilde{\omega}_{o}$ \textit{is the canonical }$2$\textit{-form in body
coordinates. The map }$\mathsf{\hat{b}}:\mathrm{H}_{c,0}\times\left(
\mathrm{N}^{\ast}\times\mathfrak{n}\right)  \longrightarrow\left(
\mathrm{N}\times\mathfrak{n}^{\ast}\right)  $ \textit{defined as}%
\begin{equation}
\mathsf{\hat{b}}\left(  \tilde{b}a,\left(  \tilde{h},X\right)  \right)
=\left(  \tilde{b}\tilde{h}_{a},Ad_{a_{\tilde{h}}}^{\mathrm{H}}X+C_{-\alpha
}\left(  a_{\tilde{h}}\right)  \right)  \label{alpha H action}%
\end{equation}
\textit{is a hamiltonian }$\mathrm{H}$\textit{-action and }$\tilde{\mu
}_{0,\alpha}^{\varphi}$ \textit{is the associated }$\widehat{Ad}%
_{0;}^{\mathrm{H}\ast}$-\textit{equivariant the momentum map}.
\end{description}

\begin{description}
\item[Remark] \textit{That} $\mathsf{\hat{b}}$ \textit{as defined above is an
action on} $\mathrm{N}^{\ast}\times\mathfrak{n}$ \textit{follows from the fact
that, when }$\left(  \ref{loop-22a}\right)  $\textit{ is satisfied, then }%
$\Pi_{\mathfrak{n}^{\ast}}C_{-\alpha}\left(  a_{\tilde{h}}\right)
=0$\textit{. Thus, the expression }$Ad_{a_{\tilde{h}}}^{\mathrm{H}%
}X+C_{-\alpha}\left(  a_{\tilde{h}}\right)  $ \textit{is always }%
$\mathfrak{n}$-\textit{valued as it should be.}
\end{description}

Consequently, we have obtained the desired third hamiltonian $\mathrm{H}%
$-space, namely $(\mathrm{N}^{\ast}\times\mathfrak{n},\tilde{\omega}%
_{o})\equiv$ $(T^{\ast}\mathrm{N}^{\ast},\tilde{\omega}_{o}),$ completing the
diagram
\begin{equation}
\begin{diagram}[h=1.9em] T^{\ast}\mathrm{N} &&& \mathrm{H}/\mathrm{H}_{\alpha} &&& T^{\ast}\mathrm{N}^{\ast} \\ & \rdTo(3,3)_{\mu_{0,0}} &&\dTo^{\hat{\Phi}_{c,0}} && \ldTo(3,3)_{\tilde{\mu}^{\varphi}_{0,\alpha}} & \\ &&&&&&\\ &\mathcal{O}_{c,0}(\alpha,1)& \hookrightarrow & (\mathfrak{h}_{c,0},\{,\}_{c,0}) &&&\\ \end{diagram} \label{diag alfa}%
\end{equation}

\paragraph{Hamiltonian $\mathrm{H}_{c,-\alpha}$-spaces}

Again, the equivalence stated in section \ref{subsec: sympl equiv thetas}
enables to seek for a similar system now hanging on\textit{ }$\mathcal{O}%
_{c,-\alpha}\left(  0,1\right)  \subset\mathfrak{h}_{c,-\alpha}^{\ast}$. In
doing so, we consider the map $\tilde{\mu}^{\alpha}:(\mathrm{N}^{\ast}%
\times\mathfrak{n},\tilde{\omega}_{o})\longrightarrow(\mathfrak{h}%
,\{,\}_{c,-\alpha}^{Aff})$ defined as
\begin{equation}
\tilde{\mu}^{\alpha}(\tilde{h},X)=Ad_{\tilde{h}}^{\mathrm{H}}X+C_{-\alpha
}\left(  \tilde{h}\right)  \label{mu-dual-alpha}%
\end{equation}
\textit{ }which is Poisson for the corresponding shifted affine Poisson
structure on $\mathfrak{h}\simeq\mathfrak{h}^{\ast}$
\[
\{\left(  \mathbf{X},-\right)  _{\mathfrak{h}},\left(  \mathbf{Y},-\right)
_{\mathfrak{h}}\}_{c,-\alpha}^{Aff}(\mathbf{Z})=-\left(  [\mathbf{X}%
,\mathbf{Y}],\mathbf{Z}\right)  _{\mathfrak{h}}-c_{-\alpha}(\mathbf{X}%
,\mathbf{Y})
\]
for $\mathbf{X},\mathbf{Y,Z}\in\mathfrak{h}$, hence it is a momentum map for
an associated action of $\mathrm{H}_{c,-\alpha}$ on $\mathrm{N}^{\ast}%
\times\mathfrak{n}$.

\section{Poisson Lie $T$-Duality}

We now translate the approach to Poisson Lie $T$-duality developed in
\cite{Ale-Hugo} to the hamiltonian \textrm{H}-spaces studied in previous section.

In order to connect hamiltonian vector fields on the phase spaces $T^{\ast
}\mathrm{N}$ and $T^{\ast}\mathrm{N}^{\ast}$, we consider some coadjoint orbit
lying in the intersection of the images of the corresponding equivariant
momentum maps associated to the \textrm{H}-actions. As explained in that
reference, PL $T$-duality holds on some subspaces of these phases spaces,
namely the $\emph{dualizable}$ \emph{subspaces}. These subspaces are
identified as the symplectic leaves of the presymplectic submanifolds obtained
by taking the pre-images, under the corresponding momentum map, of the
coadjoint orbit which we are regarding. Compatible dynamics are then
implemented by \emph{collective} \emph{hamiltonian} functions, and Poisson Lie
$T$-duality works on theses leaves mapping the solutions of these underlying
\emph{collective dynamics}.

In the following, we proceed to describe these dualizable subspaces and the PL
$T$-duality scheme for the phase spaces described above.

\subsection{Dualizable subspaces}

Let us consider first the phase spaces on $T^{\ast}\mathrm{N.}$ In these
cases, the \emph{dualizable subspaces} are the symplectic leaves in the
pre-images of the coadjoint orbit $\mathcal{O}_{c,0}\left(  \alpha,1\right)
\subset\mathfrak{h}_{c,0}^{\ast}$ by $\mu_{0,0}$, or in $\mathcal{O}%
_{c,-\alpha}\left(  0,1\right)  \subset\mathfrak{h}_{c,\mathbf{-\alpha}}%
^{\ast}$ by $\mu_{0,\alpha}$, with $\alpha\in\mathfrak{n}^{\ast}$ in both cases.

In the first case, recall the corresponding momentum map $\mu_{0,0}$ is
$\left(  \ref{prop: action on N without shift}\right)  $. The pre-image
$\mu_{0,0}^{-1}\left(  \mathcal{O}_{c,0}\left(  \alpha,1\right)  \right)  $
consists of those $\left(  g,\lambda\right)  \in\mathrm{N}\times
\mathfrak{n}^{\ast}$ such that $\mu_{0,0}\left(  g,\lambda\right)
=\psi\left(  \widehat{Ad}_{0;a\tilde{b}}^{\mathrm{H}}\left(  \alpha,1\right)
\right)  $ for some $a\tilde{b}\in\mathrm{H}$. Then, using definition $\left(
\ref{d-action-cent-ext}\right)  $, we have%
\begin{align*}
\mu_{0,0}^{-1}\left(  \mathcal{O}_{c,0}\left(  \alpha,1\right)  \right)   &
=\left\{  \left(  g,Ad_{\tilde{k}}^{\mathrm{H}}\alpha+C\left(  \tilde
{k}\right)  \right)  \in\mathrm{N}\times\mathfrak{n}^{\ast}\ /\ g\tilde{k}%
\in\mathrm{H}\right\} \\
&  =\left\{  \mathsf{\hat{d}}_{0}^{\mathrm{N}\times\mathfrak{n}^{\ast}}\left(
g\tilde{k},\left(  e,\alpha\right)  \right)  \ /\ g\tilde{k}\in\mathrm{H}%
\right\}
\end{align*}
giving rise to the following statement.

\begin{description}
\item[Proposition:] $\mu_{0,0}^{-1}\left(  \mathcal{O}_{c,0}\left(
\alpha,1\right)  \right)  \equiv\mathcal{O}_{\mathrm{N}\times\mathfrak{n}%
^{\ast}}\left(  e,\alpha\right)  $\textit{, where }$\mathcal{O}_{\mathrm{N}%
\times\mathfrak{n}^{\ast}}\left(  e,\alpha\right)  $ \textit{is the orbit
through }$\left(  e,\alpha\right)  \in$\textit{\ }$\mathrm{N}\times
\mathfrak{n}^{\ast}$\textit{\ under the action }$\mathsf{\hat{d}}%
_{0}^{\mathrm{N}\times\mathfrak{n}^{\ast}}:\mathrm{H}_{c,0}\times\left(
\mathrm{N}\times\mathfrak{n}^{\ast}\right)  \rightarrow\left(  \mathrm{N}%
\times\mathfrak{n}^{\ast}\right)  $, \textit{eq.\ }$\left(
\ref{d-action-cent-ext}\right)  $.
\end{description}

Thus, every tangent vector $\mathbf{V}\in T_{\left(  g,\lambda\right)  }%
\mu_{0,0}^{-1}\left(  \mathcal{O}_{c,0}\left(  \alpha,1\right)  \right)  $
looks like%
\[
\left.  \mathbf{V}\right\vert _{\left(  g,\lambda\right)  }=\mathsf{\hat{d}%
}_{0}^{\mathrm{N}\times\mathfrak{n}^{\ast}}\left(  \mathbf{X}\right)
_{\left(  g,\lambda\right)  }%
\]
for some $\mathbf{X}\in\mathfrak{h}$, showing that $T\mu_{0,0}^{-1}\left(
\mathcal{O}_{c,0}\left(  \alpha,1\right)  \right)  =\mu_{0,0\ast}%
^{-1}T\mathcal{O}_{c,0}\left(  \alpha,1\right)  $. Hence, following a theorem
by Kazhdan, Kostant and Sternberg \cite{KKS}, we conclude $\mu_{0,0}%
^{-1}\left(  \mathcal{O}_{c,0}\left(  \alpha,1\right)  \right)  $ is a
coisotropic submanifold, and the null distribution of the presymplectic form
is spanned the infinitesimal generators of $\left[  \mathrm{H}_{c,0}\right]
_{\left(  \alpha,1\right)  }$, the stabilizer subgroup of $\left(
\alpha,1\right)  \in\mathfrak{h}_{c,0}^{\ast}$. We then have the next statement:

\begin{description}
\item[Proposition:] $\mu_{0,0}^{-1}\left(  \mathcal{O}_{c,0}\left(
\alpha,1\right)  \right)  $ \textit{is a presymplectic submanifold with the
closed }$2$\textit{-form given by the restriction of the canonical form
}$\omega_{o}$\textit{,\ }%
\[
\left\langle \omega_{o},\mathsf{\hat{d}}_{0}^{\mathrm{N}\times\mathfrak{n}%
^{\ast}}\left(  \mathbf{X}\right)  \otimes\mathsf{\hat{d}}_{0}^{\mathrm{N}%
\times\mathfrak{n}^{\ast}}\left(  \mathbf{Y}\right)  \right\rangle _{\left(
g,\xi\right)  }=\left\langle \left(  C_{\alpha}\left(  a\tilde{b}\right)
+\alpha,1\right)  ,\widehat{ad}_{\mathbf{X}}^{\mathfrak{h}_{c,0}}%
\mathbf{Y}\right\rangle _{\mathfrak{h}_{c,0}}%
\]
\textit{for }$\mathbf{X},\mathbf{Y}\in\mathfrak{h},$\textit{and }$\left(
g,\xi\right)  =\mathsf{\hat{d}}_{\lambda\ast}^{\mathrm{N}\times\mathfrak{n}%
^{\ast}}\left(  a\tilde{b},\left(  e,\alpha\right)  \right)  \in\mu_{0,0}%
^{-1}\left(  \mathcal{O}_{c,0}\left(  \alpha,1\right)  \right)  $. \textit{Its
null distribution is spanned by the infinitesimal generators of the right
action by the stabilizer }$\left[  \mathrm{H}_{c,0}\right]  _{\left(
\alpha,1\right)  }=\ker C_{\alpha}$ \textit{of the point }$\left(
\alpha,1\right)  $ \textit{ }%
\begin{align*}
\mathsf{r}:\left[  \mathrm{H}_{c,0}\right]  _{\left(  \alpha,1\right)  }%
\times\mu_{0,0}^{-1}\left(  \mathcal{O}_{c,0}\left(  \alpha,1\right)  \right)
&  \longrightarrow\mu_{0,0}^{-1}\left(  \mathcal{O}_{c,0}\left(
\alpha,1\right)  \right) \\
\mathsf{r}\left(  l_{\circ},\mathsf{\hat{d}}_{0}^{\mathrm{N}\times
\mathfrak{n}^{\ast}}\left(  a\tilde{b},\left(  e,\alpha\right)  \right)
\right)   &  =\mathsf{\hat{d}}_{0}^{\mathrm{N}\times\mathfrak{n}^{\ast}%
}\left(  a\tilde{b}l_{\circ}^{-1},\left(  e,\alpha\right)  \right)
\end{align*}
\textit{and the null vectors at the point }$\left(  g,\xi\right)  \in\mu
_{0,0}^{-1}\left(  \mathcal{O}_{c,0}\left(  \alpha,1\right)  \right)  $
\textit{are}%
\[
\mathsf{r}_{\ast\left(  g,\xi\right)  }Z_{\circ}=-\mathsf{\hat{d}}%
_{0}^{\mathrm{N}\times\mathfrak{n}^{\ast}}\left(  Ad_{a\tilde{b}}^{D}Z_{\circ
}\right)  _{\left(  g,\xi\right)  }%
\]
\textit{for all }$Z_{o}\in\mathrm{Lie}\left(  \ker C_{\alpha}\right)
\subset\mathfrak{h}$.
\end{description}

On the other hand, in the equivalent \emph{shifted} formulation explained in
$\left(  \ref{prop: H action on N shifted formulation}\right)  $, i.e. when
considering the moment map $\mu_{0,\alpha}\ $taking values in the shifted
$\mathfrak{h}_{c,-\alpha}^{\ast}$, and when condition $\left(  \ref{loop-22a}%
\right)  $ is satisfied, the level set $\mu_{0,\alpha}^{-1}\left(
\mathcal{O}_{c,-\alpha}\left(  0,1\right)  \right)  $ is
\[
\mu_{0,\alpha}^{-1}\left(  \mathcal{O}_{c,-\alpha}\left(  0,1\right)  \right)
=\left\{  \left(  g,C_{-\alpha}\left(  \tilde{b}\right)  \right)
\in\mathrm{N}\times\mathfrak{n}^{\ast}\ /\ g\in\mathrm{N},~\tilde{b}%
\in\mathrm{N}^{\ast}\right\}
\]
It coincides with the $\mathrm{H}_{c,-\alpha}$-orbit through $\left(
e,0\right)  $ in $\mathrm{N}\times\mathfrak{n}^{\ast}$
\[
\mathcal{O}_{\mathrm{N}\times\mathfrak{n}^{\ast}}\left(  e,0\right)  =\left\{
\mathsf{\hat{d}}_{\alpha}^{\mathrm{N}\times\mathfrak{n}^{\ast}}\left(
g\tilde{b},\left(  e,0\right)  \right)  \in\mathrm{N}\times\mathfrak{n}^{\ast
}\ /\ g\in\mathrm{N},~\tilde{b}\in\mathrm{N}^{\ast}\right\}
\]
since $\mathsf{\hat{d}}_{\alpha}^{\mathrm{N}\times\mathfrak{n}^{\ast}}\left(
g\tilde{b},\left(  e,0\right)  \right)  =\left(  g,C_{-\alpha}\left(
\tilde{b}\right)  \right)  $. Tangent vectors $\mathbf{W}$ to $\mu_{0,\alpha
}^{-1}\left(  \mathcal{O}_{c,-\alpha}\left(  0,1\right)  \right)  $ at the
point $\left(  g,\xi\right)  $ are of the form
\[
\left.  \mathbf{W}\right\vert _{\left(  g,\xi\right)  }=\mathsf{\hat{d}%
}_{\alpha\ast}^{\mathrm{N}\times\mathfrak{n}^{\ast}}\left(  \mathbf{X}\right)
_{\left(  g,\xi\right)  }%
\]
for $\mathbf{X}\in\mathfrak{h}$, and then, as above, $\mu_{0,\alpha}%
^{-1}\left(  \mathcal{O}_{c,-\alpha}\left(  0,1\right)  \right)  $ is a
coisotropic submanifold yielding the analogous result:

\begin{description}
\item[Proposition:] $\mu_{0,\alpha}^{-1}\left(  \mathcal{O}_{c,-\alpha}\left(
0,1\right)  \right)  $ \textit{is a presymplectic submanifold with the closed
}$2$\textit{-form given by the restriction of the canonical form }$\omega_{o}%
$,\textit{\ }%
\[
\left\langle \omega_{o},\mathsf{\hat{d}}_{\alpha\ast}^{\mathrm{N}%
\times\mathfrak{n}^{\ast}}\left(  \mathbf{X}\right)  \otimes\mathsf{\hat{d}%
}_{\alpha\ast}^{\mathrm{N}\times\mathfrak{n}^{\ast}}\left(  \mathbf{Y}\right)
\right\rangle _{\left(  g,\xi\right)  }=\left\langle \left(  C\left(
a\tilde{b}\right)  ,1\right)  ,\widehat{ad}_{\mathbf{X}}^{\mathfrak{h}%
_{c,-\alpha}}\mathbf{Y}\right\rangle _{\mathfrak{h}_{c,-\alpha}}%
\]
\textit{for }$\mathbf{X},\mathbf{Y}\in\mathfrak{h},$\textit{and }$\left(
g,\xi\right)  =\mathsf{\hat{d}}_{\alpha\ast}^{\mathrm{N}\times\mathfrak{n}%
^{\ast}}\left(  a\tilde{b},\left(  e,0\right)  \right)  \in\mu_{0,\alpha}%
^{-1}\left(  \mathcal{O}_{c,-\alpha}\left(  0,1\right)  \right)  $.
\textit{Its null distribution is spanned by the infinitesimal generators of
the right action by the stabilizer }$\left[  \mathrm{H}_{c,-\alpha}\right]
_{\left(  0,1\right)  }:=\ker C_{-\alpha}$ \textit{of the point }$\left(
0,1\right)  $\textit{ }%
\begin{align*}
\mathsf{r}:\left[  \mathrm{H}_{c,-\alpha}\right]  _{\left(  0,1\right)
}\times\mu_{0,\alpha}^{-1}\left(  \mathcal{O}_{c,-\alpha}\left(  0,1\right)
\right)   &  \longrightarrow\mu_{0,\alpha}^{-1}\left(  \mathcal{O}_{c,-\alpha
}\left(  0,1\right)  \right) \\
\mathsf{r}\left(  l_{\circ},\mathsf{\hat{d}}_{\alpha\ast}^{\mathrm{N}%
\times\mathfrak{n}^{\ast}}\left(  a\tilde{b},\left(  e,0\right)  \right)
\right)   &  =\mathsf{\hat{d}}_{\alpha\ast}^{\mathrm{N}\times\mathfrak{n}%
^{\ast}}\left(  a\tilde{b}l_{\circ}^{-1},\left(  e,0\right)  \right)
\end{align*}
\textit{and the null vectors at the point }$\mathsf{\hat{d}}_{0}%
^{\mathrm{N}\times\mathfrak{n}^{\ast}}\left(  a\tilde{b},\left(  e,0\right)
\right)  \in\mu_{0,\alpha}^{-1}\left(  \mathcal{O}\left(  0,1\right)  \right)
$ \textit{are}%
\[
\mathsf{r}_{\ast\left(  g,\xi\right)  }Z_{o}=-\mathsf{\hat{d}}_{\alpha\ast
}^{\mathrm{N}\times\mathfrak{n}^{\ast}}\left(  Ad_{a\tilde{b}}^{D}Z_{\circ
}\right)  _{\left(  g,\xi\right)  }%
\]
\textit{for all }$Z_{\circ}\in\mathrm{Lie}\left(  \ker C\right)
\subset\mathfrak{h}$\textit{.}
\end{description}

For the phase spaces on $T^{\ast}\mathrm{N}^{\ast}$, we now consider the
symplectic manifold $\left(  \mathrm{N}^{\ast}\times\mathfrak{n},\tilde
{\omega}_{o}\right)  $ which are involved in the current scheme for
$T$-duality. Hence, the \emph{dualizable subspaces} are contained in the
submanifold $\left(  \tilde{\mu}_{0,\alpha}^{\varphi}\right)  ^{-1}\left(
\mathcal{O}_{c,0}\left(  \alpha,1\right)  \right)  $.

\begin{description}
\item[Proposition:] \textit{Let} $\alpha\in\mathfrak{n}^{\ast}$
\textit{satisfying} $\left(  \ref{loop-22a}\right)  $\textit{. Then,}
\[
\left(  \tilde{\mu}_{0,\alpha}^{\varphi}\right)  ^{-1}\left(  \mathcal{O}%
_{c,0}\left(  \alpha,1\right)  \right)  =\left\{  \mathsf{\hat{b}}\left(
\tilde{g}h,\left(  e,0\right)  \right)  \in\mathrm{N}^{\ast}\times
\mathfrak{n~}/\ \tilde{g}h\in\mathrm{H}\right\}
\]
\textit{where the} $\mathrm{H}$\textit{-action} $\mathsf{\hat{b}}$ \textit{is
given by} $\left(  \ref{alpha H action}\right)  $.
\end{description}

As in the model over $\mathrm{N}$, the restriction to $\left(  \tilde{\mu
}_{0,\alpha}^{\varphi}\right)  ^{-1}\left(  \mathcal{O}_{c,0}\left(
\alpha,1\right)  \right)  $ of the symplectic structure$\ $is degenerate.

\begin{description}
\item[Proposition:] $\left(  \tilde{\mu}_{0,\alpha}^{\varphi}\right)
^{-1}\left(  \mathcal{O}_{c,0}\left(  \alpha,1\right)  \right)  $\textit{ is a
presymplectic submanifold with closed two form obtained from the restriction
of the canonical form }$\tilde{\omega}_{o}$. \textit{The null distribution is
spanned by the infinitesimal generators of the right action of }%
$\mathrm{H}_{\left(  \alpha,1\right)  }^{c,0}:=\ker C_{-\alpha}$\textit{ }%
\begin{align*}
\mathsf{r}:\mathrm{H}_{\left(  -\eta,1\right)  }^{c,0}\times\left(  \tilde
{\mu}_{0,\alpha}^{\varphi}\right)  ^{-1}\left(  \mathcal{O}_{c,0}\left(
\alpha,1\right)  \right)   &  \longrightarrow\left(  \tilde{\mu}_{0,\alpha
}^{\varphi}\right)  ^{-1}\left(  \mathcal{O}_{c,0}\left(  \alpha,1\right)
\right) \\
\mathsf{r}\left(  l_{o},\mathsf{\hat{d}}_{0}^{\mathrm{N}\times\mathfrak{n}%
^{\ast}}\left(  \tilde{a}b,\left(  e,0\right)  \right)  \right)   &
=\mathsf{\hat{d}}_{0}^{\mathrm{N}\times\mathfrak{n}^{\ast}}\left(  \tilde
{a}bl_{o}^{-1},\left(  e,0\right)  \right)  .
\end{align*}

\end{description}

Analogous characterizations hold in the corresponding \emph{shifted}
formulation for $\mathrm{N}^{\ast}$.

\subsection{The PL $T$-duality scheme}

\label{subsec: duality}So far, we have considered independently sigma model
like models on the Lie groups $\mathrm{N}$, $\mathrm{N}^{\ast}$ and a WZNW
like model on the double $\mathrm{H}=\mathrm{N}\Join\mathrm{N}^{\ast}$. When
$\alpha\in\mathfrak{n}^{\ast}$ fulfills $\left(  \ref{loop-22a}\right)  $, the
hamiltonian $\mathrm{H}$-actions described in the previous sections give rise
to the following diagram including all the involved phase spaces%

\begin{equation}
\begin{diagram}[h=1.9em] (\mathcal{O}_{c,\mathbf{-\alpha}}\left( 0,1\right),\omega^{KK})&&\rTo^{{\hat{\Phi}_{c,\mathbf{-\alpha}}}}&&(\mathfrak{h}_{c,\mathbf{-\alpha}}^{\ast};\{ , \}_{c,\mathbf{-\alpha} })&&&&\\&&&&&&&&\\ && \ruTo^{\mu_{0,\alpha}} && && \luTo^{{\tilde{\mu}_{0,\alpha}}} &&\\ &&&& &&&&\\ (T^{\ast}\mathrm{N},\omega_o)&&&& \dTo^{\varphi} &&&&(T^{\ast}\mathrm{N}^{\ast},\tilde\omega_o)\\ &&&&&&&&\\ &&\rdTo_{\mu_{0,0}}&& &&\ldTo_{{\tilde{\mu}_{0,\alpha}^{\varphi}}}&&\\ &&&&&&&&\\ (\mathcal{O}_{c,\mathbf{0}}\left( \alpha ,1\right),\omega^{KK})& &\rTo^{{\hat{\Phi}_{c,0}}}& &(\mathfrak{h}_{c,0}^{\ast};\{ , \}_{c,0 })&&&&\\ \end{diagram} \label{dgrm: dualidad 1}%
\end{equation}

Using the \emph{geometrical} or \emph{kinematical} information of this diagram
$\left(  \ref{dgrm: dualidad 1}\right)  $, following \cite{Ale-Hugo} and
section \ref{sec: setting}, we know that by considering \emph{collective
dynamics} coming from an $\mathsf{h}:\left(  \mathfrak{h}_{c,0}\right)
^{\ast}\rightarrow\mathbb{R}$, the resulting hamiltonian systems:
\[
\left\{
\begin{array}
[c]{l}%
\left(  \mathrm{N}\times\mathfrak{n}^{\ast},\omega_{o},\mathrm{H}_{c,0}%
,\mu_{0,0},\mathsf{h}\circ\mu_{0,0}\right) \\
\\
\left(  \mathcal{O}_{c,0}\left(  \alpha,1\right)  ,\omega^{KK},\mathrm{H}%
_{c,0},\hat{\Phi}_{c,0},\mathsf{h}\circ\hat{\Phi}_{c,0}\right) \\
\\
\left(  \mathrm{N}^{\ast}\times\mathfrak{n},\omega_{o},\mathrm{H}_{c,0}%
,\tilde{\mu}_{0,\alpha}^{\varphi},\mathsf{h}\circ\tilde{\mu}_{0,\alpha
}^{\varphi}\right)
\end{array}
\right.
\]
become dual to each other with respect to $\mathcal{O}_{c,0}\left(
\alpha,1\right)  $.

\begin{description}
\item[Remark:] \emph{(The equivalent shifted formulation)} \textit{In view of
the equivalence stated in section} \ref{subsec: sympl equiv thetas}\textit{,
the collective hamiltonian systems corresponding to the }$\mathfrak{h}%
_{c,-\alpha}^{\ast}$-\textit{valued momentum maps lead to duality with respect
to} $\mathcal{O}_{c,\mathbf{-\alpha}}\left(  0,1\right)  $.
\end{description}

As stated in \ref{sec: setting}, solutions lying on the \emph{dualizable
subspaces }$\mu_{0,0}^{-1}\left(  \mathcal{O}_{c,0}\left(  \alpha,1\right)
\right)  $, $\left(  \tilde{\mu}_{0,\alpha}^{\varphi}\right)  ^{-1}\left(
\mathcal{O}_{c,0}\left(  \alpha,1\right)  \right)  $ (equiv., in the shifted
formulation, $\mu_{0,\alpha}^{-1}\left(  \mathcal{O}_{c,\mathbf{-\alpha}%
}\left(  0,1\right)  \right)  $ and $\left(  \tilde{\mu}^{\alpha}\right)
^{-1}\left(  \mathcal{O}_{c,\mathbf{-}\alpha}\left(  0,1\right)  \right)  $)
are generated by the same curve $l(t)\in\mathrm{H}_{c,0}$ defined by%
\begin{align}
\left(  \frac{dl}{dt}~l^{-1}\right)  (t)  &  =\mathcal{L}_{\mathsf{h}}%
(\gamma(t))\label{coll-trayect-1}\\
l(0)  &  =e\nonumber
\end{align}
where $\gamma(t)\in\mathcal{O}_{c,0}\left(  \alpha,1\right)  \subset
\mathfrak{h}_{c,0}^{\ast}$ is the integral curve of $\mathsf{h}$ for the
WZNW\ like model of Section \ref{sec: Hamiltonian WZ H}. The corresponding
initial values are related by
\[
\mathcal{O}_{c,0}\left(  \alpha,1\right)  \ni\gamma(0)=\mu_{0,0}\left(
g_{o},\eta_{o}\right)  =\tilde{\mu}_{0,\alpha}^{\varphi}\left(  \tilde{h}%
_{o},Z_{o}\right)
\]
where $\left(  g_{o},\eta_{o}\right)  \in\mathrm{N}\times\mathfrak{n}^{\ast}$
and $\left(  \tilde{h}_{o},Z_{o}\right)  \in\mathrm{N}^{\ast}\times
\mathfrak{n}$ stand for the initial conditions for the $\mathrm{N}$ and
$\mathrm{N}^{\ast}$ sigma models of sects. \ref{sec: ham model N} and
\ref{sec: ham model N*}, respectively. Thus, finally, the dual solutions can
be written as:
\[
\left\{
\begin{array}
[c]{l}%
\mathsf{\hat{d}}\left(  l(t),\left(  g_{o},\eta_{o}\right)  \right)
\in\mathrm{N}\times\mathfrak{n}^{\ast}\\
\\
\left[  l(t)l_{o}\right]  \in\mathcal{O}_{c,0}\left(  \alpha,1\right)
\simeq\mathrm{H}/\mathrm{H}_{\mathbf{\alpha}}\\
\\
\mathsf{\hat{b}}\left(  l(t),\left(  \tilde{h}_{o},Z_{o}\right)  \right)
\in\mathrm{N}^{\ast}\times\mathfrak{n}%
\end{array}
\right.
\]

Notice that duality transformations between $T^{\ast}\mathrm{N}$ and $T^{\ast
}\mathrm{N}^{\ast}$ involve finding the curve $l(t)$ of $\left(
\ref{coll-trayect-1}\right)  $ and applying the two factorizations
$\mathrm{H}=\mathrm{N}\bowtie\mathrm{N}^{\ast}\sim\mathrm{N}^{\ast}%
\bowtie\mathrm{N}$. As mentioned before, this duality transformations hold on
the \emph{dualizable} subspaces described above.

\begin{description}
\item[Remark: (Plurality)] \textit{In this context, it is clear that another
decomposition \cite{Plurality}} $\mathrm{H}=\mathrm{M}\bowtie\mathrm{M}^{\ast
}$ \textit{of the same group shall yield \textbf{plurality} between the
corresponding collective models on }$\mathrm{M}$\textit{, }$\mathrm{M}^{\ast}%
$\textit{, }$\mathrm{N}$\textit{ and }$\mathrm{N}^{\ast}$\textit{, as long as
the moment maps associated to the }$\mathrm{H}$-\textit{action intersect
(recall sec. \ref{sec: setting}). Also notice that, if }$P$\textit{ is another
hamiltonian H-space whose moment map intersects the others, then a collective
hamiltonian system on }$P$\textit{ will also be dual to the previous ones. In
the first cases, being the phase spaces cotangent bundles }$T^{\ast}%
\mathrm{N}$\textit{ (}$T^{\ast}\mathrm{N}^{\ast}$\textit{, etc.), it allows
for a lagrangian description in terms of curves (or fields, see next sections)
lying in }$\mathrm{N}$\textit{, (}$\mathrm{N}^{\ast}$\textit{, etc.).}
\end{description}

\section{Associated hamiltonian Systems}

\label{sec: hamiltonian asociados}

In the previous section, we have been concentrated on the kinematic-geometric
aspects of Poisson Lie $T$-duality. In the current section, we describe the
hamiltonian systems' equations associated to the dual phase spaces we have
constructed endowed with the corresponding \emph{collective} dynamics.

\subsection{Master WZNW-type model on\ $(T^{\ast}\mathrm{H},\omega
_{c,\mathbf{\theta}})$}

\label{sec: Hamiltonian WZ H}The equation of motion on the phase space
$(T^{\ast}\mathrm{H},\omega_{c,\mathbf{\theta}})$ described in Section
\ref{sec: WZW phase space}, for some\textit{ }$\mathcal{H}\in C^{\infty
}\left(  \mathrm{H}\times\mathfrak{h}^{\ast}\right)  $, are%
\begin{equation}%
\begin{array}
[c]{l}%
l^{-1}\dot{l}=\delta\mathcal{H}\\
~~~~~~\dot{\eta}=ad_{\delta H}^{\ast}\left(  \eta-C_{\mathbf{\theta}}\left(
l^{-1}\right)  \right)  -\hat{c}_{\theta}\left(  \delta\mathcal{H}\right)
-l\mathbf{d}\mathcal{H}%
\end{array}
\label{ext-eq-mov}%
\end{equation}
for $\left(  l,\eta\right)  \in\mathrm{H}\times\mathfrak{h}^{\ast}$,
with\textit{ }$\left.  d\mathcal{H}\right\vert _{\left(  l,\eta\right)
}=\left(  \mathbf{d}\mathcal{H},\delta\mathcal{H}\right)  _{\left(
l,\eta\right)  }\in T_{\left(  l,\eta\right)  }^{\ast}\left(  \mathrm{H}%
\times\mathfrak{h}^{\ast}\right)  $. By setting $\mathbf{\theta}=0$ one gets
the equation of motion for the \textbf{system 1 }(S1) and, for $\mathbf{\theta
}=-\alpha$, the equation of motion for the \textbf{system 2 }(S2), as
described in that Section.

Collective dynamics warranties Lax type equations, so we consider some
examples with quadratic Hamilton functions including some arbitrary linear
self adjoint operators $\mathbb{L}_{i}:\mathfrak{h}\longrightarrow
\mathfrak{h}$ and $\mathbb{L}_{i}^{\ast}\equiv\psi\circ\mathbb{L}_{i}\circ
\bar{\psi}:\mathfrak{h}^{\ast}\longrightarrow\mathfrak{h}^{\ast}$, $i=1,2,3$,
where $\psi:\mathfrak{h}\longrightarrow\mathfrak{h}^{\ast}$ denotes the
identification induced by the symmetric Ad-invariant non degenerate bilinear
form $\left(  ,\right)  _{\mathfrak{h}}$, and $\bar{\psi}$ means the inverse map.

\begin{description}
\item[S1-] For the symplectic manifold $\left(  \mathrm{H}\times
\mathfrak{h}^{\ast},\omega_{c,0}\right)  $, and their reduced space
$M_{c,0}^{\left(  \mathbf{\alpha},1\right)  }$, we consider
\begin{align}
\mathcal{H}_{c,0}\left(  l,\eta;\alpha\right)   &  =\dfrac{1}{2}\left(
Ad_{l^{-1}}^{\ast}\eta,\mathbb{L}_{3}^{\ast}Ad_{l^{-1}}^{\ast}\eta\right)
_{\mathfrak{h}^{\ast}}+\left(  Ad_{l^{-1}}^{\ast}\eta,\mathbb{L}_{2}^{\ast
}\left(  C_{-\alpha}\left(  l\right)  +\alpha\right)  \right)  _{\mathfrak{h}%
^{\ast}}\label{wznw-ham-1}\\
&  ~~~~~~~~~~~~-\dfrac{1}{2}\left(  \left(  C_{-\alpha}\left(  l\right)
+\alpha\right)  ,\mathbb{L}_{2}^{\ast}\left(  C_{-\alpha}\left(  l\right)
+\alpha\right)  \right)  _{\mathfrak{h}^{\ast}}\nonumber
\end{align}
To realize the collective form of this hamiltonian function, we follow the
results of item \textbf{S1}\textrm{ }in Sec.\textrm{ }%
\ref{sec: WZW phase space} in order to restrict the system to the
Marsden-Weinstein reduced spaces $M_{c,0}^{\left(  \alpha,1\right)  }$. This
means to make $\eta=Ad_{l}^{\ast}C\left(  l\right)  +\alpha$ and, because of
the\ residual left action of $\mathrm{H}$\ on $\mathrm{H}/\mathrm{H}%
_{\mathbf{\alpha}}$ we have the associated momentum map $\hat{\Phi}%
_{c,0}:\mathrm{H}/\mathrm{H}_{\mathbf{\alpha}}\longrightarrow\mathfrak{h}%
_{c,0}^{\ast}$, $\hat{\Phi}_{c,0}\left(  l\cdot\mathrm{H}_{\mathbf{\alpha}%
}\right)  =\left(  C_{-\alpha}\left(  l\right)  +\alpha,1\right)  $, which
allow us retrieve the collective form%
\[
\left.  \mathcal{H}_{c,0}\left(  l;\alpha\right)  \right\vert _{M_{c,0}%
^{\left(  \alpha,1\right)  }}=\dfrac{1}{2}\left(  \hat{\Phi}_{c,0}\left(
l\cdot\mathrm{H}_{\alpha}\right)  ,\left(  \mathbb{L}_{3}^{\ast}%
+\mathbb{L}_{2}^{\ast}\right)  \hat{\Phi}_{c,0}\left(  l\cdot\mathrm{H}%
_{\alpha}\right)  \right)  _{\mathfrak{h}^{\ast}}%
\]
Hence, the Hamilton equations a restricted to
\[
\left[  \hat{J}_{c,0}^{R}\right]  ^{-1}\left(  \alpha,1\right)  =\left\{
\left(  l,Ad_{l}^{\mathrm{H}\ast}C\left(  l\right)  +\alpha\right)
/l\in\mathrm{H}\right\}
\]
are%
\[%
\begin{array}
[c]{l}%
l^{-1}\dot{l}=\bar{\psi}\left(  \left(  \mathbb{L}_{3}^{l\ast}+\mathbb{L}%
_{2}^{l\ast}\right)  \eta\right) \\
\dot{\eta}=ad_{\bar{\psi}\left(  \left(  \mathbb{L}_{3}^{l\ast}+\mathbb{L}%
_{2}^{l\ast}\right)  \eta\right)  }^{\ast}\left(  \eta-\alpha\right)  -\hat
{c}\left(  \bar{\psi}\left(  \left(  \mathbb{L}_{3}^{l\ast}+\mathbb{L}%
_{2}^{l\ast}\right)  \eta\right)  \right)
\end{array}
\]
The Legendre transformation from the first equation leads to the lagrangian
function $L_{c,0}=\left\langle \eta,l^{-1}\dot{g}\right\rangle -H_{c,0}$,%
\begin{align*}
\mathcal{L}_{c,0}=  &  \dfrac{1}{2}\left\langle \mathbb{\bar{L}}_{3}^{\ast
}\psi\left(  \dot{l}l^{-1}\right)  ,\dot{l}l^{-1}\right\rangle -\left\langle
\mathbb{\bar{L}}_{3}^{\ast}\mathbb{L}_{2}^{\ast}\left(  C_{-\alpha}\left(
l\right)  +\alpha\right)  ,l^{-1}\dot{l}\right\rangle \\
&  +\dfrac{1}{2}\left(  \left(  C_{-\alpha}\left(  l\right)  +\alpha\right)
,\left(  \mathbb{L}_{2}^{\ast}+\mathbb{L}_{2}^{\ast}\mathbb{\bar{L}}_{3}%
^{\ast}\mathbb{L}_{2}^{\ast}\right)  \left(  C_{-\alpha}\left(  l\right)
+\alpha\right)  \right)  _{\mathfrak{h}^{\ast}}%
\end{align*}

\item[S2-] For the symplectic manifold $\left(  \mathrm{H}\times
\mathfrak{h}^{\ast},\omega_{c,\mathbf{-\alpha}}\right)  $ and their reduced
space $M_{c,\mathbf{-\alpha}}^{\left(  0,1\right)  }$, we consider the
quadratic Hamilton functions
\begin{align}
\mathcal{H}_{c,-\alpha}\left(  l,\eta;-\alpha\right)   &  =\dfrac{1}{2}\left(
Ad_{l^{-1}}^{\ast}\eta,\mathbb{L}_{3}^{\ast}Ad_{l^{-1}}^{\ast}\eta\right)
_{\mathfrak{h}^{\ast}}+\left(  Ad_{l^{-1}}^{\ast}\eta,\mathbb{L}_{2}^{\ast
}C_{-\alpha}\left(  l\right)  \right)  _{\mathfrak{h}^{\ast}}%
\label{wznw-ham-2}\\
&  ~~~~~~~~~~-\dfrac{1}{2}\left(  C_{-\alpha}\left(  l\right)  ,\mathbb{L}%
_{2}^{\ast}C_{-\alpha}\left(  l\right)  \right)  _{\mathfrak{h}^{\ast}%
}\nonumber
\end{align}
When restricted to $M_{c,-\alpha}^{\left(  0,1\right)  }$, $\eta=Ad_{l}^{\ast
}C_{-\alpha}\left(  l\right)  $, see item \textbf{S2}\textrm{ }in\textrm{
}Sec. \ref{sec: WZW phase space}, it takes the form%
\[
\left.  \mathcal{H}_{c,-\alpha}\left(  l,\eta;-\alpha\right)  \right\vert
_{M_{c,-\alpha}^{\left(  0,1\right)  }}=\dfrac{1}{2}\left(  C_{-\alpha}\left(
l\right)  ,\left(  \mathbb{L}_{2}^{\ast}+\mathbb{L}_{3}^{\ast}\right)
C_{-\alpha}\left(  l\right)  \right)  _{\mathfrak{h}^{\ast}}%
\]
which is collective for the momentum map $\hat{\Phi}_{c,-\alpha}\left(
l\cdot\mathrm{H}_{\alpha}^{c,-\alpha\symbol{94}}\right)  =\left(  C_{-\alpha
}\left(  l\right)  ,1\right)  $. The Hamilton equations of motion reduced to
\[
\left[  \hat{J}_{c,-\alpha}^{R}\right]  ^{-1}\left(  0,1\right)  =\left\{
\left(  l,-C_{-\alpha}\left(  l^{-1}\right)  \right)  /l\in\mathrm{H}\right\}
\]
are%
\[%
\begin{array}
[c]{l}%
l^{-1}\dot{l}=\bar{\psi}\left(  \left(  \mathbb{L}_{3}^{l\ast}+\mathbb{L}%
_{2}^{l\ast}\right)  \eta\right) \\
\dot{\eta}=ad_{\bar{\psi}\left(  \left(  \left(  \mathbb{L}_{3}^{l\ast
}+\mathbb{L}_{2}^{l\ast}\right)  \eta\right)  \right)  }^{\ast}\eta-\hat
{c}_{-\alpha}\left(  \bar{\psi}\left(  \left(  \mathbb{L}_{3}^{l\ast
}+\mathbb{L}_{2}^{l\ast}\right)  \eta\right)  \right)
\end{array}
\]
We use the first equation of motion to invert the Legendre transformation in
order to obtain the lagrangian function $\mathcal{L}_{c,-\alpha}=\left\langle
\eta,l^{-1}\dot{l}\right\rangle -\mathcal{H}_{c,-\alpha}$. Thus, we get%
\begin{align*}
\mathcal{L}_{c,-\alpha}\left(  l,\dot{l}\right)  =  &  \dfrac{1}%
{2}\left\langle \mathbb{\bar{L}}_{3}^{\ast}\psi\left(  \dot{l}l^{-1}\right)
,\dot{l}l^{-1}\right\rangle -\left\langle \mathbb{\bar{L}}_{3}^{\ast
}\mathbb{L}_{2}^{\ast}C_{-\alpha}\left(  l\right)  ,\dot{l}l^{-1}\right\rangle
\\
&  +\dfrac{1}{2}\left(  C_{-\alpha}\left(  l\right)  ,\left(  \mathbb{L}%
_{2}^{\ast}+\mathbb{L}_{2}^{\ast}\mathbb{\bar{L}}_{3}^{\ast}\mathbb{L}%
_{2}^{\ast}\right)  C_{-\alpha}\left(  l\right)  \right)  _{\mathfrak{h}%
^{\ast}}%
\end{align*}

\end{description}

\subsection{Collective hamiltonian on the factor $\left(  \mathrm{N}%
\times\mathfrak{n}^{\ast},\omega_{o}\right)  $}

\label{sec: ham model N}

\subsubsection{Collective system for $\mu_{0,0}:\mathrm{N}\times
\mathfrak{n}^{\ast}\longrightarrow\mathfrak{h}_{c,0}$}

Sigma models are now regarded as collective systems on $\mathrm{N}%
\times\mathfrak{n}^{\ast}$. $T$-duality scheme can be applied to the
hamiltonian space $(\mathrm{N}\times\mathfrak{n}^{\ast},\omega_{o}%
,\mathrm{H}_{c,0},\mu_{0,0},\mathsf{h}\circ\mu_{0,0})$ , for an arbitrary
function $\mathsf{h}:\mathfrak{h}_{c,0}^{\ast}\longrightarrow\mathbb{R}$.

For instance, a specific type of quadratic hamiltonian for $\mu_{0,0}%
:\mathrm{N}\times\mathfrak{n}^{\ast}\longrightarrow\mathfrak{h}_{c,0}$ was
built up in \cite{Ale-Hugo} for $\mathrm{N}\equiv LG$, so we refer to it and
give here a brief recall. Using the invariant bilinear form $\left(  ,\right)
_{\mathfrak{h}}$ we propose the collective hamiltonian%
\begin{equation}
\mathcal{H}_{\sigma}^{0}\left(  g,\eta\right)  =\dfrac{1}{2}\left(
Ad_{g}^{\mathrm{H}}\eta+C\left(  g\right)  ,\mathcal{E}\left(  Ad_{g}%
^{\mathrm{H}}\eta+C\left(  g\right)  \right)  \right)  _{\mathfrak{h}%
}\label{sigma-colham-2}%
\end{equation}
where $\mathcal{E}:\mathfrak{h}\rightarrow\mathfrak{h}$ is a self adjoint
linear operator. Motivated by standard Poisson-Lie T-duality \cite{Ale-Hugo},
we further assume that $\mathcal{E}^{2}=Id$. Now, following Appendix 2, we
call $\mathcal{E}_{g}=$ $Ad_{g^{-1}}^{\mathrm{H}}\mathcal{E}Ad_{g}%
^{\mathrm{H}}$, $\mathcal{G}_{g}=(\Pi_{\mathfrak{n}}\mathcal{E}_{g}%
\Pi_{\mathfrak{n}^{\ast}})^{-1}:\mathfrak{n}\longrightarrow\mathfrak{n}^{\ast
}$ and $\mathcal{B}_{g}=-\mathcal{G}_{g}\circ\Pi_{\mathfrak{n}}\mathcal{E}%
_{g}\Pi_{\mathfrak{n}}:\mathfrak{n}\longrightarrow\mathfrak{n}^{\ast}$. Then,
Hamilton equation for $g\in\mathrm{N}$ is%
\[
g^{-1}\dot{g}=\left(  \mathcal{G}_{g}\right)  ^{-1}\mathcal{B}_{g}C\left(
g^{-1}\right)  +\left(  \mathcal{G}_{g}\right)  ^{-1}\eta
\]
The lagrangian coming from the collective hamiltonian given by such symmetric
operator $\mathcal{E}$\textit{\ }defines a sigma model since, as a block
matrix in $\mathfrak{g}\oplus\mathfrak{g}^{\ast}$, we have%
\begin{equation}
\mathcal{E}_{g}=\left(
\begin{array}
[c]{cc}%
-\mathcal{G}_{g}^{-1}\mathcal{B}_{g} & \mathcal{G}_{g}^{-1}\\
\mathcal{G}_{g}-\mathcal{B}_{g}\mathcal{G}_{g}^{-1}\mathcal{B}_{g} &
\mathcal{B}_{g}\mathcal{G}_{g}^{-1}%
\end{array}
\right)  \label{GBmatrix}%
\end{equation}
Following ref. \cite{Ale-Hugo} we arrive to the sigma model lagrangian (see
also Appendix 2),
\begin{equation}
\mathcal{L}_{\sigma}^{0}=\frac{1}{2}\left\langle \left(  \mathcal{G}%
_{g}+\mathcal{B}_{g}\right)  \left(  g^{-1}\dot{g}-C\left(  g^{-1}\right)
\right)  ,g^{-1}\dot{g}+C\left(  g^{-1}\right)  \right\rangle
~.\label{eq: Lag 00}%
\end{equation}

\subsubsection{Collective system for $\mu_{0,\alpha}:\mathrm{N}\times
\mathfrak{n}^{\ast}\longrightarrow\mathfrak{h}_{c,-\alpha}^{\ast}$}

Let us now consider the hamiltonian space $(\mathrm{N}\times\mathfrak{n}%
^{\ast},\omega_{o,},\mathrm{H}_{c,-\alpha},\mu_{0,\alpha},\mathsf{h}\circ
\hat{\varphi}\circ\mu_{0,\alpha})$ where $\mu_{0,\alpha}:\mathrm{N}%
\times\mathfrak{n}^{\ast}\longrightarrow\mathfrak{h}_{c,-\alpha}^{\ast}$ was
defined in $\left(  \ref{loop-22b}\right)  $ for $\alpha\ $satisfying
condition $\left(  \ref{loop-22a}\right)  $, $\varphi:\mathfrak{h}%
_{c,-\mathbf{\alpha}}^{\ast}$ $\longrightarrow$ $\mathfrak{h}_{c,0}^{\ast}$ is
the Poisson diffeomorphism introduced in \ref{subsec: sympl equiv thetas}. For
$\mathsf{h}$ being the quadratic function as in $\left(  \ref{sigma-colham-2}%
\right)  $, we have%
\[
\mathcal{H}_{\sigma}^{\alpha}=\mathsf{h}\circ\varphi\circ\mu_{0,\alpha}%
=\frac{1}{2}\left(  \varphi\circ\mu_{0,\alpha},\mathcal{E}\varphi\circ
\mu_{0,\alpha}\right)  _{\mathfrak{h}}%
\]
The non degenerate symmetric bilinear form $\left(  ,\right)  _{\mathfrak{h}}$
induces the identification $\psi$ between $\mathfrak{h}$ and $\mathfrak{h}%
^{\ast}$, so that the hamiltonian turns out to be%
\[
\mathcal{H}_{\sigma}^{\alpha}=\frac{1}{2}\left(  \varphi\circ\widehat
{Ad}_{-\alpha;g^{-1}}^{\mathrm{H}\ast}\left(  \psi\left(  \lambda\right)
,1\right)  ,\mathcal{E}\bar{\psi}\left(  \varphi\circ\widehat{Ad}%
_{-\alpha;g^{-1}}^{\mathrm{H}\ast}\left(  \psi\left(  \lambda\right)
,1\right)  \right)  \right)  _{\mathfrak{h}}%
\]
Hamilton equations of motion yield%
\[
g^{-1}\dot{g}=\Pi_{\mathfrak{n}}\psi\left(  \mathcal{E}_{g}\left(
\lambda+Ad_{g^{-1}}^{\mathrm{H}}\bar{\psi}\left(  C\left(  g\right)  \right)
+\alpha\right)  \right)
\]
Taking into account the form of the operator $\mathcal{E}_{g}$ of eq. $\left(
\ref{GBmatrix}\right)  $,%
\[
g^{-1}\dot{g}=\mathcal{G}_{g}^{-1}\mathcal{B}_{g}C\left(  g^{-1}\right)
+\mathcal{G}_{g}^{-1}\left(  \lambda+\alpha\right)
\]
Once again, to retrieve the \emph{lagrangian function}, we use the first
Hamilton equation to invert the Legendre transformation%
\[
\lambda=\mathcal{G}_{g}g^{-1}\dot{g}-\mathcal{B}_{g}C\left(  g^{-1}\right)
-\alpha
\]
Thus, by Appendix 2,%

\[
\mathcal{L}_{\sigma}^{\alpha}=\dfrac{1}{2}\left(  \left(  \mathcal{G}%
_{g}+\mathcal{B}_{g}\right)  \left(  g^{-1}\dot{g}-C\left(  g^{-1}\right)
\right)  ,g^{-1}\dot{g}+C\left(  g^{-1}\right)  \right)  _{\mathfrak{h}%
}-\left\langle g^{-1}\dot{g},\alpha\right\rangle ~.
\]

\subsection{Collective hamiltonian on the factor $\left(  \mathrm{N}^{\ast
}\times\mathfrak{n},\tilde{\omega}_{o}\right)  $}

\label{sec: ham model N*}

Let us now consider the collective dynamic system on $T^{\ast}\mathrm{N}%
^{\ast}\cong\mathrm{N}^{\ast}\times\mathfrak{n}$ defined by,
\[
\mathcal{\tilde{H}}^{\alpha}=\mathsf{h}\circ\tilde{\mu}_{0,\alpha}^{\varphi}%
\]
for the momentum map $\tilde{\mu}_{0,\alpha}^{\varphi}:\mathrm{N}^{\ast}%
\times\mathfrak{n}\longrightarrow\mathfrak{h}_{c,0}^{\ast}$, defined in
$\left(  \ref{mu-dual}\right)  $, and $\mathsf{h}$ being the quadratic
function as above. This yields%
\[
\mathcal{\tilde{H}}^{\alpha}=\frac{1}{2}\left(  Ad_{\tilde{h}}^{D}\left(
X+\alpha\right)  +C\left(  \tilde{h}\right)  ,\mathcal{E}\left(  Ad_{\tilde
{h}}^{D}\left(  X+\alpha\right)  +C\left(  \tilde{h}\right)  \right)  \right)
_{\mathfrak{h}}%
\]
so that the Hamilton equation of motion for $\tilde{h}$ is%
\[
\tilde{h}^{-1}\overset{\centerdot}{\tilde{h}}=\mathcal{E}_{\tilde{h}}\left(
X+\alpha-C\left(  \tilde{h}^{-1}\right)  \right)
\]
where $\mathcal{E}_{\tilde{h}}=Ad_{\tilde{h}^{-1}}^{\mathrm{H}}\mathcal{E}%
Ad_{\tilde{h}}^{\mathrm{H}}$. Using the dual decomposition $\mathfrak{h}%
^{\top}=\mathfrak{n}^{\ast}\oplus\mathfrak{n}$, we can write%

\[
\mathcal{E}_{\tilde{h}}=\left(
\begin{array}
[c]{cc}%
-\mathcal{\tilde{G}}_{\tilde{h}}^{-1}\mathcal{\tilde{B}}_{\tilde{h}} &
\mathcal{\tilde{G}}_{\tilde{h}}^{-1}\\
\mathcal{\tilde{G}}_{\tilde{h}}-\mathcal{\tilde{B}}_{\tilde{h}}\mathcal{\tilde
{G}}_{\tilde{h}}^{-1}\mathcal{\tilde{B}}_{\tilde{h}} & \mathcal{\tilde{B}%
}_{\tilde{h}}\mathcal{\tilde{G}}_{\tilde{h}}^{-1}%
\end{array}
\right)
\]
with the operators $\mathcal{\tilde{G}}_{\tilde{h}}=(\Pi_{\mathfrak{n}^{\ast}%
}\mathcal{E}_{\tilde{h}}\Pi_{\mathfrak{n}})^{-1}:\mathfrak{n}^{\ast
}\longrightarrow\mathfrak{n}$ and $\mathcal{\tilde{B}}_{\tilde{h}%
}=-\mathcal{\tilde{G}}_{\tilde{h}}\circ\Pi_{\mathfrak{n}^{\ast}}%
\mathcal{E}_{\tilde{h}}\Pi_{\mathfrak{n}^{\ast}}:\mathfrak{n}^{\ast
}\longrightarrow\mathfrak{n}$. The equation for $\tilde{h}$ yields%
\[
X=\mathcal{\tilde{G}}_{\tilde{h}}\left(  \tilde{h}^{-1}\overset{\centerdot
}{\tilde{h}}\right)  +\mathcal{\tilde{B}}_{\tilde{h}}\left(  \alpha-C\left(
\tilde{h}^{-1}\right)  \right)
\]
We thus obtain, following Appendix 2, the following \emph{lagrangian function
}%

\[
\mathcal{\tilde{L}}^{\alpha}\left(  \tilde{h},\overset{\centerdot}{\tilde{h}%
}\right)  =\frac{1}{2}\left(  \tilde{h}^{-1}\overset{\centerdot}{\tilde{h}%
}+C\left(  \tilde{h}^{-1}\right)  -\alpha,\left(  \mathcal{\tilde{G}}%
_{\tilde{h}}+\mathcal{\tilde{B}}_{\tilde{h}}\right)  \left(  \tilde{h}%
^{-1}\overset{\centerdot}{\tilde{h}}-C\left(  \tilde{h}^{-1}\right)
+\alpha\right)  \right)
\]

\begin{description}
\item[Remark:] \emph{(The equivalent shifted formulation) }Analogously, a
model on $T^{\ast}\mathrm{N}^{\ast}\cong\mathrm{N}^{\ast}\times\mathfrak{n}$
in corresponding \emph{shifted} version can be obtained by proposing a
hamiltonian function
\[
\mathcal{\tilde{H}}^{\alpha}=\mathsf{h}\circ\varphi\circ\tilde{\mu}^{\alpha}%
\]
for $\tilde{\mu}^{\alpha}:(\mathrm{N}^{\ast}\times\mathfrak{n},\tilde{\omega
}_{o})\longrightarrow(\mathfrak{h},\{,\}_{c,-\alpha}^{Aff})$ defined in eq.
$\left(  \ref{mu-dual-alpha}\right)  $, leading to the same Hamilton equations
and lagrangian since $\varphi\circ\tilde{\mu}^{\alpha}=\tilde{\mu}_{0,\alpha
}^{\varphi}$.
\end{description}

\section{Loop groups and Poisson Lie T-duality for trivial monodromies}

\label{sec: Loops}In the classical field theory context, Sigma and WZNW models
can be built up on the cotangent bundle of loop groups. In particular, for the
WZNW model, a pure cocycle is added to the canonical symplectic form in order
to produce the decoupling of the chiral modes \cite{Harnad-1}, leading to the
fact that a general solution for the equation of motion can be written as a
non linear combination of both modes. Finally, on the \emph{chiral phase
space}, solutions related to non-trivial extended coadjoint orbits are the
ones associated to non trivial monodromies.

In this Section, we apply the results obtained in the previous sections to the
case of underlying loop groups, thus addressing the case in which non trivial
monodromies appear. More precisely, we consider $\mathrm{H}=LD$, for some Lie
group $D$ with Lie algebra $\mathfrak{d}$ such that $\mathfrak{h}%
=L\mathfrak{d}$. Moreover, $\mathfrak{d}$ will be regarded also as a double
Lie group $D=G\Join G^{\ast}$ with Lie bialgebra $\mathfrak{d}=\mathfrak{g}%
\oplus\mathfrak{g}$, so that $\mathrm{N}=LG$, $\mathfrak{n}=L\mathfrak{g}$,
and the same for the corresponding duals.

For $l\in LD$, $l^{\prime}$ denotes the derivative in the loop parameter $s\in
S^{1}$, and we write $vl^{-1}$ and $l^{-1}v$ for the right and left
translation of any vector field $v\in TD$. Let $\mathfrak{d}$ be the Lie
algebra of $D$ equipped with a non degenerate symmetric $Ad$-invariant
bilinear form $\left(  ,\right)  _{\mathfrak{d}}$. Frequently we will work
with the subset $L\mathfrak{d}^{\ast}\subset$ $(L\mathfrak{d})^{\ast}$ instead
of $(L\mathfrak{d})^{\ast}$ itself, and we identify it with $L\mathfrak{d}$
through the map $\psi:L\mathfrak{d}\rightarrow L\mathfrak{d}^{\ast}$ provided
by the bilinear form
\[
\left(  ,\right)  _{L\mathfrak{d}}\equiv\dfrac{1}{2\pi}\int_{S^{1}}\left(
,\right)  _{\mathfrak{d}}%
\]
on $L\mathfrak{d}$. In this framework, the two cocycle $c:\mathfrak{h}%
\times\mathfrak{h}\longrightarrow\mathbb{R}$ of Section
\ref{sec: WZW phase space} is given by the bilinear form $\Gamma_{\mathrm{k}%
}:L\mathfrak{d}\times L\mathfrak{d}\rightarrow\mathbb{R}$ \cite{Pres-Seg},
\[
c(\mathbf{X},\mathbf{Y})\equiv\Gamma_{\mathrm{k}}(\mathbf{X},\mathbf{Y}%
)=\frac{k}{2\pi}\int_{S^{1}}\left(  \mathbf{X}\left(  s\right)  ,\mathbf{Y}%
^{\prime}\left(  s\right)  \right)  _{\mathfrak{d}}\,ds
\]
with $\mathbf{X},\mathbf{Y}\in L\mathfrak{d}$. It is derived from the one
cocycle $C_{\mathrm{k}}:LD\rightarrow L\mathfrak{d}^{\ast}$,
\[
C_{\mathrm{k}}\left(  l\right)  =\mathrm{k}\psi\left(  l^{\prime}%
l^{-1}\right)  ~~.
\]

As in section $\ref{sec: centrally exended Lie}$, when the corresponding
central extension $LD_{\Gamma_{\mathrm{k}}}$ of $LD$ does exist, the centrally
extended affine adjoint and coadjoint actions of $LD_{\Gamma_{\mathrm{k}}}$ on
$L\mathfrak{d}_{\Gamma_{\mathrm{k}}}$ and $L\mathfrak{d}_{\Gamma_{\mathrm{k}}%
}^{\ast}$ are defined as
\[
\widehat{Ad}_{l^{-1}}^{\ast}\left(  \mathbf{\xi},b\right)  =\left(
Ad_{l^{-1}}^{\ast}\mathbf{\xi}+b\mathrm{k}\psi\left(  l^{\prime}l^{-1}\right)
\,,\,b\right)
\]
Note that the $S^{1}\subset LD_{\Gamma_{\mathrm{k}}}$ action is trivial and
the embedding $\xi\hookrightarrow(\xi,1)$ is a Poisson map from $L\mathfrak{d}%
_{Aff}^{\ast}$ to $L\mathfrak{d}_{\Gamma_{\mathrm{k}}}\sim L\mathfrak{d}%
\times\mathbb{R}$ which maps the \emph{affine} coadjoint action of $LD$ to the
centrally extended one of $LD\hookrightarrow LD_{\Gamma_{\mathrm{k}}}$.

We now recall some facts about the stabilizer subgroup of a point $\left(
\eta,1\right)  \in L\mathfrak{d}_{\Gamma_{\mathrm{k}}}^{\ast}$. If\textit{
}$\left[  LD\right]  _{\left(  \eta,1\right)  }$ denotes the stabilizer,
with\textit{ }$\eta\in L\mathfrak{g}^{\ast}$, consider $\tilde{h}_{\eta
}\left(  s\right)  \in C^{\infty}\left(  \mathbb{R},G^{\ast}\right)  $,
$\tilde{h}_{\alpha}\left(  0\right)  =e$, and $\tilde{M}_{\eta}\in G^{\ast}$
such that%
\[%
\begin{array}
[c]{c}%
\eta\left(  s\right)  =\tilde{h}_{\eta}^{\prime}\left(  s\right)  \tilde
{h}_{\eta}^{-1}\left(  s\right) \\
\tilde{h}_{\eta}\left(  s+2\pi\right)  =\tilde{h}_{\eta}\left(  s\right)
~\tilde{M}_{\eta}%
\end{array}
\]
where $\tilde{M}_{\eta}\in G^{\ast}$ is the monodromy of the\textit{
}$\mathfrak{g}^{\ast}$-valued map\textit{ }$\eta\left(  s\right)  $.
Then\textit{ }$\left[  LD\right]  _{\left(  \eta,1\right)  }$ consists of
loops of the form%
\[
l\left(  s\right)  =\tilde{h}_{\eta}\left(  s\right)  l\left(  0\right)
\tilde{h}_{\eta}^{-1}\left(  s\right)
\]
\textit{\ }based at some point\textit{ }$l\left(  0\right)  \ $within the
stabilizer $D_{\tilde{M}_{\eta}}$ of\textit{ }$\tilde{M}_{\eta}$ \textit{in
}$D$%
\[
\tilde{M}_{\eta}=l^{-1}\left(  0\right)  \tilde{M}_{\eta}l\left(  0\right)
\]
For any other point $\left(  \mathbf{\beta},1\right)  =\widehat{Ad}%
_{m}^{LD\ast}\left(  \eta,1\right)  \in\mathcal{O}\left(  \eta,1\right)  $,
$m\in LD$, the isotropy group is $m^{-1}\left(  LD_{\left(  \eta,1\right)
}\right)  m$. Writing $\mathbf{\beta}\left(  s\right)  =-l_{\mathbf{\beta}%
}^{\prime}\left(  s\right)  l_{\mathbf{\beta}}^{-1}\left(  s\right)  $ with
monodromy $M_{\mathbf{\beta}}=l_{\mathbf{\beta}}^{-1}\left(  s\right)
l_{\mathbf{\beta}}\left(  s+2\pi\right)  $, it is related with $\tilde
{M}_{\eta}$ as%
\[
M_{\mathbf{\beta}}=m^{-1}\left(  s\right)  \tilde{M}_{\eta}m\left(  s\right)
\]

As a symplectic manifold, the coadjoint orbit $\mathcal{O}_{\Gamma
_{\mathrm{k}},0}(\alpha,1)$ $\subset$ $L\mathfrak{d}_{\Gamma_{\mathrm{k}}%
,0}^{\ast}$ passing through $(\alpha,1)\in$ $L\mathfrak{d}_{\Gamma
_{\mathrm{k}},0}^{\ast}$, for $\alpha\in L\mathfrak{g}^{\ast}$, is endowed
with the Kirillov-Kostant symplectic form
\[
\left\langle \omega_{KK}^{\alpha},\widehat{ad}_{(\mathbf{X},a)}^{\ast
}(\mathbf{\beta},1)\otimes\widehat{ad}_{(\mathbf{Y},b)}^{\ast}(\mathbf{\beta
},1)\right\rangle _{(\beta,1)}=\left\langle [\mathbf{X},\mathbf{Y}%
],\mathbf{\beta}\right\rangle +\Gamma_{\mathrm{k}}(\mathbf{X},\mathbf{Y})
\]
with $(\mathbf{\beta},1)=\widehat{Ad}_{l}^{LD\ast}\left(  \alpha,1\right)
\in\mathcal{O}_{\Gamma_{\mathrm{k}},0}(\alpha,1)$, for some $l\in LD$.

Let us define
\begin{equation}%
\begin{array}
[c]{c}%
\phi:LD\longrightarrow\mathcal{O}_{\Gamma_{\mathrm{k}},0}(\alpha,1)\subset
L\mathfrak{d}_{\Gamma_{\mathrm{k}}}^{\ast}\\
\\
\phi(l)=\widehat{Ad}_{l}^{LD\ast}(\alpha,1)
\end{array}
\label{phi}%
\end{equation}
and consider the pull back of $\omega_{KK}^{\alpha}$ through $\phi$%
\[
\left.  \phi^{\ast}\omega_{KK}^{\alpha}\right\vert _{l}=\Gamma_{\mathrm{k}%
}(l^{-1}dl\overset{\otimes}{,}l^{-1}dl)+\left\langle \alpha,[l^{-1}%
dl\overset{\otimes}{,}l^{-1}dl]\right\rangle
\]
The null distribution of this form is spanned by the Lie algebra of the
stabilizer subgroup $LD_{(\alpha,1)}$ of $(\alpha,1)$. Thus, if
$LD/LD_{(\alpha,1)}$\ is a smooth manifold, there is a symplectic
diffeomorphism $(LD/LD_{(\alpha,1)},\phi^{\ast}\omega_{KK}^{\alpha}%
)\overset{\phi}{\longrightarrow}(\mathcal{O}_{\Gamma_{\mathrm{k}},0}%
(\alpha,1),\omega_{KK}^{\alpha}).$

As a consequence we have,

\begin{description}
\item[Remark:] \textit{The momentum map associated to the residual symmetry of
}$LD$\textit{ on} $LD/\left[  LD\right]  _{(\alpha,1)}$ \textit{is} $\phi.$
\end{description}

By also recalling the map $\mu_{0,0}$ from \cite{Ale-Hugo} (called just $\mu$
there), so far we have%
\begin{equation}
\begin{diagram}[h=1.9em] T^{\ast}LG && \\ &\rdTo^{\mu_{0,0}}&\\ LD/LD_{\alpha(x)} \cong \mathcal{O}_{\Gamma_{\mathrm{k}},0}(\alpha,1)&\rTo_{\phi}& L\mathfrak{d}_{\Gamma_{\mathrm{k}},0}^{\ast}\\ \end{diagram} \label{t dual lg ld}%
\end{equation}
We are going now to describe the other phase spaces presented in section
\ref{sec: centrally exended Lie} which are also related to the orbit
$\mathcal{O}_{\Gamma_{\mathrm{k}},0}(\alpha,1)$.

\subsection{The chiral WZNW phase space}

The phase spaces described in section \ref{sec: Hamiltonian WZ H}, for
$\mathrm{H}=LD$, correspond to a chiral sector of the WZNW model
\cite{Feheretal}. Consider the unique curve $\tilde{h}_{\alpha}\in C^{\infty
}\left(  \mathbb{R},G^{\ast}\right)  $ such that%
\[
\alpha\left(  s\right)  =\tilde{h}_{\alpha}^{\prime}\left(  s\right)
\tilde{h}_{\alpha}^{-1}\left(  s\right)
\]
with $\tilde{h}_{\alpha}\left(  0\right)  =e$ so that $\tilde{h}_{\alpha
}\left(  2\pi\right)  =\tilde{M}_{\alpha}=Hol(\alpha(s))$ is the\emph{
holonomy} of $\alpha(s)\subset\mathfrak{g}^{\ast}$. The solutions in $LD$ to
the classical equation of motion of the WZNW model
\[
\partial_{-}(\partial_{+}l~l^{-1})=0
\]
are $l(s,t)=l_{L}(x_{+})l_{R}^{-1}(x_{-})$, which combines the chiral
solutions $l_{L}(x_{-})$ and $l_{R}(x_{+})$, with $x_{\pm}=s\pm t$. In spite
of the fact that $l(s,t)$ is periodic in $s$, the chiral modes are not
restricted to be closed loops. In fact, for chiral modes satisfying
\[
l_{\frac{L}{R}}(x_{\pm}+2\pi)=l_{\frac{L}{R}}(x_{\pm})\ M
\]
the solutions $l(s,t)$ remains periodic. Let $PD$ be the \emph{extended
}chiral phase space of the WZNW model (see \cite{Feheretal})
\[
PD=\{l:R\longrightarrow D\;/~l(x+2\pi)=l(x)M,\;\text{for some}~M\in D\}.
\]
Here, each $M\in D$ is the monodromy of path $l(x)\ $satisfying
\[
l(0)Ml^{-1}(0)=Hol(l^{\prime}l^{-1})=l(2\pi)l^{-1}(0)\;
\]
since $l(0)^{-1}l(2\pi)=M$.

In order to study separately the chiral degree of freedom, the following
symplectic form on $PD$ must be considered
\[
\omega^{PD}(l)=-\Gamma_{k}(l^{-1}dl\overset{\otimes}{,}l^{-1}dl)-\frac{1}%
{2}\left\langle (l^{-1}dl)(0)\overset{\wedge}{,}dMM^{-1}\right\rangle
+\rho(M)
\]
with $\rho$ being a $2$-form just depending on the holonomy $M$ so that
\[
d\rho(M)=\frac{1}{6}\left\langle dMM^{-1}\overset{\wedge}{,}[dMM^{-1}%
\overset{\wedge}{,}dMM^{-1}]\right\rangle
\]
It is worth to remark that $\rho$ is defined just locally in some neighborhood
of the identity. In ref. \cite{Feheretal}, it is shown that such $\rho$ is
given by generalized $r$ matrices solutions of the Generalized Dynamically
Modified Classical Yang-Baxter equation, defining Poisson structure chiral
space $PD$.

Let us define the map
\begin{equation}%
\begin{array}
[c]{c}%
J:PD\longrightarrow L\mathfrak{d}_{\Gamma_{\mathrm{k}}}^{\ast}\\
\\
J(l)=(l^{\prime}l^{-1},1)
\end{array}
\label{J}%
\end{equation}
The loop group $LD$ acts on $PD$ by left multiplication with associated
momentum map $J$. On the other side, the group $D$ acts on $PD$ by right
multiplication $l(x)\overset{d}{\longrightarrow}l(x).d$, which is a
Poisson-Lie symmetry in relation to the Poisson-Lie structure on $D\ $given by
the canonical $r$-matrix in the double Lie algebra $\mathfrak{d}$ and a
particular election of $\rho(M)$, as it is shown in \cite{Feheretal}. In that
case, the map $\sigma:LD\longrightarrow D^{\ast}$ defined as $\sigma
(l(x))=l^{-1}(x)l(x+2\pi)=M\;$defines the corresponding $D^{\ast}$-valued
momentum map.

Then, we have the following diagram%

\[
\begin{diagram}[h=1.9em]
&&PD&&\\
&\ldTo^{J} &&\rdTo^{\sigma} & \\
L\mathfrak{d}_{\Gamma_{\mathrm{k}}}^{\ast}&&&& D^{\ast}\\
\end{diagram}
\]

For each value $M$, we consider the pre image $\sigma^{-1}(M)\subset PD$ and
define the subspace
\[
E_{\alpha}(D)=LD\cdot\tilde{h}_{\alpha}=\{l\cdot\tilde{h}_{\alpha}~/~l\in
LD\}
\]
which is injected into $\sigma^{-1}(M)\subset PD$ as follows
\[
E_{\alpha}(D)\overset{i}{\hookrightarrow}\sigma^{-1}(M=Hol(\alpha))\subset
PD.
\]
Because it can be identified with $LD$, the restriction of the symplectic form
on $PD$ to $E_{\alpha}(D)$ coincides with $\left.  \mu^{\alpha\ast}\omega
_{KK}^{\alpha}\right\vert _{l}$, and
\[
\mu^{\alpha}:=J\circ i
\]
is a presymplectic map. Hence, we have following phase spaces related by
symplectic morphisms with the orbit $\mathcal{O}_{\Gamma_{\mathrm{k}}%
,0}(\alpha,1)$,
\[
\begin{diagram}[h=1.9em]
T^{\ast}LG &&&& E_{\alpha}(D)&&&&\\
&&&&&&&&\\
&&\rdTo_{\mu_{0,0}} &&\dTo_{\mu^\alpha}&& && \\
&&&&&&&&\\
&& \mathcal{O}_{\Gamma_{\mathrm{k}},0}(\alpha,1)&\hookrightarrow&(L{\mathfrak{d}}_{\Gamma_{\mathrm{k}},0}^{\ast}\;\{,\}_{KK})&&&&\\ \end{diagram}
\]

\subsection{A symplectic $LD$ action on $LG^{\ast}\times L\mathfrak{g}$}

From Section \ref{sec: dual factor phase space}, it follows that $\tilde{\mu
}_{0,\alpha}^{\varphi}:(LG^{\ast}\times L\mathfrak{g},\tilde{\omega}%
_{o})\longrightarrow(L\mathfrak{d},\{,\}_{Aff})\subset\left(  L\mathfrak{d}%
_{\Gamma_{\mathrm{k}},0}^{\ast},\{,\}_{\Gamma_{\mathrm{k}},0}^{KK}\right)  $,
given by
\begin{equation}
\tilde{\mu}_{0,\alpha}^{\varphi}(\tilde{g},Z)=(\tilde{g}^{\prime}\tilde
{g}^{-1}+Ad_{\tilde{g}}^{LD}(Z+\alpha))~, \label{mu 0 lg star}%
\end{equation}
is a Poisson map provided that condition $\left(  \ref{loop-22a}\right)
$\textit{ }is satisfied. In the current loop group case, this means that%
\begin{equation}
ad_{\alpha(s)}^{\mathfrak{d}}X\in\mathfrak{g}\subset\mathfrak{d}
\label{Eq: Condicion Poisson}%
\end{equation}
for all $X\in\mathfrak{g}$ and all $s\in S^{1}$.

\begin{description}
\item[Remark:] \textit{When }$G^{\ast}=K$\textit{ is a compact simple real Lie
group, e.g. }$SU(N)$\textit{, any constant element }$\alpha\in\mathfrak{t}%
$\textit{, where }$\mathfrak{t}$\textit{ is the Cartan subalgebra of
}$\mathfrak{k}=Lie(K)$\textit{, condition }$\left(
\ref{Eq: Condicion Poisson}\right)  $\textit{ is satisfied (see Appendix }%
$1$\textit{).}
\end{description}

Under this condition, the loop group $LD$ acts on $LG^{\ast}\times
L\mathfrak{g}$ via
\[
(\tilde{g},Z)\overset{l=\tilde{b}a}{\longmapsto}(\tilde{b}\tilde{g}%
^{a},\ Ad_{a^{\tilde{g}}}^{LD}(Z+\alpha)+(a^{\tilde{g}})^{\prime}(a^{\tilde
{g}})^{-1}-\alpha)
\]
where $\tilde{g}^{a}a^{\tilde{g}}=a\tilde{g}$. The corresponding infinitesimal
generator $\left.  \mathbf{X}_{LG^{\ast}\times L\mathfrak{g}}\right\vert
_{(\tilde{g},Z)}\in T_{(\tilde{g},Z)}(LG^{\ast}\times L\mathfrak{g})$
associated to an $\mathbf{X}=(\xi,X)\in L\mathfrak{d}$ at the point
$(\tilde{g},Z)\in LG^{\ast}\times L\mathfrak{g}$ is%
\[%
\begin{array}
[c]{l}%
\left.  \mathbf{X}_{LG^{\ast}\times L\mathfrak{g}}\right\vert _{(\tilde{g}%
,Z)}\\
\qquad=\left(  \tilde{g}\Pi_{\mathfrak{g}^{\ast}}(Ad_{\tilde{g}^{-1}}%
^{LD}\mathbf{X}),\Pi_{\mathfrak{g}}\left(  Ad_{\tilde{g}^{-1}}^{LD}%
\mathbf{X}^{\prime}-\left[  \tilde{g}^{-1}\tilde{g}^{\prime}+Z+\alpha
,\Pi_{\mathfrak{g}}(Ad_{\tilde{g}^{-1}}^{LD}\mathbf{X})\right]  \right)
\right)
\end{array}
\]
where $\Pi_{\mathfrak{g}^{\ast}},\Pi_{\mathfrak{g}}$ are the projectors on the
Lie subalgebras $\mathfrak{g}^{\ast},\mathfrak{g}$, respectively. Moreover,
this action is hamiltonian and the corresponding moment map is precisely
$\tilde{\mu}_{0,\alpha}^{\varphi}$.

\subsection{T-duality diagram for non trivial monodromies}

Thus, considering the results of this section, we may assemble a
PL\ $T$-duality diagram based on the coadjoint orbit $\mathcal{O}%
_{\Gamma_{\mathrm{k}},0}(\alpha,1)$ $\subset$ $L\mathfrak{d}_{\Gamma
_{\mathrm{k}},0}^{\ast}$.

This fact doesn't affect the side corresponding to the phase space $T^{\ast
}LG$: there we have a $\sigma$-model describing the dynamics of a closed
string valued on the target $G$, symmetric under the hamiltonian action of the
group $LD_{\Gamma_{\mathrm{k}},0}$ given in eq. $\left(
\ref{d-action-cent-ext}\right)  $and with associated momentum map $\mu_{0,0}$
of Proposition $\left(  \ref{prop: action on N without shift}\right)  $, as
depicted in diagram $\left(  \ref{t dual lg ld}\right)  $.

The double Lie group $LD$ appears in $T$-duality diagrams through the subspace
$E_{\alpha}(D)$ bringing the open string models into the scheme. It is
injected in the \emph{extended }chiral phase space of the WZNW model, $PD$,
and relates to the orbit $\left(  L\mathfrak{d}_{\Gamma_{\mathrm{k}},0}^{\ast
},\{,\}_{\Gamma_{\mathrm{k}},0}^{KK}\right)  $ by the map $\mu^{\alpha
}:=J\circ i$.

Finally, on the dual side things become subtler: as explained just above, and
in section $\ref{sec: dual factor phase space}$, the connection of the
cotangent bundle $T^{\ast}LG^{\ast}\cong LG^{\ast}\times L\mathfrak{g}$ to the
coadjoint orbit $\left(  L\mathfrak{d}_{\Gamma_{\mathrm{k}},0}^{\ast
},\{,\}_{\Gamma_{\mathrm{k}},0}^{KK}\right)  $ does exist provided $\alpha$ be
such that $\left(  \ref{Eq: Condicion Poisson}\right)  $ is fulfilled. Under
this condition, the map $\tilde{\mu}_{0,\alpha}^{\varphi}:(LG^{\ast}\times
L\mathfrak{g},\tilde{\omega}_{o})\longrightarrow(L\mathfrak{d},\{,\}_{Aff}%
)\subset\left(  L\mathfrak{d}_{\Gamma_{\mathrm{k}},0}^{\ast},\{,\}_{\Gamma
_{\mathrm{k}},0}^{KK}\right)  $ turns in a Poisson one, enabling to assemble
the complete $T$-duality diagram%

\[
\begin{diagram}[h=1.9em]
T^{\ast}LG &&&& E_{\alpha}(D)&&&&T^{\ast}L{G^{\ast}}\\
&&&&&&&&\\
&&\rdTo_{\mu_{0,0}} &&\dTo_{\mu^\alpha}&& \ldTo_{\tilde{\mu}_{0,\alpha}^{\varphi}} && \\
&&&&&&&&\\
&& \mathcal{O}_{\Gamma_{\mathrm{k}},0}(\alpha,1)&\hookrightarrow&(L{\mathfrak{d}}_{\Gamma_{\mathrm{k}},0}^{\ast};\;\{,\}_{\Gamma_{\mathrm{k}},0}^{KK})&&&&\\ \end{diagram}
\]

\subsection{Induced hamiltonian systems on loops groups}

Following section $\ref{sec: hamiltonian asociados}$, we now choose a
quadratic hamiltonian function $H:L\mathfrak{d}_{c,0}^{\ast}\sim
L\mathfrak{d}_{c,0}\longrightarrow\mathbb{R}$ given by%

\[
H(\mathbf{X},1)=\frac{1}{2}\left\langle \mathbf{X},\mathcal{E}{(}%
\mathbf{X})\right\rangle
\]
with $\mathcal{E}:\mathfrak{d}\longrightarrow\mathfrak{d}\ $a linear operator.
The lifts to $LD/D_{\alpha(x)}$ and $PD$ are, respectively,%
\[
\left\{
\begin{array}
[c]{l}%
H\circ\phi(l)=\frac{1}{2}\left\langle \widehat{Ad}_{l}^{\ast}(\alpha
,1),\mathcal{E}{(}\widehat{Ad}_{l}^{\ast}(\alpha,1))\right\rangle \\
\\
H\circ J(l)=\frac{1}{2}\left\langle l^{\prime}l^{-1},\mathcal{E}{(}l^{\prime
}l^{-1})\right\rangle
\end{array}
\right.
\]

We notice that $H\circ J$ in $PD$ is invariant under the right (PL) action of
$D$ and then, the corresponding momentum map shall give conserved quantities.
This implies that the monodromy $\tilde{M}$ is constant and that we can
restrict to the subspace within $m^{-1}(\tilde{M})\subset PD$ consisting of
paths with the same monodromy, for example $E_{\alpha}(D)$ defined above. The
resulting dynamics is the chiral WZNW-type one corresponding to section
$\ref{sec: Hamiltonian WZ H}$.

In $L\mathfrak{d}_{c}^{\ast}\sim L\mathfrak{d}_{c}$, writing the corresponding
integral curve as $\gamma(t)=\widehat{Ad}_{l(t)}^{\ast}(\alpha,1)$ for the
initial value $\gamma(0)=(\alpha,1)$, we have that the corresponding equation
for $l(t)\in LD$ is
\begin{align}
\frac{d}{dt}ll^{-1}  &  =\mathcal{E}(l^{\;\prime}l^{-1}+Ad_{l}\alpha
)=\mathcal{E}((l\tilde{b})^{\prime}(l\tilde{b})^{-1})\label{ec Ld}\\
l(t  &  =0)=e\nonumber
\end{align}
with $\alpha=\tilde{b}^{\prime}\tilde{b}^{-1}$, $\tilde{b}(0)=e$, $\tilde
{b}\in PG^{\ast}\subset PD$. Since the hamiltonian functions are in collective
form (recall sec. \ref{subsec: duality}), the integral curves in
$LD/D_{\alpha(x)}$ and $PD$ will be determined by $l(t)$.

\begin{description}
\item[Remark:] \emph{(dual factorizations)\ }Writing $l=g\tilde{h}\;$or
$l=\tilde{g}h$, we alternatively obtain
\begin{align*}
\frac{d}{dt}gg^{-1}+Ad_{g}\frac{d}{dt}\tilde{h}\tilde{h}^{-1}  &
=\mathcal{E}{(}g^{\prime}g^{-1}+Ad_{g}(Ad_{\widetilde{h}}\alpha+\tilde
{h}^{\prime}\tilde{h}^{-1}))\\
&  =\mathcal{E}(\;(g\tilde{h}\tilde{b})^{\prime}(g\tilde{h}\tilde{b})^{-1})
\end{align*}%
\begin{align*}
\frac{d}{dt}\tilde{g}\tilde{g}^{-1}+Ad_{\tilde{g}}\frac{d}{dt}hh^{-1}  &
=\mathcal{E}{(}\tilde{g}^{\prime}\tilde{g}^{-1}+Ad_{\tilde{g}}(Ad_{h}%
\alpha+h^{\prime}h^{-1}))\\
&  =\mathcal{E}((\tilde{g}h\tilde{b})^{\prime}(\tilde{g}h\tilde{b})^{-1}).
\end{align*}
In the first case, since $\alpha\in\mathfrak{g}^{\ast}$ and, thus, $\tilde
{b}\in G^{\ast}$, the terms corresponding to fields in $G$ and $G^{\ast}$
split, allowing to lift the dynamics from $\mathcal{O}_{\Gamma_{\mathrm{k}}%
,0}(\alpha,1)$ to $LT^{\ast}G$ via $\mu$, as we already knew \cite{Ale-Hugo}.
In turn, in the second case above, this cannot be done generally because the
$G$ and $G^{\ast}$ variables are mixed up.
\end{description}

Now, if $\left(  \ref{Eq: Condicion Poisson}\right)  $ is satisfied for a
constant $\alpha\in{\mathfrak{g}}^{\ast}$, following Sections
$\ref{sec: ham model N*}$ and $\ref{subsec: duality}$, we find that the
resulting model pulls back to $(LG^{\ast}\times L\mathfrak{g},\tilde{\omega
}_{o},\mathsf{h}\circ\tilde{\mu}_{0,\alpha}^{\varphi})$ and that it
becomes\emph{ dual }to that on $LT^{\ast}G$. The equations of motion on
$LG^{\ast}\times L\mathfrak{g}$ are%
\begin{align*}
\tilde{g}^{-1}\frac{d}{dt}\tilde{g}  &  =\Pi_{G^{\ast}}(\mathcal{E}_{g}%
(\tilde{g}^{-1}\tilde{g}^{\prime}+Z+\alpha))\\
& \\
\frac{d}{dt}Z  &  =\Pi_{G}\{[Z,\tilde{g}^{-1}\frac{d}{dt}\tilde{g}%
]+\mathcal{E}_{g}ad_{\tilde{g}^{-1}\tilde{g}^{\prime}}^{D}(Z+\alpha
)+\mathcal{E}_{g}(\tilde{g}^{-1}\tilde{g}^{\prime}+Z+\alpha)^{\prime}\\
&  ~~~~~~-ad_{\tilde{g}^{-1}\tilde{g}^{\prime}}^{D}\mathcal{E}_{g}(\tilde
{g}^{-1}\tilde{g}^{\prime}+Z+\alpha)-ad_{(Z+\alpha)}^{D}\mathcal{E}_{g}%
(\tilde{g}^{-1}\tilde{g}^{\prime}+Z+\alpha)\}
\end{align*}
whose solutions are given by%
\[
(\tilde{g},Z)(t)=l_{t}\cdot(\tilde{g}_{0},Z_{0})
\]
when $l_{t}\in LD$ is a solution of equation $\left(  \ref{ec Ld}\right)  $
and $(\tilde{g}_{0},Z_{0})\in(\tilde{\mu}_{0,\alpha}^{\varphi})^{-1}%
(\alpha,1)$.

\subsubsection{The induced lagrangians}

\label{non trivial monodromy models}

The action functional corresponding to the above dynamics in $LD/D_{\alpha
(x)}$, expressed in terms of $D$-valued field $l$, is%
\[
S=\int\theta^{\alpha}(l)-H\circ\phi(l)
\]
where $H$ is the hamiltonian given in the previous section and $\theta
^{\alpha}$ is a potential for the $2$-form $\omega_{KK}^{\alpha}$, which can
be written as%
\[
\theta^{\alpha}(l)=\left\langle l^{\prime}l^{-1},dll^{-1}\right\rangle
+\frac{1}{6}d^{-1}\left\langle dll^{-1},[dll^{-1},dll^{-1}]\right\rangle
+2\left\langle l^{-1}dl,\alpha\right\rangle .
\]
This agrees with the action proposed in ref. \cite{Monodromic} for a
particular choice of $H$. By construction, this WZW-like model on $D$ is
\emph{dual} to the corresponding sigma models with targets $G$ and $G^{\ast}$,
as described in the previous sections.

The sigma model on $G$, for a particular choice of $\mathcal{E}$, can be found
in ref. \cite{Ale-Hugo}. The same operator $\mathcal{E}$, following section
\ref{sec: ham model N*} and assuming $\left(  \ref{Eq: Condicion Poisson}%
\right)  $ is satisfied for a constant $\alpha\in{\mathfrak{g}}^{\ast}$,
induces a lifted hamiltonian on $(LG^{\ast}\times L\mathfrak{g},\tilde{\omega
}_{o})$. The Legendre transformation turns out to be non singular and the
resulting model is given by the lagrangian (see Appendix $2$)%
\begin{equation}
\mathcal{L}_{\alpha}(\tilde{g},\frac{d\tilde{g}}{dt})=\left\langle
\partial_{-}\tilde{g}\tilde{g}^{-1}-Ad_{\tilde{g}}^{D}\alpha,\left(  \left(
\mathcal{B}_{e}+\mathcal{G}_{e}\right)  +\tilde{\pi}(\tilde{g})\right)
^{-1}(\partial_{+}\tilde{g}\tilde{g}^{-1}+Ad_{\tilde{g}}^{D}\alpha
)\right\rangle \label{Lag dual}%
\end{equation}
where $\tilde{\pi}(\tilde{g})$ is the Poisson bivector of $G^{\ast}$. Using
the fact that $\tilde{\pi}(\tilde{g})=\tilde{\pi}(\tilde{g}e^{x\alpha})$ since
$\left(  \ref{Eq: Condicion Poisson}\right)  $ is satisfied, then this
lagrangian can be expressed in terms of the \emph{open monondromic string}
variable $\tilde{m}=\tilde{g}e^{x\alpha},$ $x\in\lbrack0,2\pi],$%
\[
\mathcal{L}_{\alpha}(\tilde{m},\frac{d\tilde{m}}{dt})=\left\langle
\partial_{-}\tilde{m}\tilde{m}^{-1},\left(  \left(  \mathcal{B}_{e}%
+\mathcal{G}_{e}\right)  +\tilde{\pi}(\tilde{m})\right)  ^{-1}\partial
_{+}\tilde{m}\tilde{m}^{-1}\right\rangle
\]

\begin{description}
\item[Remark:] \textit{This lagrangian corresponds to the one given in}
\textit{\cite{Monodromic}, but in our case we know by construction that the
dynamics preserves the monodromy (it restricts to }$\left(  \tilde{\mu
}_{0,\alpha}^{\varphi}\right)  ^{-1}\left(  \mathcal{O}_{\Gamma_{\mathrm{k}%
},0}(\alpha,1)\right)  \subset$\textit{ }$LG^{\ast}\times L\mathfrak{g}$
\textit{when the initial value lies there) and no further constraints are
needed.}
\end{description}

In the above case, for each $\alpha$ we have a dual model $\mathcal{L}%
_{\alpha}$. One may ask if we can glue all this models together making a
unique $L$ which is dual to the one in $G$ for any $\alpha.$ Note that to that
end, we should consider $\alpha$ as varying in a space of all possible ($log$
of) monodromies $t$ (e.g.: in a torus $\mathfrak{t}$ inside a compact group
$K$ \cite{Monodromic}) and then consider a structure on $LG^{\ast}\times
L\mathfrak{g}\times\{\alpha\}$ giving the correct dynamics.

If $\alpha\in\mathfrak{t}$ is considered as a variable, the correct dynamics
is given by one which sets $\alpha=const$. To that end, we enlarge the phase
space from $LG^{\ast}\times L\mathfrak{g}$ to the (also symplectic) $\left(
LG^{\ast}\times L\mathfrak{g}\times T^{\ast}\mathfrak{t},\ \tilde{\omega}%
_{o}(\tilde{g},Z)\oplus\omega_{o}(\alpha,\lambda)\right)  $. There, we define
\begin{align*}
\tilde{\mu}  &  :LG^{\ast}\times L\mathfrak{g}\times T^{\ast}{\mathfrak{t}%
}\longrightarrow L{\mathfrak{d}}\\
&  :(\tilde{g},Z,\alpha,\lambda)\mapsto\tilde{g}^{\prime}\tilde{g}%
^{-1}+Ad_{\tilde{g}}^{D}(Z+\alpha)
\end{align*}
which is still a Poisson map. The action it generates is the same on
$LG^{\ast}\times L\mathfrak{g}$, it is trivial on $\mathfrak{t}$ and is non
trivial on $\mathfrak{t}^{\ast}.$ Now, the image of $\tilde{\mu}$ covers all
the orbits $\mathcal{O}_{\Gamma_{\mathrm{k}},0}(\alpha,1)$ for $\alpha
\in\mathfrak{t}.$The lagrangian associated to the (now singular since
$\mathsf{h}\circ\tilde{\mu}$ does not depend on $\lambda$) hamiltonian system
$(LG^{\ast}\times L\mathfrak{g}\times T^{\ast}\mathfrak{t},\tilde{\omega}%
_{o}\oplus\omega_{o},\mathsf{h}\circ\tilde{\mu})$ is%
\[%
\begin{array}
[c]{l}%
\mathcal{L}(\tilde{g},\frac{\partial\tilde{g}}{\partial t},\alpha
,\frac{\partial\alpha}{\partial t},\lambda)\\
\\
=\left\langle \left(  \partial_{-}\tilde{g}\right)  \tilde{g}^{-1}%
-Ad_{\tilde{g}}^{D}\alpha,\left(  \left(  \mathcal{B}_{e}+\mathcal{G}%
_{e}\right)  +\tilde{\pi}(\tilde{g})\right)  ^{-1}(\left(  \partial_{+}%
\tilde{g}\right)  \tilde{g}^{-1}+Ad_{\tilde{g}}^{D}\alpha)\right\rangle
+\left\langle \lambda,\dfrac{\partial\alpha}{\partial t}\right\rangle
\end{array}
\]
where in the last added term $\lambda$ plays the role of a Lagrange
multiplier, and so the dynamics for the new variables is%
\begin{align*}
\frac{\partial\alpha}{\partial t}  &  =0\\
\frac{\partial\lambda}{\partial t}  &  =\nabla_{\alpha}L_{\alpha}.
\end{align*}

\begin{description}
\item[Example:] \emph{(Lu-Weinstein doubles)}\textit{ Let }$K$\textit{ be a
real simple and compact Lie group. Then }$D=AN\times K$\textit{ where
}$G=AN\;$\textit{and }$G^{\ast}=K$\textit{ are the subgroups given by the
Iwasawa decomposition of }$K^{\mathbb{C}}$\textit{. Now, let }$T\subset
K$\textit{ be a maximal torus and }$\mathfrak{t}$\textit{ its Lie algebra. By
choosing a constant }$\alpha\in\mathfrak{t}$\textit{ and a particular
hamiltonian on }$L\mathfrak{d}_{c}^{\ast}$\textit{, we obtain a resulting
model in phase space }$(LG^{\ast}\times L\mathfrak{g},\tilde{\omega}_{o}%
)$\textit{. This model can be expressed in terms of }$\tilde{g}(x).e^{x\alpha
}$\textit{ exclusively, yielding a monodromic strings model similar to that of
}\cite{Monodromic}\textit{. This follows from the above considerations
because, in this case, }$\tilde{b}=e^{x\alpha}$\textit{ and }$\tilde{g}%
h\tilde{b}=(\tilde{g}\tilde{b})(h^{\tilde{b}})$\textit{ (see Appendix }%
$1$\textit{). The resulting lagrangian is }$\left(  \ref{Lag dual}\right)
$\textit{ (compare to the ad hoc constrained one of }\cite{Monodromic}%
\textit{).}
\end{description}

\section{Conclusions}

\label{sec: conclusions}

We carried out an enlargement of the $T$-duality scheme developed in
\cite{Ale-Hugo} in order to include coadjoint orbits with non trivial
monodromy as pivotal phase space. To that end, we considered a general
framework for studying duality between different phase spaces which share the
same symmetry group $\mathrm{H}$. Solutions corresponding to collective
dynamics become dual in the sense that they are generated by the same curve in
$\mathrm{H}$. Explicit examples of dual phase spaces in the above sense were
constructed on the cotangent bundles of the factors of a double Lie group
$\mathrm{H}=\mathrm{N}\Join\mathrm{N}^{\ast}$.

When considering duality over non trivial extended orbits $\mathcal{O}%
_{c,0}\left(  \alpha,1\right)  $ with $\alpha\in\mathfrak{n}^{\ast}$, some
important new facts appeared, being the most significative the loss of the
symmetry between the role played by the factors of the double Lie group. Also,
a condition on $\alpha$, eq. $\left(  \ref{loop-22a}\right)  $, has to be
imposed in order for the momentum maps intersecting in $\mathcal{O}%
_{c,0}\left(  \alpha,1\right)  $ to be compatible with the underlying
hamiltonian structure (i.e. to be Poisson maps). In the loop group case,
standard sigma models are now T-dually related to models with non trivial
$\alpha$-monodromies, namely open string models \cite{Monodromic}, as showed
in subsection \ref{non trivial monodromy models}\textit{. }It is worth to
remark that, in the present framework, the dynamics of open string models are
monodromy preserving by construction, no further constraints need to be added.

On the other hand, since a non trivial orbit becomes related to a trivial one
by changing the cocycle extension by coboundary, we introduced a second
pivotal vertex in the T-duality scheme. This trivial orbit can be regarded as
the phase space chiral modes of a WZNW type model, with shifted collective
lagrangian, and allows for additional collective models on the cotangent
bundles of the factors $\mathrm{N}$ and $\mathrm{N}^{\ast}$ in the T-duality
class of those hanged from the non trivial monodromy orbit.

Thus, we succeeded in to generalize the symplectic geometry approach to
Poisson Lie T-duality of ref. \cite{Ale-Hugo}, stressing the fundamental role
played by coadjoint orbits of double Lie groups, central extensions and
collective dynamics, also analyzing some hamiltonian an lagrangian models on
the involved phase spaces.

\section{Appendix 1}

We are going to study the structure of the brackets in the double Lie algebra
for $\mathfrak{\tilde{g}}$ being a simple compact real Lie algebra (e.g.
$su(n)$). Recall from \cite{Chari-Pressley}, that if $(H_{i},e_{i},f_{i})$
with $i=1,...,rank(\mathfrak{\tilde{g}})$ denotes the elements of the
Chevalley basis of $\mathfrak{\tilde{g}}$ for a fixed Borel subalgebra
$\mathfrak{b},$ then%
\begin{align}
\lbrack H_{i},H_{j}] &  =0\label{eq en g}\\
\lbrack H_{i},e_{j}] &  =a_{ij}e_{j}\nonumber\\
\lbrack H_{i},f_{j}] &  =-a_{ij}f_{j}\nonumber
\end{align}
where $(a_{ij})$ denotes the Cartan matrix of $\mathfrak{\tilde{g}}$. Note
that $span\{H_{i}\}$ is the abelian Lie algebra of a maximal torus
$T\subset\tilde{G}$. The standard bialgebra structure $\delta:\mathfrak{\tilde
{g}}\longrightarrow\mathfrak{\tilde{g}}\wedge\mathfrak{\tilde{g}}$ on
$\mathfrak{\tilde{g}}$ is defined by
\begin{align}
\delta(H_{i}) &  =0\label{cocorchete}\\
\delta(e_{i}) &  =d_{i}H_{i}\wedge e_{i}\nonumber\\
\delta(f_{i}) &  =d_{i}H_{i}\wedge f_{i}\nonumber
\end{align}
with $d_{i}$ being the length of the $i$-th. root.

Now, let $(H^{i},e^{i},f^{i})$ denote the basis of $\mathfrak{g}%
=\mathfrak{\tilde{g}}^{\ast}$ dual to $(H_{i},e_{i},f_{i})$. We want to
express the commutation relations in terms of this dual basis of the bracket
on $\mathfrak{g}$ induced by $\delta$. Using $\left(  \ref{eq en g}\right)  $,
$\left(  \ref{cocorchete}\right)  $ and the definition
\[
\left\langle \lbrack X,Y],\alpha\right\rangle =\left\langle X\otimes
Y,\delta(\alpha)\right\rangle
\]
for all $X,Y\in\mathfrak{g}$, $\alpha\in\mathfrak{\tilde{g}}$, it is easy to
see that in $\mathfrak{g}$
\[
\left\langle \lbrack X,Y],H_{i}\right\rangle =0.
\]
Thus, in the double $\mathfrak{d}=\mathfrak{\tilde{g}}\oplus\mathfrak{g}$, we
have that%
\[
\lbrack(H_{i},0),(0,X)]_{\mathfrak{d}}=(0,ad_{H_{i}}^{\ast}X)
\]
hence, if $\tilde{b}=exp(\Sigma_{i}c_{i}H_{i})\in T\subset\tilde{G}$, $h\in
G$,%
\[
\tilde{b}\cdot h=h^{\tilde{b}}\cdot\tilde{b}%
\]
so the dressing action of $h$ on $\tilde{b}$ is \emph{trivial}, i. e.
$\tilde{b}^{h}=\tilde{b}$.

\section{Appendix 2}

\label{lagrang}

Here we give some details on the algebra involved in passing from a 1$^{st.}$
order lagrangian to a 2$^{nd.}$ order \emph{sigma model lagrangian}.

Let us start with a 1$^{st.}$ order lagrangian, $\dot{q},q%
%TCIMACRO{\U{b4}}%
%BeginExpansion
\acute{}%
%EndExpansion
\in\tilde{V},$ $p\in V,$%
\[
L(q,p)=\left\langle p,\dot{q}\right\rangle -\frac{1}{2}\left\langle p+q%
%TCIMACRO{\U{b4}}%
%BeginExpansion
\acute{}%
%EndExpansion
,\mathcal{E}(p+q%
%TCIMACRO{\U{b4}}%
%BeginExpansion
\acute{}%
%EndExpansion
)\right\rangle
\]
for $\mathcal{E}$ linear operator on $\tilde{V}\oplus$ $V\ $satisfying
$\mathcal{E}^{2}=Id$ and being self adjoint with respect to the pairing
$\left\langle ,\right\rangle $. From these conditions, it is easy to see that
the operator $\mathcal{G}^{-1}:=\rho_{\tilde{V}}\mathcal{E}\rho_{V}%
:V\longrightarrow\tilde{V}$ is invertible and that%
\[
\left\langle \mathcal{G}\tilde{v},\tilde{w}\right\rangle =\left\langle
\tilde{v},\mathcal{G}\tilde{w}\right\rangle
\]
$\tilde{v},\tilde{w}\in\tilde{V}$. Also, $\mathcal{B}:=(\rho_{V}%
\mathcal{E}\rho_{V})\circ(\rho_{V^{\ast}}\mathcal{E}\rho_{V})^{-1}:\tilde
{V}\longrightarrow V$ satisfies%
\[
\left\langle \mathcal{B}\tilde{v},\tilde{w}\right\rangle =-\left\langle
\tilde{v},\mathcal{B}\tilde{w}\right\rangle
\]
with $\rho_{V},\ \rho_{\tilde{V}}$ denoting the projections on $V\ $and
$\tilde{V}$, respectively. We also have the relations%
\begin{align*}
\rho_{\tilde{V}}\mathcal{E}\rho_{\tilde{V}}  &  =-\mathcal{G}^{-1}%
\mathcal{B}\\
\rho_{V}\mathcal{E}\rho_{\tilde{V}}  &  =\mathcal{G}-\mathcal{BG}%
^{-1}\mathcal{B}.
\end{align*}

Euler-Lagrange equations for the variable $p$ imply%
\begin{align*}
\frac{\partial L}{\partial p}  &  =0\\
\dot{q}  &  =\mathcal{E}(p+q%
%TCIMACRO{\U{b4}}%
%BeginExpansion
\acute{}%
%EndExpansion
)
\end{align*}
thus,
\[
p=\mathcal{G}\dot{q}+\mathcal{B}q%
%TCIMACRO{\U{b4}}%
%BeginExpansion
\acute{}%
%EndExpansion
.
\]
Hence, using the above equations to rewrite the lagrangian in terms of
$\dot{q},q%
%TCIMACRO{\U{b4}}%
%BeginExpansion
\acute{}%
%EndExpansion
$ we obtain%
\begin{align*}
L(q,\dot{q})  &  =\frac{1}{2}\left\langle \dot{q},\mathcal{G}\dot
{q}\right\rangle -\frac{1}{2}\left\langle q%
%TCIMACRO{\U{b4}}%
%BeginExpansion
\acute{}%
%EndExpansion
,\mathcal{G}q%
%TCIMACRO{\U{b4}}%
%BeginExpansion
\acute{}%
%EndExpansion
\right\rangle -\left\langle q%
%TCIMACRO{\U{b4}}%
%BeginExpansion
\acute{}%
%EndExpansion
,\mathcal{B}\dot{q}\right\rangle \\
&  =\frac{1}{2}\left\langle \dot{q}-q%
%TCIMACRO{\U{b4}}%
%BeginExpansion
\acute{}%
%EndExpansion
,(\mathcal{G}+\mathcal{B})(\dot{q}+q%
%TCIMACRO{\U{b4}}%
%BeginExpansion
\acute{}%
%EndExpansion
)\right\rangle
\end{align*}
the sigma model like 2$^{nd.}$ order lagrangian.

\subsection*{Acknowledgments}

H.M. and A.C. thanks to CONICET for financial support.

%\section{References}

\end{document}